\documentclass[aps,prd,twocolumn,10pt,groupedaddress]{revtex4-1}
\usepackage{amssymb}
\usepackage{graphicx}
\usepackage{amsmath}
\usepackage{hyperref}
\usepackage{subfigure}
\usepackage{float}
\usepackage[utf8]{inputenc}

\begin{document}

\title{The GWs from the S-stars revolving around the SMBH at Sgr A*}
\author{Rong-Gen Cai}
\email{cairg@itp.ac.cn}
\author{Tong-Bo Liu}
\email{liutongbo@itp.ac.cn}
\author{Shao-Jiang Wang}
\email{schwang@itp.ac.cn}
\affiliation{CAS Key Laboratory of Theoretical Physics, Institute of Theoretical Physics, Chinese Academy of Sciences, Beijing 100190, China}
\affiliation{School of Physical Sciences, University of Chinese Academy of Sciences, No.19A Yuquan Road, Beijing 100049, P.R. China}
\date{\today}

\begin{abstract}
A preliminary estimation of gravitational waves (GWs) from the extreme-mass-ratio-inspirals (EMRIs) system in the Galactic Centre (GC) is given for the 37 observed S-stars revolving around the supermassive black hole (SMBH) at Sagittarius (Sgr) A*. Within this century, the total strain of the gravitational waveform calculated from the post-Newtonian (PN) method with eccentricity is well below the current planned sensitivity of pulsar-timing-array (PTA). New technology might be required in order to extract GW signal from this EMRIs system for future PTA detections.
\end{abstract}
\maketitle

\section{Introduction}

There is a considerable observational evidence for the existence of a supermassive black hole (SMBH) with roughly four million solar masses at Sagittarius (Sgr) A* in the Galactic Centre (GC) \citep[For review, see][]{Eckart:2017bhq}, supported by the observed orbits of many S-stars from close monitoring over the past few decades \citep[See, for example,][]{1607.05726}. It has been expected in \cite{Iorio:2017med,Grould:2017bsw} that around 2018 a full data of 16-year-period orbit of the star S2 would reveal much information on relativistic effect. It turns out that, the recent result \cite{Abuter:2018drb} from the GRAVITY Collaboration has eliminated the pure Newtonian dynamics from S2 data, which has also been used for testing some modified gravity theories in the previous literatures \citep[See, for example,][]{Borka:2013dba,Capozziello:2014rva,Zakharov:2016lzv,Zakharov:2018efo}. Therefore, our Milky Way centre is a valuable play yard for the test of new physics \cite{Alexander:2005jz}. The current means to explore our GC are electromagnetic-wave (EMW) observations, for example, the Event Horizon Telescope (EHT) Very Long Baseline Interferometry (VLBI) observations and the infrared interferometry experiment GRAVITY at the Very Large Telescope Interferometer (VLTI) and other experiments like the X-ray observations. 

Meanwhile, the coming era of gravitational-wave (GW) astronomy \cite{Abbott:2016blz,TheLIGOScientific:2017qsa} provides us a new way approaching the center of Milky Way. A stellar-mass object revolving around a Sgr A* SMBH could be served as an extreme-mass-ratio-inspirals (EMRIs) system in the GC \cite{AmaroSeoane:2007aw}. Compact object like dwarfs, neutron stars and stellar BHs are not subject to tidal disruption, therefore they can reach very close to the SMBH and thus emit GW signals during the last coalescence phase of merger \cite{Barack:2003fp,Gair:2004iv,Hopman:2006qr,Freitag:2006qf,Hopman:2006xn}, which can be detected by the future space-borne GW detectors like LISA \cite{Audley:2017drz}, Taiji \cite{Taiji,Cai:2017cbj,Guo:2018npi}, DECIGO \cite{Kawamura:2006up} and BBO \cite{Corbin:2005ny}. Detections of GWs events from such EMRIs system could probe the spacetime metric around the centre SMBH, which gives rise to the independent measurements on the mass and spin \cite{Berry:2012im}. The modified gravity theories can also be constrained by measuring any deviation from Kerr metric through the peribothron as well as Lense-Thirring precession \cite{AmaroSeoane:2012tx}.

However, the current observations on the Sgr A* SMBH also give us many S-stars that are not compact enough to survive the tidal disruption \cite{Perets:2009jw,Madigan:2010pg,Antonini:2012jm}, therefore we can only detect such GW signals during current inspiral phase through pulsar-timing-array (PTA) observations, which to our best knowledge has not been specifically discussed in the literatures. Nevertheless, we still point out here some previous literatures addressing similar issues. In \cite{Freitag:2002nm}, it was claimed by simulatioins that there must be one to a few low-mass main sequence stars sufficiently bound to the Sgr A* SMBH to be conspicuous sources in LISA observations. In \cite{Kocsis:2011ch}, the prospect of detecting GWs background by PTA observations from the intermediate-mass-black-holes (IMBHs) and stellar BHs orbiting around Sgr A* SMBH was estimated, although we have yet no convincing evidence for identifying any of such BHs in GC. In \cite{Linial:2017hep}, it was shown that mass leakage occurs through the outer Lagrange point L2 could accelerate the time-evolution of the EMRIs system of a main-sequence star and a SMBH, resulting in a GW signal detectable by future space-borne GW detectors.

The arrangement of this paper is as follows: In Sec.\ref{sec:PN}, the post-Newtonian (PN) method with eccentricity is outlined for binary BHs (BBHs) with arbitrary mass ratio; In Sec.\ref{sec:GW}, the total strain of the gravitational waveforms from 37 observed S-stars is given. Sec.\ref{sec:con} is devoted to conclusion. For reference, the dataset and gravitational waveforms for each S-star are given in the Appendix \ref{subsec:stars} and \ref{subsec:GWs}, respectively.

\section{The GWs from eccentric BBHs system}\label{sec:PN}

In this section, we follow closely the paper \cite{Hinder:2008kv} \citep[See also][for the 0PN, 1PN, 1.5PN, 2PN, 3PN and 3.5PN results]{Peters:1964zz,Peters:1963ux,Wagoner:1976am,Blanchet:1989cu,1992MNRAS.254..146J,Blanchet:1993ec,Rieth:1997mk,SCHAFER1993196,0264-9381-12-4-009,Damour:1988mr,Will:1996zj,Gopakumar:1997bs,Gopakumar:2001dy,Damour:2004bz,Memmesheimer:2004cv,Konigsdorffer:2006zt,Arun:2007rg,Arun:2007sg,Arun:2009mc,Blanchet:2006zz} on the eccentric PN model since only the early-stage inspiral phase is considered in our case. In what follows, the 3PN conservative quasi-Keplerian orbit equations are solved with the solutions from the 2PN truncated adiabatic evolution equations of the orbital elements to construct the adiabatic PN waveform to leading Newtonian order.

The system under consideration is made up of two point particles of masses $m_1$ and $m_2$. The total mass , the reduced mass and the symmetric mass are denoted as $M=m_1+m_2$, $\mu=m_1m_2/M$ and $\eta=\mu/M$, respectively. The unit convention is $G=c=1$.

In Newtonian case, the conservations of energy $E$ and angular momentum $J$ can be expressed in terms of the eccentricity $e$ and mean motion $n$, which can be related to the orbital period $P$ and the semi-major axis $a$ by $n=2\pi/P=a^{-3/2}M^{1/2}$. Then the conservative orbital dynamics of the relative orbital radius $r$ and angular frequency $\dot{\phi}$ is given by
\begin{align}
r&=a(1-e\cos u);\\
\dot{\phi}&=\frac{n\sqrt{1-e^2}}{(1-e\cos u)^2},
\end{align}
where the eccentric anomaly $u$ satisfies the Kepler's equation $l=u-e\sin u$ with the mean anomaly $l$ given by $\dot{l}=n$. For constant $n$ and $e$, the above conservative orbital dynamics can be solved for the input initial conditions $\phi_0\equiv\phi(t_0)$ and $l_0\equiv l(t_0)$.

In the PN case, the quasi-Keplerian parametrization gives rise to three eccentricities, $e_t, e_r$, and $ e_{\phi}$ signifying the deviations from circlar motion in $t, r,$ and $\phi$. As stated in \cite{Hinder:2008kv}, it is sufficient to consider just $e_t$, or $e$ for short. In \cite{Hinder:2008kv}, it is also shown that the dubbed $x$-model can match the numerical relativity much better than the usual $n$-model.
Here $x\equiv (M\omega)^{2/3}$ with $\omega\equiv(2\pi+\Delta\phi)/P$, where $\Delta\phi$ is the angle of precession of the pericenter during one orbit of period $P$.

First, one solves the 2PN truncated adiabatic evolution equations of the orbital elements
\begin{align}\label{eq:xdot}
M\dot{x}=&\frac{2\eta}{15(1-e^2)^{7/2}}(96+292e^2+37e^4)x^5+\dot{x}_{\rm 1PN}x^{6}\notag\\
&+\dot{x}_{\rm 1.5PN}x^{13/2}+\dot{x}_{\rm 2PN}x^{7}+\mathcal{O}(x^{15/2}),
\end{align}
\begin{align}\label{eq:edot}
M\dot{e}=&\frac{-e\eta}{15(1-e^2)^{5/2}}(304+121e^2)x^4+\dot{e}_{\rm 1PN}x^{5}\notag\\
&+\dot{e}_{\rm 1.5PN}x^{11/2}+\dot{e}_{\rm 2PN}x^{6}+\mathcal{O}(x^{13/2}),
\end{align}
for the given initial conditions $x_0\equiv x(t_0)$ and $e_0\equiv e(t_0)$. Here the PN coefficients are given in the appendix of \cite{Hinder:2008kv}.

Then, one solves the 3PN conservative quasi-Keplerian orbit equations
\begin{align}\label{eq:r}
r/M=&(1-e\cos u)x^{-1}+r_{\rm 1PN}+r_{\rm 2PN}x+r_{\rm 3PN}x^2\notag\\
&+\mathcal{O}(x^3),
\end{align}
\begin{align}\label{eq:phidot}
M\dot{\phi}=&\frac{\sqrt{1-e^{2}}}{(1-e\cos u)^2}x^{3/2}+\dot{\phi}_{\rm 1PN} x^{5/2}+\dot{\phi}_{\rm 2PN} x^{7/2}\notag\\
&+\dot{\phi}_{\rm 3PN} x^{9/2}+\mathcal{O}(x^{11/2}),
\end{align}
\begin{align}\label{eq:u}
l=u-e\sin u+l_{\rm 2PN}x^2+l_{\rm 3PN}x^3+\mathcal{O}(x^4),
\end{align}
\begin{align}\label{eq:ldot}
M\dot{l}&=Mn=x^{3/2}+n_{\rm 1PN}x^{5/2}+n_{\rm 2PN}x^{7/2}\notag\\
&+n_{\rm 3PN}x^{9/2}+\mathcal{O}(x^{11/2}),
\end{align}
with input time-varying solutions $x(t)$ and $e(t)$. Here the PN coefficients are given in the appendix of \cite{Hinder:2008kv}. To be more specific, one first integrates \eqref{eq:ldot} to obtain $l(t)$ with integration constant $l_0\equiv l(t_0)$, and then one solves $u(t)$ from \eqref{eq:u} by root-finding with $l(t)$, and next one can compute $r(t)$ directly from \eqref{eq:r}, and $\dot{\phi}(t)$ directly from \eqref{eq:phidot}. At last, the $\dot{r}(t)$ and $\phi(t)$ can be obtained by numerically differentiating and integrating the $r(t)$ and $\dot{\phi}(t)$, respectively, with integration constant $\phi_0\equiv \phi(t_0)$.

Finally, with the solutions of $r(t)$ and $\dot{\phi}(t)$ (and also $\dot{r}(t), \phi(t)$) in hand, one can express the complex PN waveform strain to the leading Newtonian order as
\begin{align}
h=h_+-ih_{\times}
\end{align}
\begin{align}\label{eq:hplus}
h_+=&-\frac{M\eta}{R}\bigg\{(1+\cos^2\theta)\bigg[\cos2\phi^{\prime}\bigg(-\dot{r}^2+r^2\dot{\phi}^2+\frac{M}{r}\bigg)\notag\\
&+2r\dot{r}\dot{\phi}\sin2\phi^{\prime}\bigg]+\bigg(-\dot{r}^2-r^2\dot{\phi}^2+\frac{M}{r}\bigg)\sin^2\theta\bigg\},
\end{align}
\begin{align}\label{eq:hcross}
h_{\times}=&-\frac{2M\eta}{R}\cos\theta\bigg\{\bigg(-\dot{r}^2+r^2\dot{\phi}^2+\frac{M}{r}\bigg)\sin2\phi^{\prime}\notag\\
&-2r\dot{r}\dot{\phi}\cos2\phi^{\prime}\bigg\},
\end{align}
where $\phi^{\prime}=\phi-\varphi$, $R$ is the radial distance to the binary, and $\theta$ and $\varphi$ are the spherical polar angles of the observer in the orbital plane.

\section{The total GW strain from 37 S-stars}\label{sec:GW}

\begin{figure*}
\centering
\includegraphics[width=0.8\textwidth]{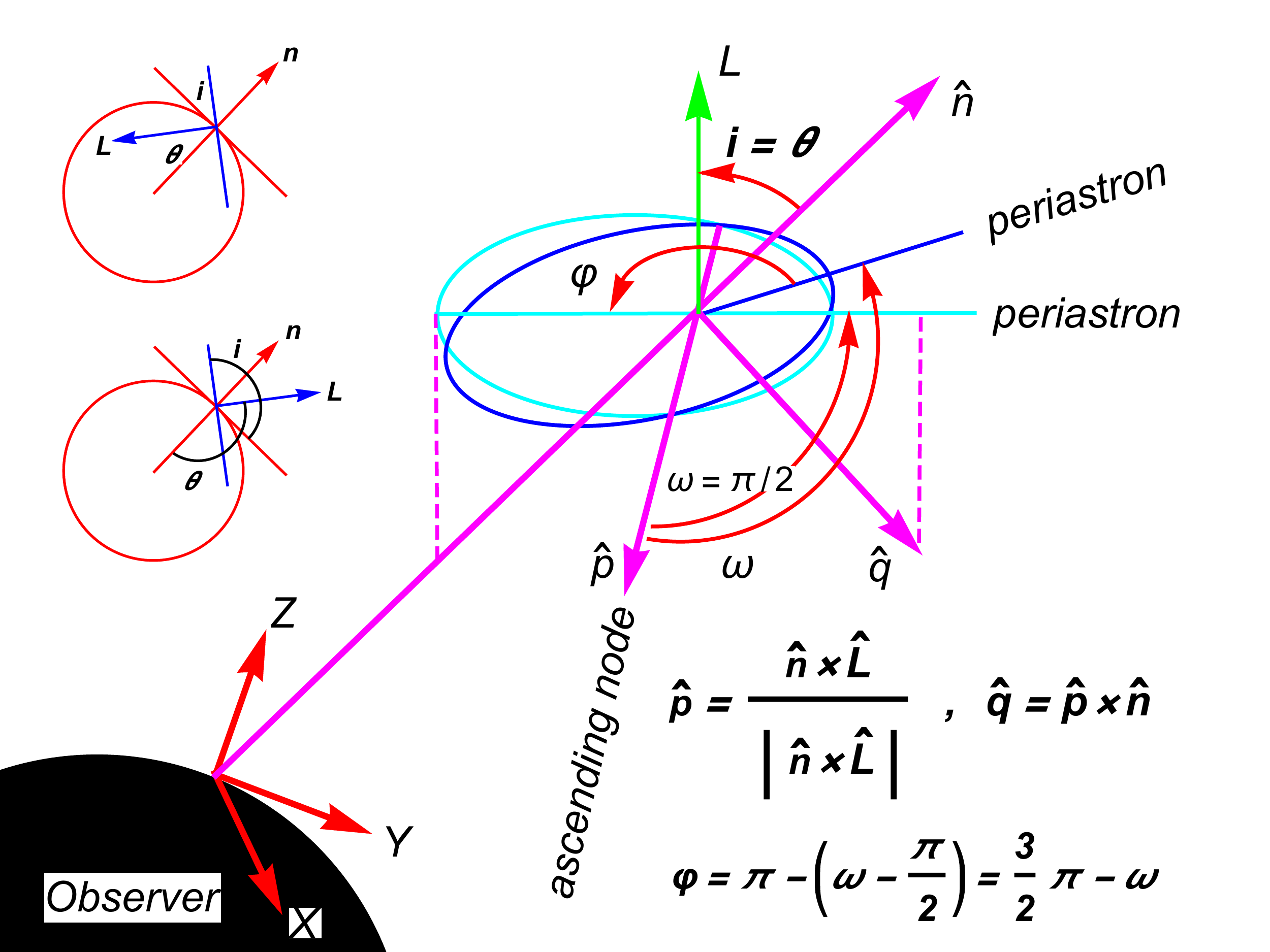}\\
\caption{An illustration of the conventions used in our calculations. The observation frame (red arrow lines) is the usual elliptical coordinate, where the direction of SMBH is along the unit vector $\hat{n}$. The source frame (magenta arrow lines) contains two polarization unit vectors $\hat{p}$ and $\hat{q}$ in addition to $\hat{n}$. $\hat{p}$ goes through the ascending node, which is the endpoint of the intersection line between the orbit plane and the tangent plane to the elliptical sphere. Two different orbits (cyan ellipse and blue ellipse) are shown with the same orbit angular momentum $L$ labeled by the green arrow line, whose inclination angle from the orbit plane to the tangent plane is labeled by $i$. The cyan ellipse is speciall since its minor axis is exactly along $\hat{p}$, thus the projections of both $\hat{n}$ and $\hat{q}$ on the cyan orbit plane exactly go through the periastron. In this case, the angle $\omega$ from the ascending node to the periastron is exactly $\pi/2$.  The blue ellipse with arbitrary angle $\omega$ is shown for a general orbit. The direction of observer is specified by the two polar anle $\theta$ and $\varphi$, which can be determined by $\theta=i$ and $\varphi=3\pi/2-\omega$ as shown from the illustration.}\label{fig:EMRIs}
\end{figure*}

\begin{figure*}
\includegraphics[width=0.49\textwidth]{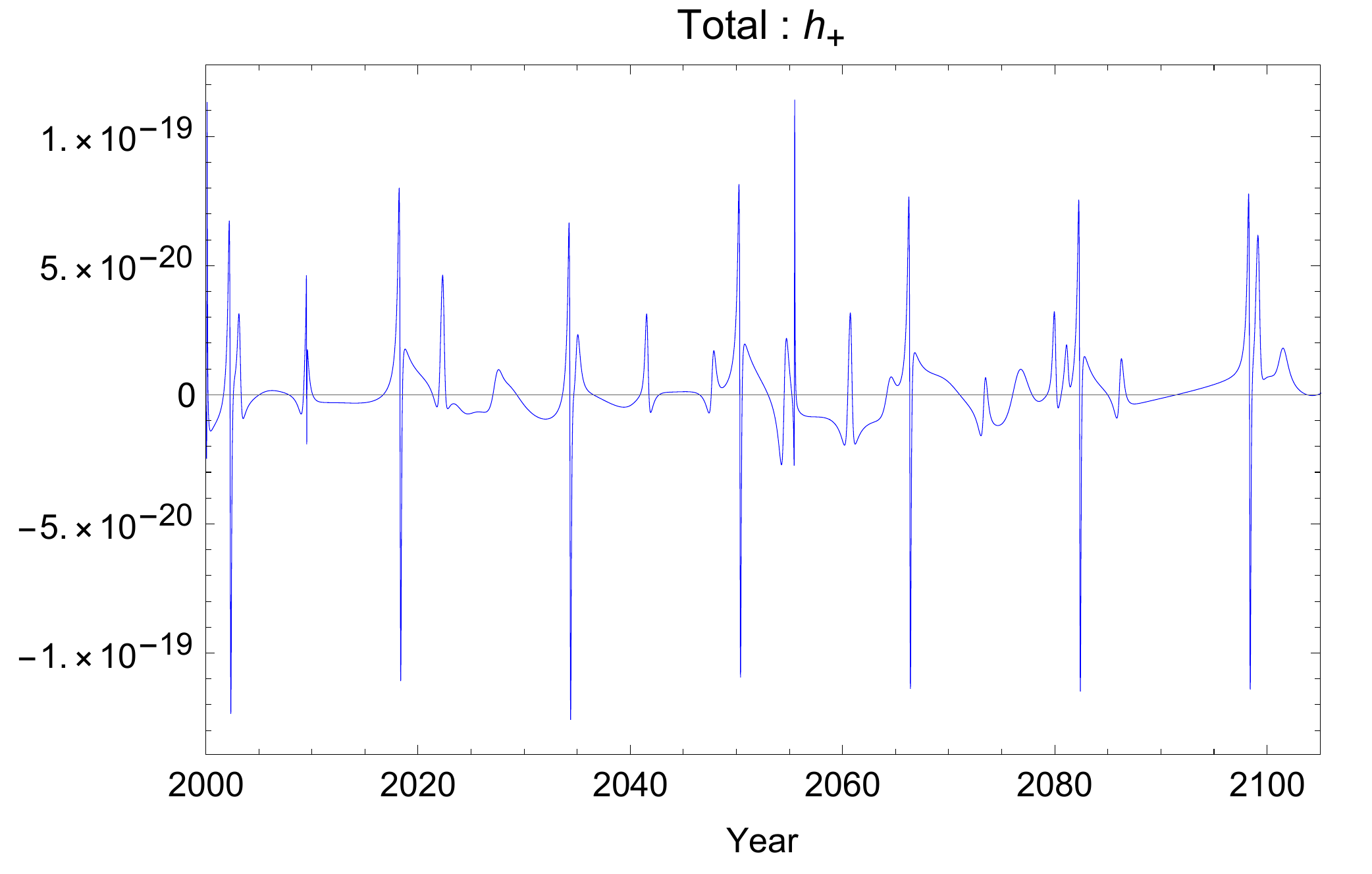}
\includegraphics[width=0.49\textwidth]{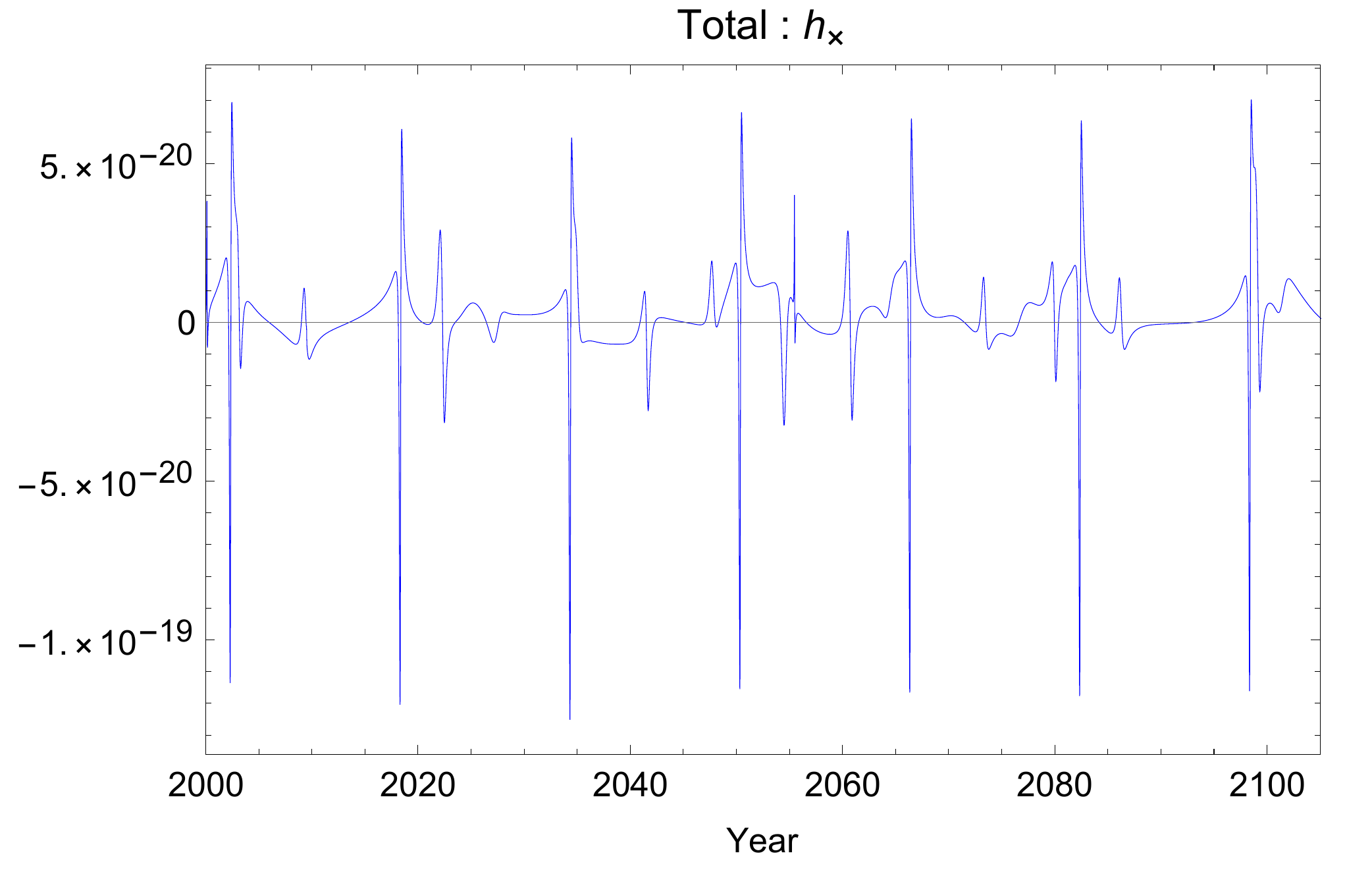}\\
\caption{The total amplitudes of the two polarizations for 37 observed S-stars orbiting around the Sgr A* SMBH in GC with respect to the observation year within this century.}\label{fig:htot}
\end{figure*}

The mass $M_{\rm SMBH}=4.02\times10^6\,M_\odot$ and distance $R=7.86\,\mathrm{kpc}$ of Sgr A* SMBH are adopted from \cite{1607.05726}. The data set of S-stars orbiting around the Sgr A* SMBH is taken from \cite{1611.09144}, where the semi-major axis $a$ [arcsec], the numerical eccentricity $e$, the inclination angle $i$ [degrees], the position angle of the ascending node $\Omega$ [degrees], the longitude of periastron $\omega$ [degrees], the epoch of periastron passage $t_P$ [year] and the K-band apparent magnitude $m_K$ are provided as referred here in the Table \ref{tab:stars}. The masses for each S-star are also not specified in \cite{1611.09144}, which will be estimated later in the last column of Table \ref{tab:stars}. 

The intial time is chosen at $t_0=t_P$ so that $l_0=0$ and $\phi_0=0$. The spherical polar angle $\theta$ of the observer in the orbital plane is simply equal to the inclination angle $i$, which measures the angle between the orbital plane and the plane tangential to the celestial sphere as shown in the upper left part of Fig.\ref{fig:EMRIs}. It is between $0^\circ$ and $90^\circ$ for counter-clockwise apparent motion, and between $90^\circ$ and $180^\circ$ for clockwise apparent motion. As explicitly shown in Fig.\ref{fig:EMRIs}, the other spherical polar angle is given by 
\begin{align}
\varphi=\pi\pm(\pm\frac{\pi}{2}\mp\omega)=\frac32\pi-\omega
\end{align}
with the upper sign for counter-clockwise apparent motion, and the lower sign for clockwise apparent motion. This simple relation is limited to the case without the rotation of SMBH \cite{Damour:2004bz} (See \cite{Barack:2003fp,Kraniotis:2006ux} for the general case). We hope to report elsewhere the general case of a spining SMBH.

To estimate the masses for each S-star, one can first combine the definition of bolometric magnitude 
\begin{align}
M=M_\odot^{\rm bol}-2.5\lg\frac{L}{L_\odot},\quad M_\odot^{\rm bol}=4.74
\end{align}
with the definition of K-band BC correction
\begin{align}
BC_K=m-K=M-M_K
\end{align}
to give rise to
\begin{align}
\lg\frac{L}{L_\odot}=-0.4(M_K+BC_K-M_\odot^{\rm bol}).
\end{align}
Then one can replace the bolometric magnitude $M$ with K-band apparent magnitude $m_K$ by
\begin{align}
m_K=M_K+5\lg\frac{D}{10\,\mathrm{pc}}+\delta_K,
\end{align}
where the distance could be chosen roughly as $D=8\,\mathrm{kpc}$ and the K-band extinction $\delta_K=2.4$. Next, one can also replace the BC correction $BC_K$ by the semiempirical relation adopted from \cite{Martins:2006vf},
\begin{align}
BC_K=A_K\lg T_{\rm eff}+B_K,\quad A_K=-7.24, B_K=28.8
\end{align}
where the effective temperature $T_{\rm eff}$ of star can be estimated by the Stepan's law $L\propto RT^4$, namely
\begin{align}
\lg\frac{L}{L_\odot}=2\lg\frac{R}{R_\odot}+4\lg\frac{T_\mathrm{eff}}{T_\mathrm{eff}^\odot},\quad T_\mathrm{eff}^\odot=5800\,\mathrm{K}.
\end{align}
Finally, after inserting the semiempirical relations
\begin{align}
\lg\frac{L}{L_\odot}&=A_L\lg\frac{M}{M_\odot}+B_L;\\
\lg\frac{R}{R_\odot}&=A_R\lg\frac{M}{M_\odot}+B_R,
\end{align}
one eventually obtains
\begin{align}
\lg\frac{M}{M_\odot}=km_K+b,
\end{align}
where
\begin{align}
k&=-\frac{2}{5a};\\
b&=\frac{1}{a}\left(B_L+\frac25\left(-5\lg\frac{D}{10\mathrm{pc}}-\delta_K-M_\odot^{\rm bol}\right.\right.\notag\\
&\left.\left.+A_K\left(\frac{B_L}{4}-\frac{B_R}{2}\right)+B_K+A_K\lg T_\odot\right)\right)
\end{align}
with
\begin{align}
a\equiv A_L+\frac25A_K\left(\frac{A_L}{4}-\frac{A_R}{2}\right)
\end{align}
The linear coefficients $k=-0.192$ and $b=3.885$ can be simply calibrated from a smaller data set \cite{1708.06353} by naively fitting the determined masses and K-band magnitudes. Although the mass refereed from the K-band magnitude is generally smaller than the true mass \cite{1708.06353}, however, the EMRIs system is actually dominated by the SMBH, whose mass is much larger than the mass of stars. Therefore the precise values in the last column of Table \ref{tab:stars} are not important since it makes little difference on the final results of gravitational wavefroms.

Using above data for the 37 observed S-stars orbiting around the Sgr A* SMBH, one can directly calculate the gravitational waveforms for each S-star as presented in the appendix \ref{subsec:GWs}. The GW frequency is at the order of decades, namely nHz band for PTA observations. The total gravitational waveforms for the amplitudes of the two polarizations are presented in Fig.\ref{fig:htot}, where the GW signal is dominated by the S2 star, but the total amplitude $10^{-20}\sim10^{-19}$ is well below the sensitivity of current planned PTA observations, say $10^{-15}\sim10^{-14}$ as reviewed in \cite{Hobbs:2017oam}.

\section{Conclusions}\label{sec:con}

Our GC may be the most important laboratory for testing the EMRIs system against new physics since there are much appealing evidences for the presence of a SMBH at Sgr A*. Previous studies have revealed the exciting possibility to detect the GW signals from the compact objects orbiting around this SMBH within the sensitivity range of future space-borne GW detectors like LISA, Taiji, DECIGO and BBO. In this paper, we explore the possibility that if one can detect the GW signal from the non-compact main sequence stars orbiting around the Sgr A* SMBH. The gravitational waveforms are explicitly calculate from an eccentric PN model for 37 observed S-stars, and the total amplitudes for each polarization are well below the sensitivity of current planned PTA observations, which might require new technology other than extending the observation time in order to extract GW signal from this EMRIs system for future PTA detections.

\begin{acknowledgments}
We would like to thank Xian Chen, Li-Wei Ji, You-Jun Lu for helpful discussion during the Workshop on Gravitational Wave Physics and Detection at the Institute of Theoretical Physics, Chinese Academy of Sciences, Beijing between 15-28th July, 2018. We also want to thank Li-Ming Cao, Wen-Biao Han for helpful discussion and Hu Bin and Qi Guo for helpful correspondences. This work is supported by the National Natural Science Foundation of China Grants No.11690022, No.11375247, No.11435006, and No.11647601, and by the Strategic Priority Research Program of CAS Grant No.XDB23030100 and by the Key Research Program of Frontier Sciences of CAS. We acknowledge the use of HPC Cluster of ITP-CAS.

\end{acknowledgments}

\section{Appendix}


\subsection{The data set of 37 S-stars}\label{subsec:stars}
\newpage

\begin{table*}[!ht]
\caption{The dataset of 37 S-stars we used in this paper for estimating the gravitational wavefrom adopted from \cite{1611.09144}, where the semi-major axis $a$ [arcsec], the numerical eccentricity $e$, the inclination angle $i$ [degrees], the position angle of the ascending node $\Omega$ [degrees], the longitude of periastron $\omega$ [degrees], the epoch of periastron passage $t_P$ [year], the K-band apparent magnitude $m_K$, and the star mass $m_\mathrm{star}$ referred from the K-band apparent magnitude are shown below.}
\label{tab:stars} 
{\scriptsize
\begin{tabular}{lcccccccccc}
Star& $a$['']& $e$& $i\,[^\circ]$& $\Omega\,[^\circ]$& $\omega\,[^\circ]$& $t_P$[yr]& $T$[yr]&$m_K$&$m_\mathrm{star}[M_\odot]$\\
\hline
S1&$0.595\pm0.024$&$0.556\pm0.018$&$119.14\pm0.21$&$342.04\pm0.32$&$122.3\pm1.4$&$2001.80\pm0.15$&$166.0\pm5.8$&14.7&11.36\\
S2&$0.1255\pm0.0009$&$0.8839\pm0.0019$&$134.18\pm0.40$&$226.94\pm0.60$&$65.51\pm0.57$&$2002.33\pm0.01$&$16.00\pm0.02$&13.95&15.84\\
S4&$0.3570\pm0.0037$&$0.3905\pm0.0059$&$80.33\pm0.08$&$258.84\pm0.07$&$290.8\pm1.5$&$1957.4\pm1.2$&$77.0\pm1.0$&14.4&12.98\\
S6&$0.6574\pm0.0006$&$0.8400\pm0.0003$&$87.24\pm0.06$&$85.07\pm0.12$&$116.23\pm0.07$&$2108.61\pm0.03$&$192.0\pm0.17$&15.4&8.33\\
S8&$0.4047\pm0.0014$&$0.8031\pm0.0075$&$74.37\pm0.30$&$315.43\pm0.19$&$346.70\pm0.41$&$1983.64\pm0.24$&$92.9\pm0.41$&14.5&12.42\\
S9&$0.2724\pm0.0041$&$0.644\pm0.020$&$82.41\pm0.24$&$156.60\pm0.10$&$150.6\pm1.0$&$1976.71\pm0.92$&$51.3\pm0.70$&15.1&9.52\\
S12&$0.2987\pm0.0018$&$0.8883\pm0.0017$&$33.56\pm0.49$&$230.1\pm1.8$&$317.9\pm1.5$&$1995.59\pm0.04$&$58.9\pm0.22$&15.5&7.97\\
S13&$0.2641\pm0.0016$&$0.4250\pm0.0023$&$24.70\pm0.48$&$74.5\pm1.7$&$245.2\pm2.4$&$2004.86\pm0.04$&$49.00\pm0.14$&15.8&6.98\\
S14&$0.2863\pm0.0036$&$0.9761\pm0.0037$&$100.59\pm0.87$&$226.38\pm0.64$&$334.59\pm0.87$&$2000.12\pm0.06$&$55.3\pm0.48$&15.7&7.29\\
S17&$0.3559\pm0.0096$&$0.397\pm0.011$&$96.83\pm0.11$&$191.62\pm0.21$&$326.0\pm1.9$&$1991.19\pm0.41$&$76.6\pm1.0$&15.3&8.71\\
S18&$0.2379\pm0.0015$&$0.471\pm0.012$&$110.67\pm0.18$&$49.11\pm0.18$&$349.46\pm0.66$&$1993.86\pm0.16$&$41.9\pm0.18$&16.7&4.68\\
S19&$0.520\pm0.094$&$0.750\pm0.043$&$71.96\pm0.35$&$344.60\pm0.62$&$155.2\pm2.3$&$2005.39\pm0.16$&$135\pm14$&16.&6.39\\
S21&$0.2190\pm0.0017$&$0.764\pm0.014$&$58.8\pm1.0$&$259.64\pm0.62$&$166.4\pm1.1$&$2027.40\pm0.17$&$37.00\pm0.28$&16.9&4.29\\
S22&$1.31\pm0.28$&$0.449\pm0.088$&$105.76\pm0.95$&$291.7\pm1.4$&$95\pm20$&$1996.9\pm10.2$&$540\pm63$&16.6&4.90\\
S23&$0.253\pm0.012$&$0.56\pm0.14$&$48.0\pm7.1$&$249\pm13$&$39.0\pm6.7$&$2024.7\pm3.7$&$45.8\pm1.6$&17.8&2.88\\
S24&$0.944\pm0.048$&$0.8970\pm0.0049$&$103.67\pm0.42$&$7.93\pm0.37$&$290\pm15$&$2024.50\pm0.03$&$331\pm16$&15.6&7.62\\
S29&$0.428\pm0.019$&$0.728\pm0.052$&$105.8\pm1.7$&$161.96\pm0.80$&$346.5\pm5.9$&$2025.96\pm0.94$&$101.0\pm2.0$&16.7&4.68\\
S31&$0.449\pm0.010$&$0.5497\pm0.0025$&$109.03\pm0.27$&$137.16\pm0.30$&$308.0\pm3.0$&$2018.07\pm0.14$&$108.\pm1.2$&15.7&7.29\\
S33&$0.657\pm0.026$&$0.608\pm0.064$&$60.5\pm2.5$&$100.1\pm5.5$&$303.7\pm1.6$&$1928\pm12$&$192.0\pm5.2$&16.&6.39\\
S38&$0.1416\pm0.0002$&$0.8201\pm0.0007$&$171.1\pm2.1$&$101.06\pm0.24$&$17.99\pm0.25$&$2003.19\pm0.01$&$19.2\pm0.02$&17.&4.10\\
S39&$0.370\pm0.015$&$0.9236\pm0.0021$&$89.36\pm0.73$&$159.03\pm0.10$&$23.3\pm3.8$&$2000.06\pm0.06$&$81.1\pm1.5$&16.8&4.48\\
S42&$0.95\pm0.18$&$0.567\pm0.083$&$67.16\pm0.66$&$196.14\pm0.75$&$35.8\pm3.2$&$2008.24\pm0.75$&$335\pm58$&17.5&3.28\\
S54&$1.20\pm0.87$&$0.893\pm0.078$&$62.2\pm1.4$&$288.35\pm0.70$&$140.8\pm2.3$&$2004.46\pm0.07$&$477\pm199$&17.5&3.28\\
S55&$0.1078\pm0.0010$&$0.7209\pm0.0077$&$150.1\pm2.2$&$325.5\pm4.0$&$331.5\pm3.9$&$2009.34\pm0.04$&$12.80\pm0.11$&17.5&3.28\\
S60&$0.3877\pm0.0070$&$0.7179\pm0.0051$&$126.87\pm0.30$&$170.54\pm0.85$&$29.37\pm0.29$&$2023.89\pm0.09$&$87.1\pm1.4$&16.3&5.59\\
S66&$1.502\pm0.095$&$0.128\pm0.043$&$128.5\pm1.6$&$92.3\pm3.2$&$134\pm17$&$1771\pm38$&$664\pm37$&14.8&10.87\\
S67&$1.126\pm0.026$&$0.293\pm0.057$&$136.0\pm1.1$&$96.5\pm6.4$&$213.5\pm1.6$&$1705\pm22$&$431\pm10$&12.1&35.97\\
S71&$0.973\pm0.040$&$0.899\pm0.013$&$74.0\pm1.3$&$35.16\pm0.86$&$337.8\pm4.9$&$1695\pm21$&$346\pm11$&16.1&6.11\\
S83&$1.49\pm0.19$&$0.365\pm0.075$&$127.2\pm1.4$&$87.7\pm1.2$&$203.6\pm6.0$&$2046.8\pm6.3$&$656\pm69$&13.6&18.50\\
S85&$4.6\pm3.30$&$0.78\pm0.15$&$84.78\pm0.29$&$107.36\pm0.43$&$156.3\pm6.8$&$1930.2\pm9.8$&$3580\pm2550$&15.6&7.62\\
S87&$2.74\pm0.16$&$0.224\pm0.027$&$119.54\pm0.87$&$106.32\pm0.99$&$336.1\pm7.7$&$611\pm154$&$1640\pm105$&13.6&18.50\\
S89&$1.081\pm0.055$&$0.639\pm0.038$&$87.61\pm0.16$&$238.99\pm0.18$&$126.4\pm4.0$&$1783\pm26$&$406\pm27$&15.3&8.71\\
S91&$1.917\pm0.089$&$0.303\pm0.034$&$114.49\pm0.32$&$105.35\pm0.74$&$356.4\pm1.6$&$1108\pm69$&$958\pm50$&12.2&34.41\\
S96&$1.499\pm0.057$&$0.174\pm0.022$&$126.36\pm0.96$&$115.66\pm0.59$&$233.6\pm2.4$&$1646\pm16$&$662\pm29$&10.&91.24\\
S97&$2.32\pm0.46$&$0.35\pm0.11$&$113.0\pm1.3$&$113.2\pm1.4$&$28\pm14$&$2132\pm29$&$1270\pm309$&10.3&79.88\\
S145&$1.12\pm0.18$&$0.50\pm0.25$&$83.7\pm1.6$&$263.92\pm0.94$&$185\pm16$&$1808\pm58$&$426\pm71$&17.5&3.28\\
S175&$0.414\pm0.039$&$0.9867\pm0.0018$&$88.53\pm0.60$&$326.83\pm0.78$&$68.52\pm0.40$&$2009.51\pm0.01$&$96.2\pm5.0$&17.5&3.28\\
\end{tabular}
}
\end{table*}

\subsection{The waveforms of 37 S-stars}\label{subsec:GWs}
\newpage

\begin{figure*}
\includegraphics[width=0.4\textwidth]{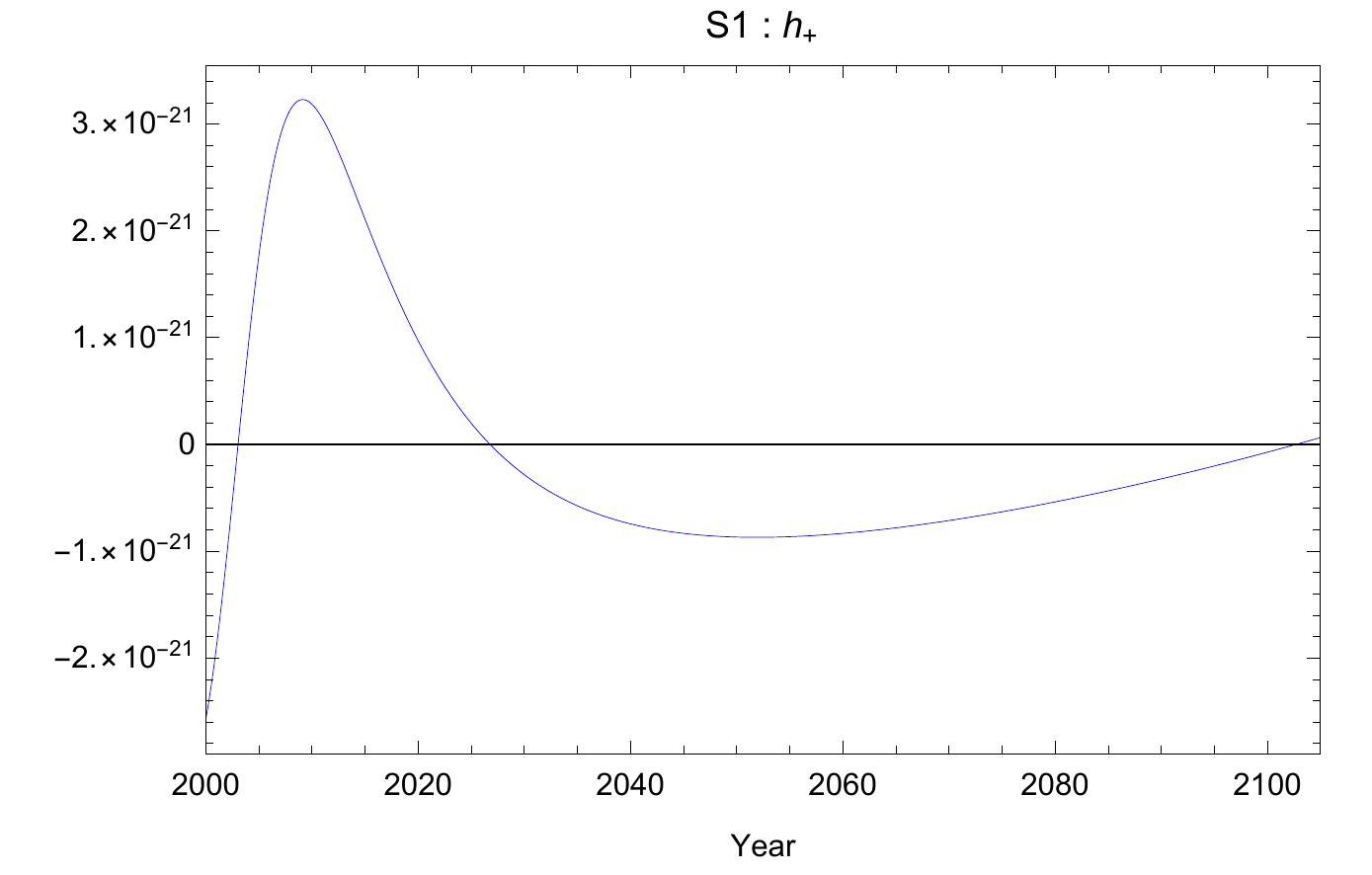}
\includegraphics[width=0.4\textwidth]{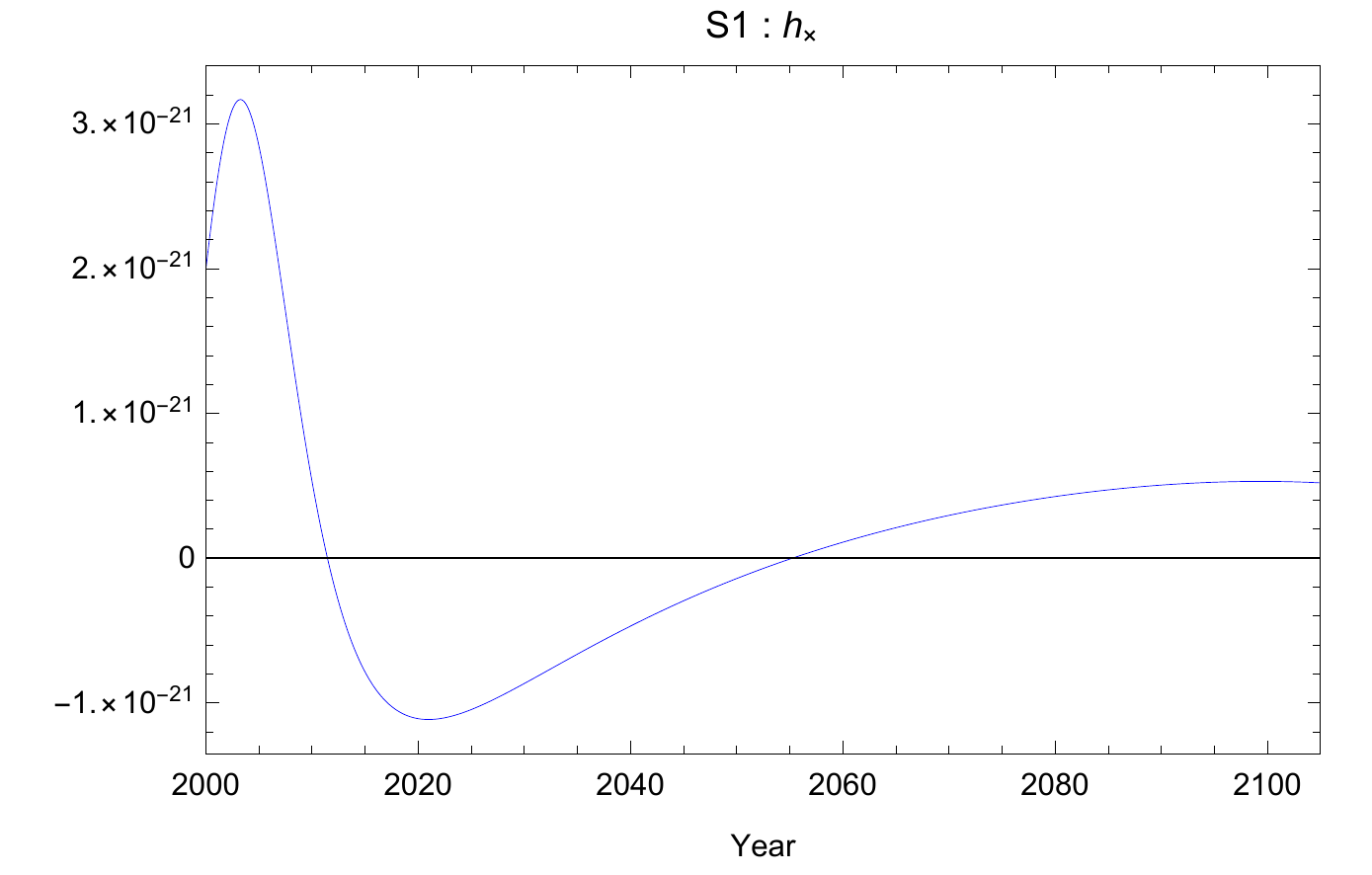}\\
\includegraphics[width=0.4\textwidth]{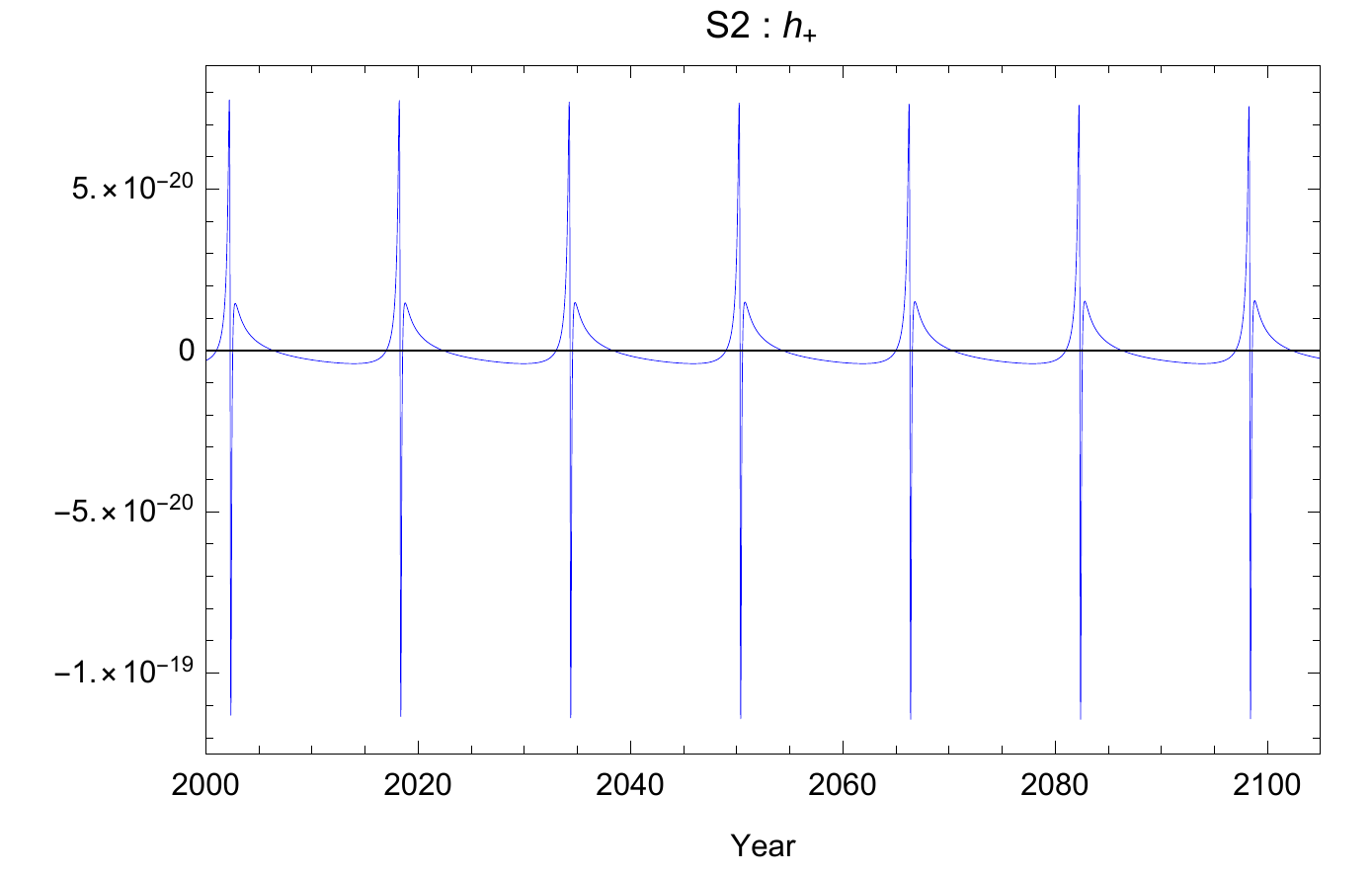}
\includegraphics[width=0.4\textwidth]{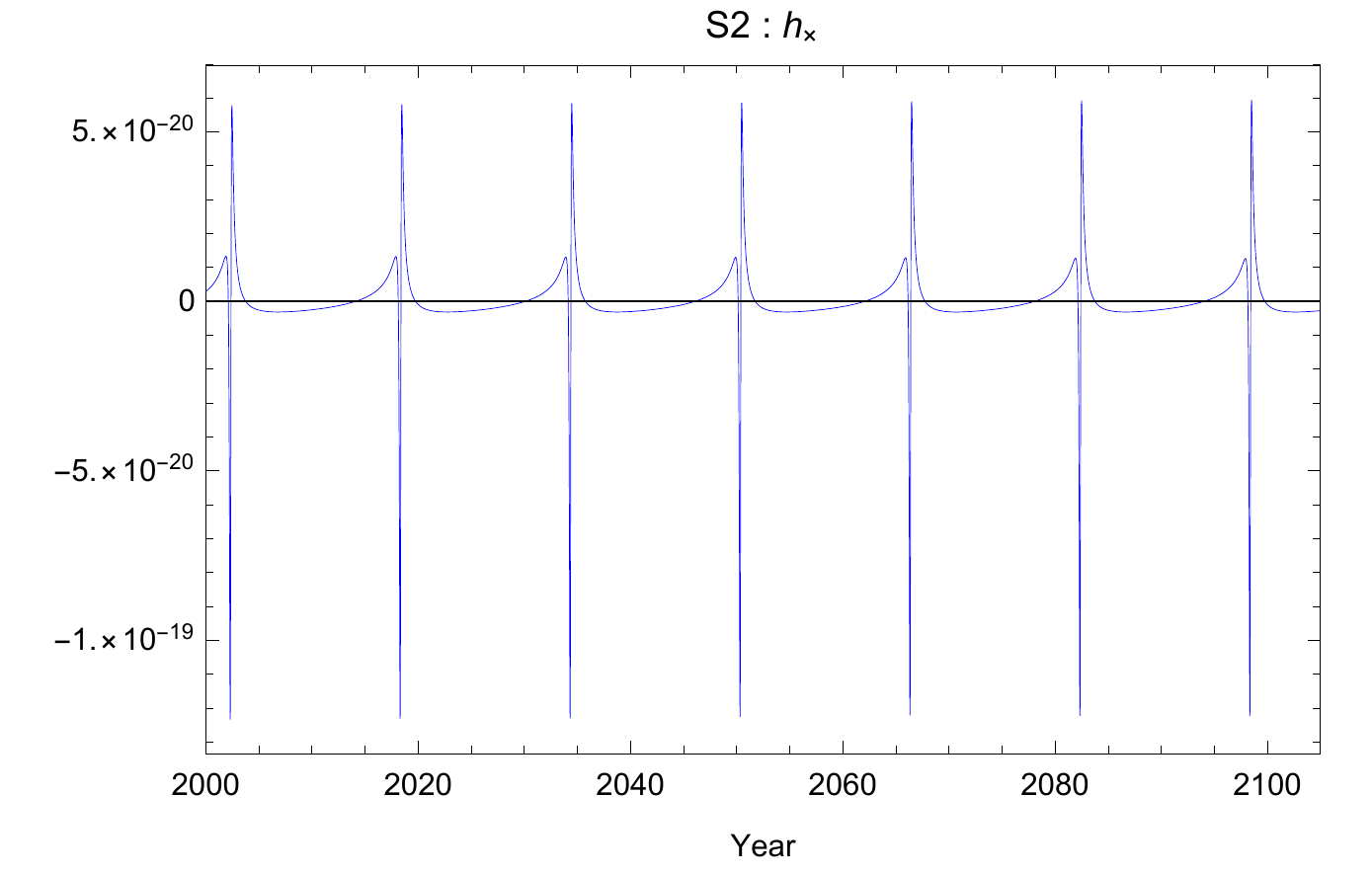}\\
\includegraphics[width=0.4\textwidth]{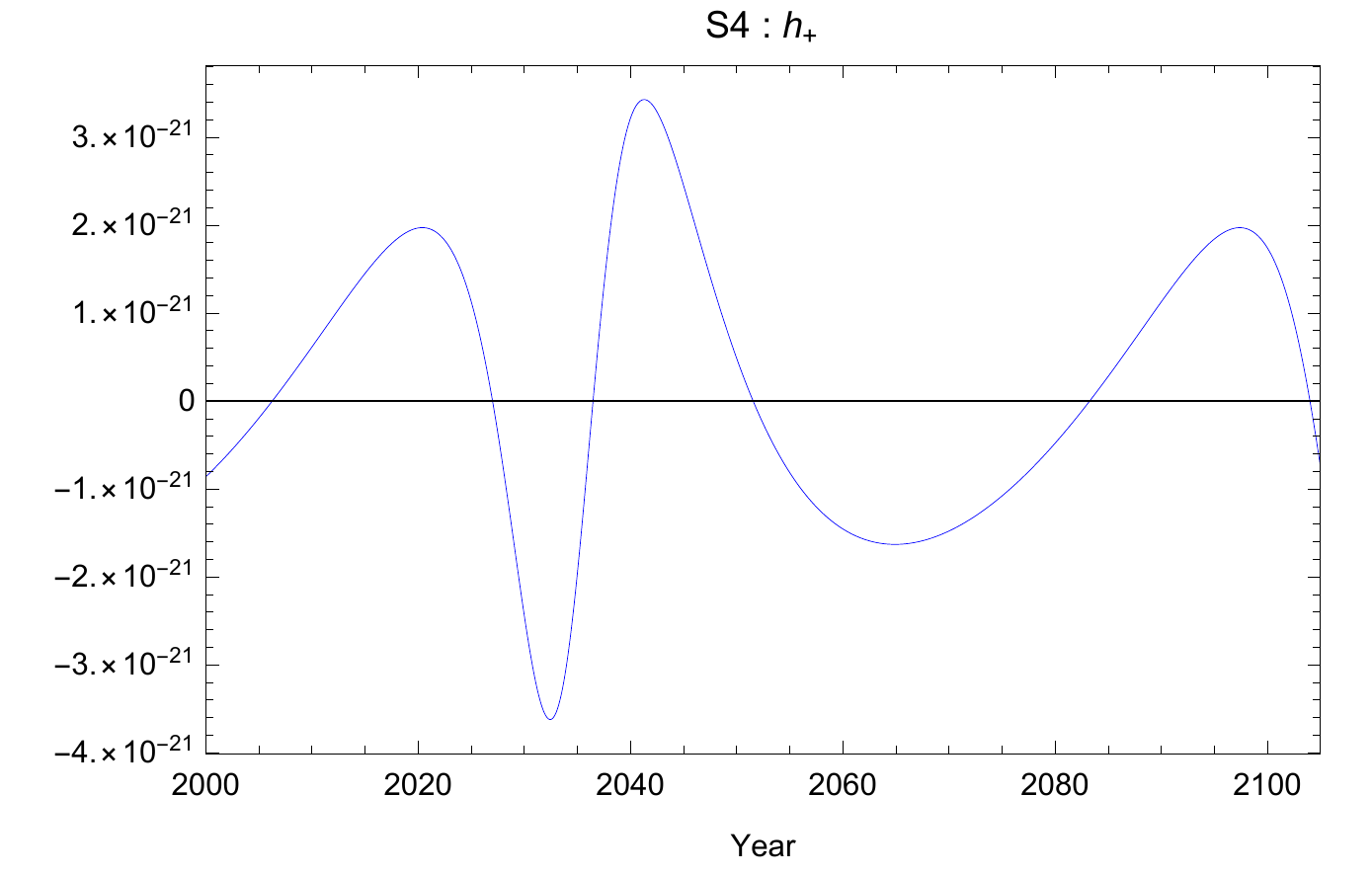}
\includegraphics[width=0.4\textwidth]{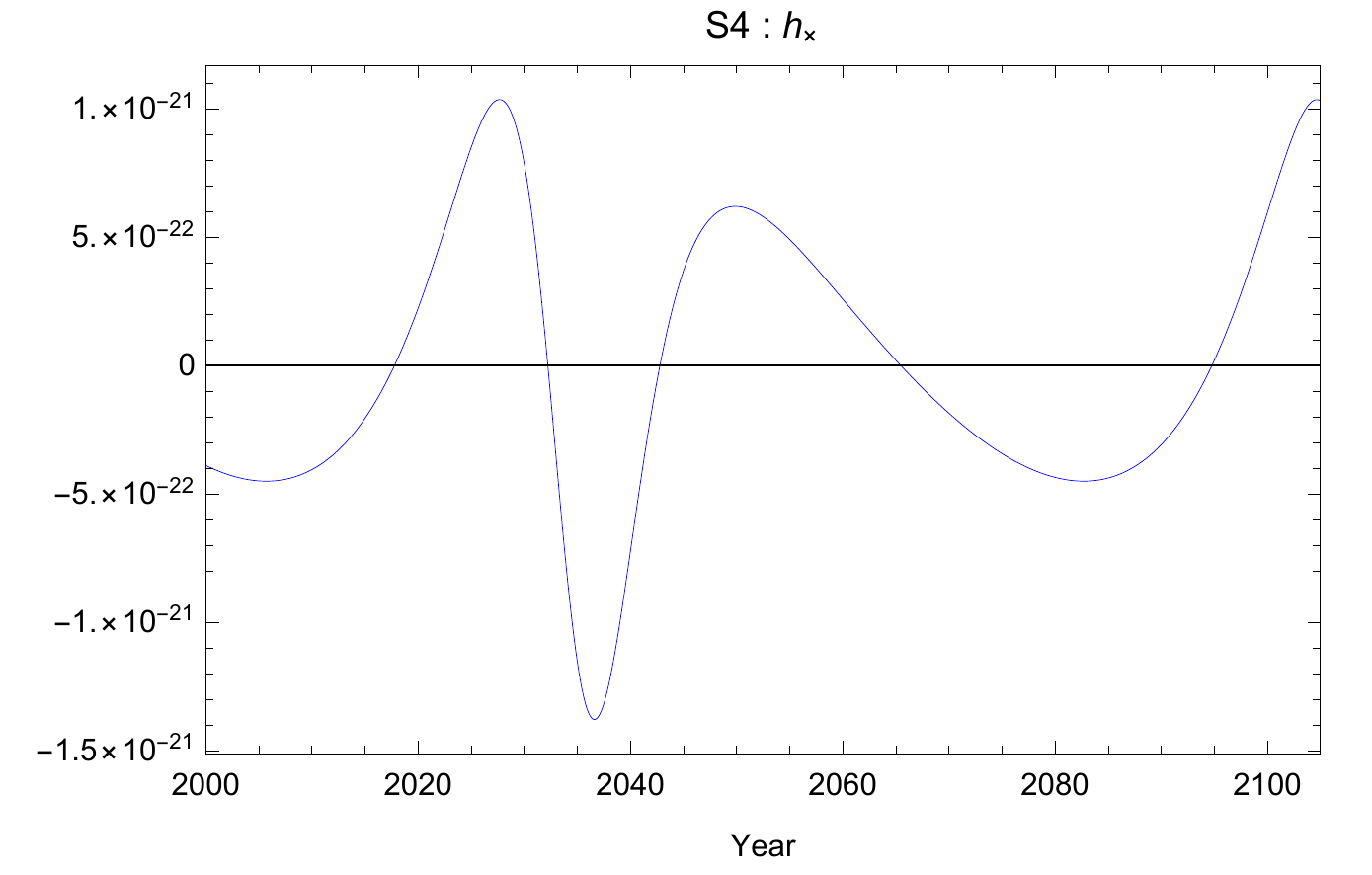}\\
\includegraphics[width=0.4\textwidth]{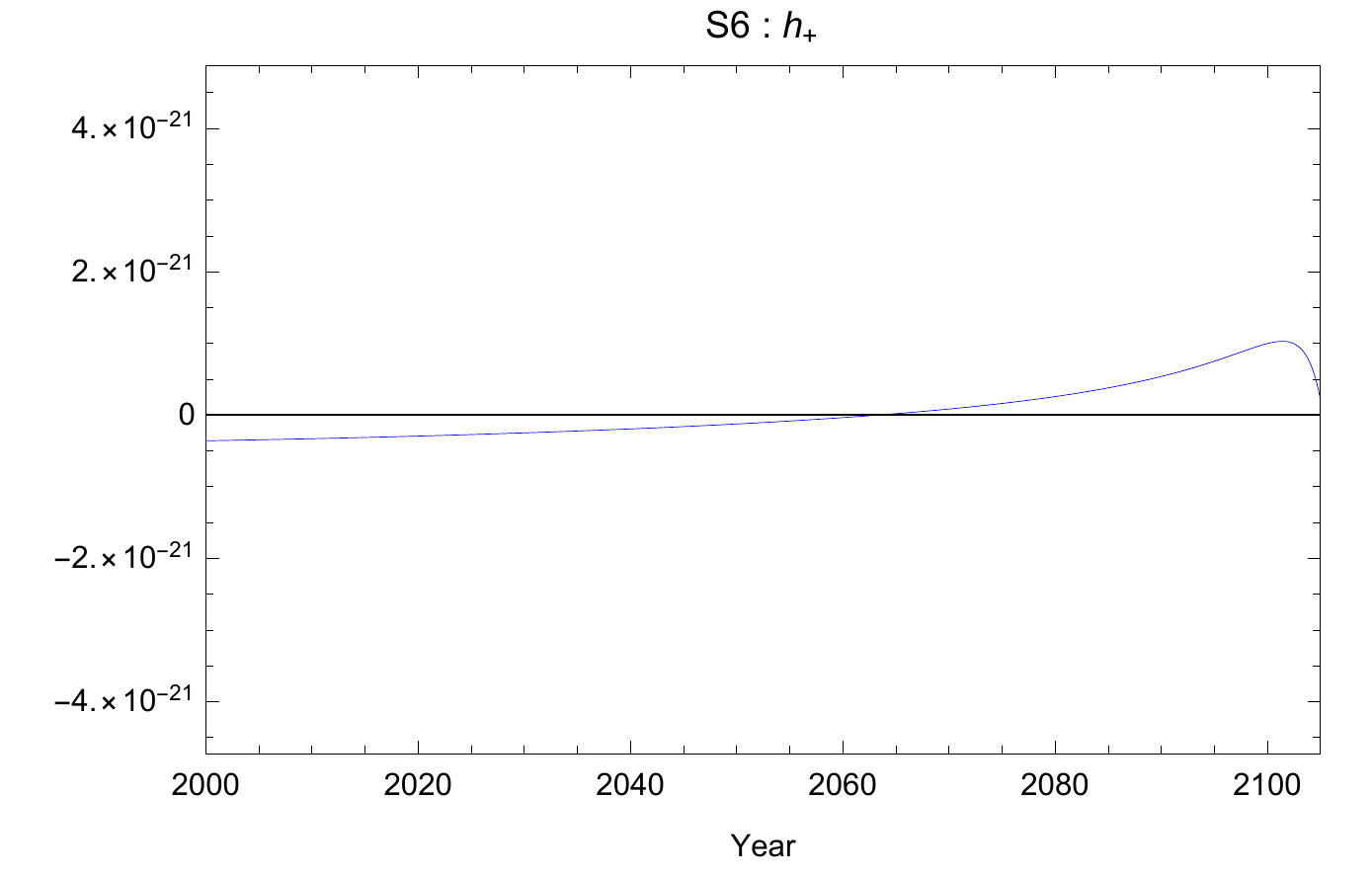}
\includegraphics[width=0.4\textwidth]{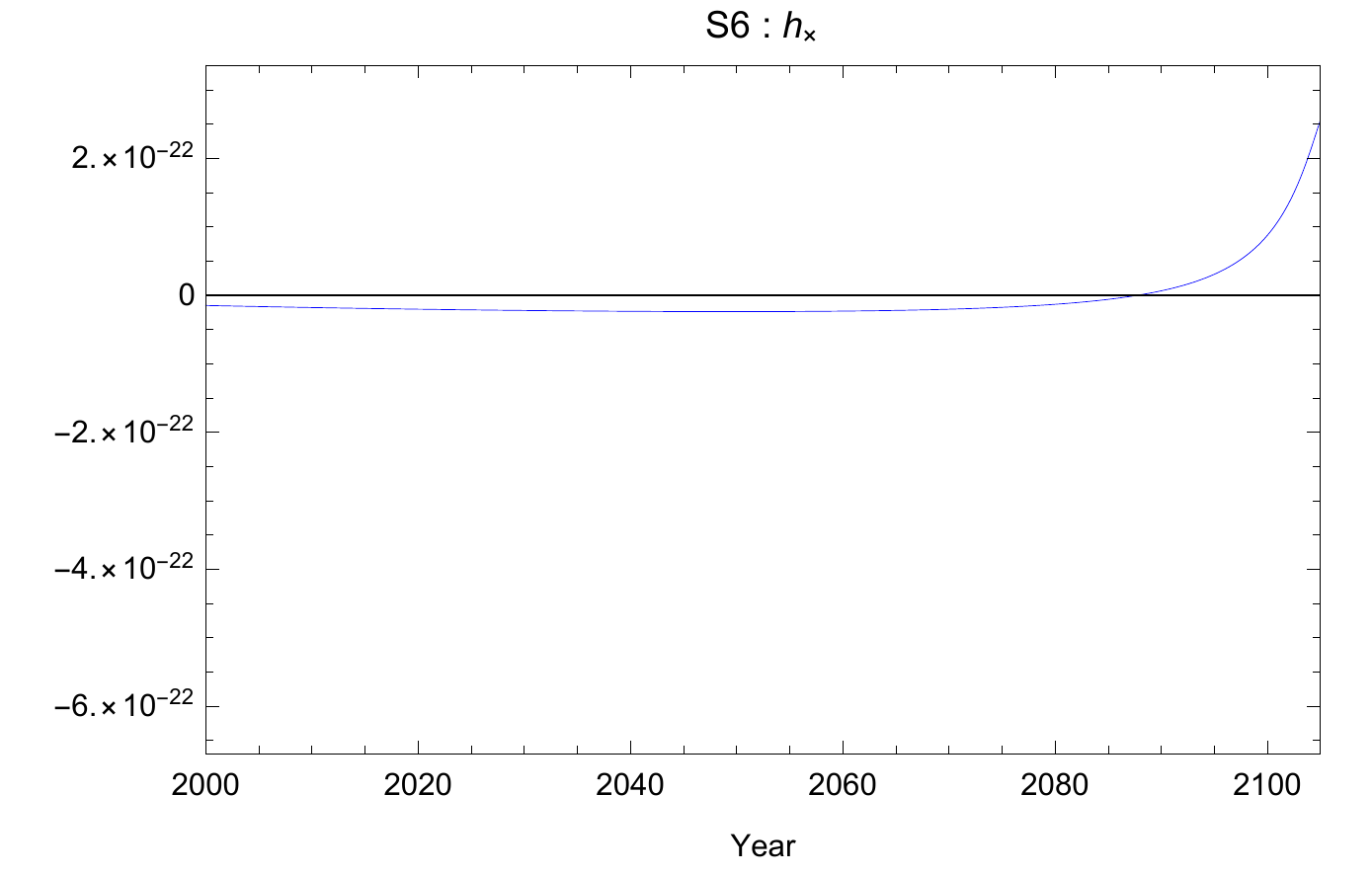}\\
\includegraphics[width=0.4\textwidth]{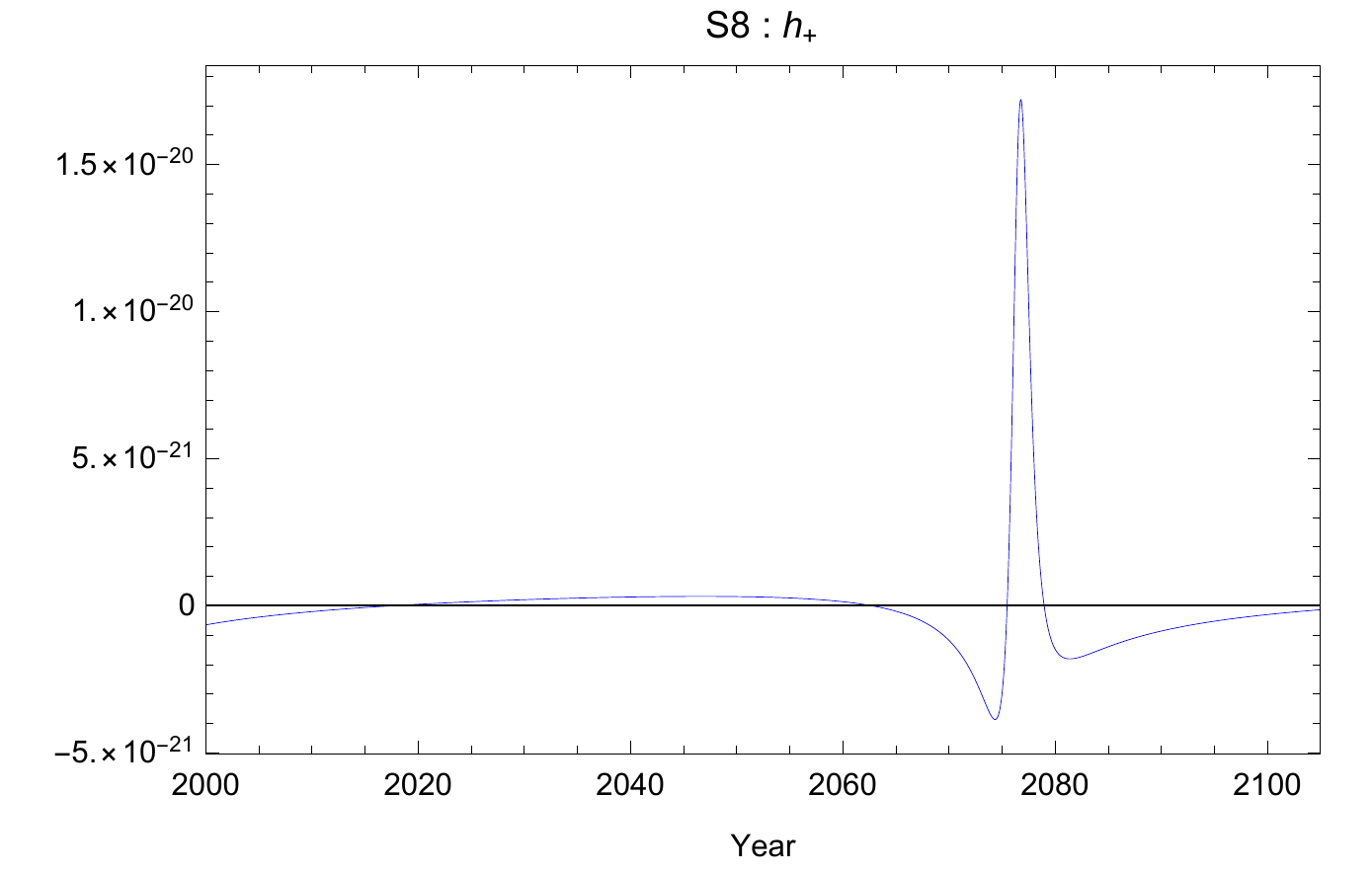}
\includegraphics[width=0.4\textwidth]{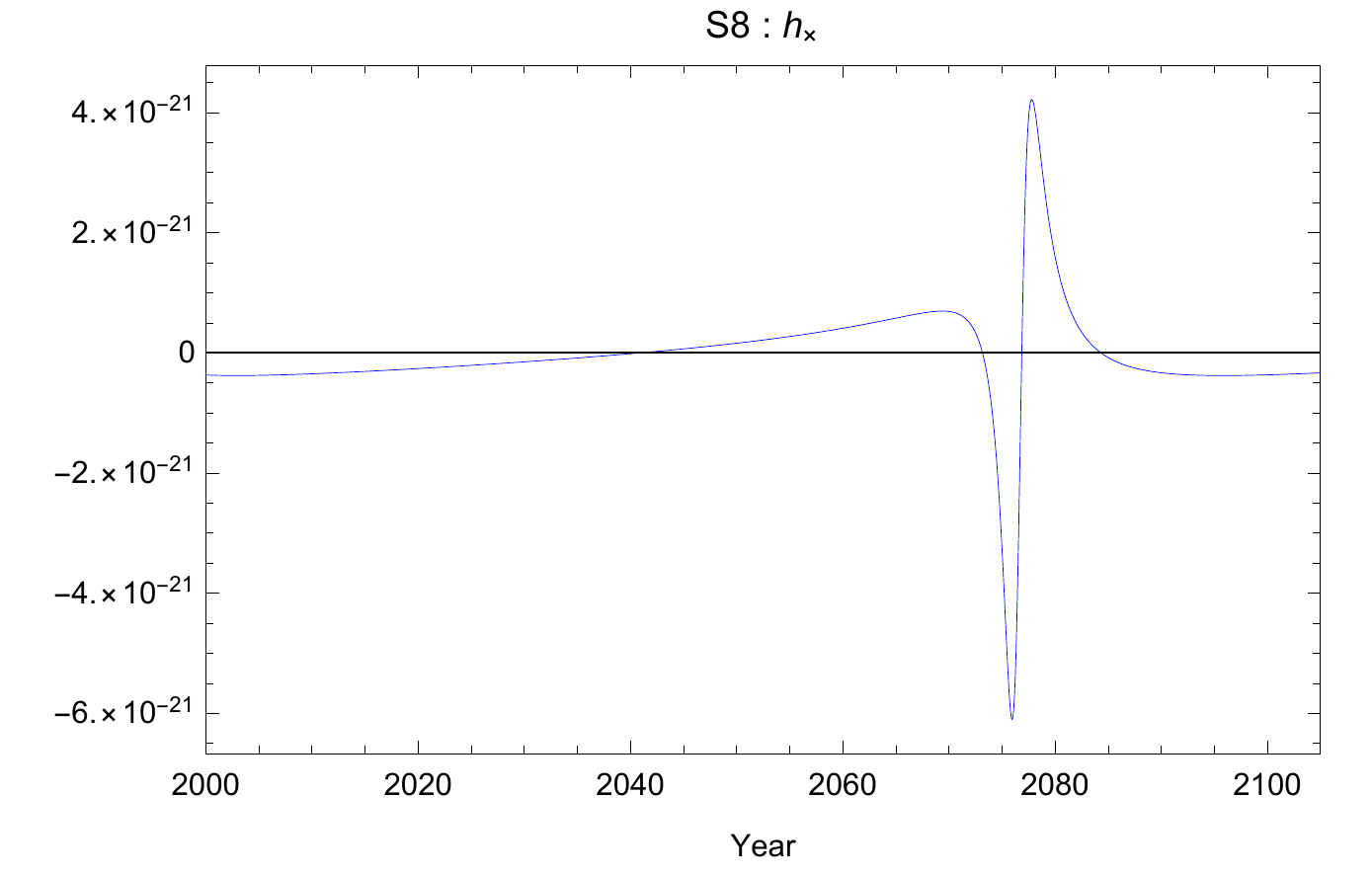}
\end{figure*} 

\begin{figure*}
\includegraphics[width=0.4\textwidth]{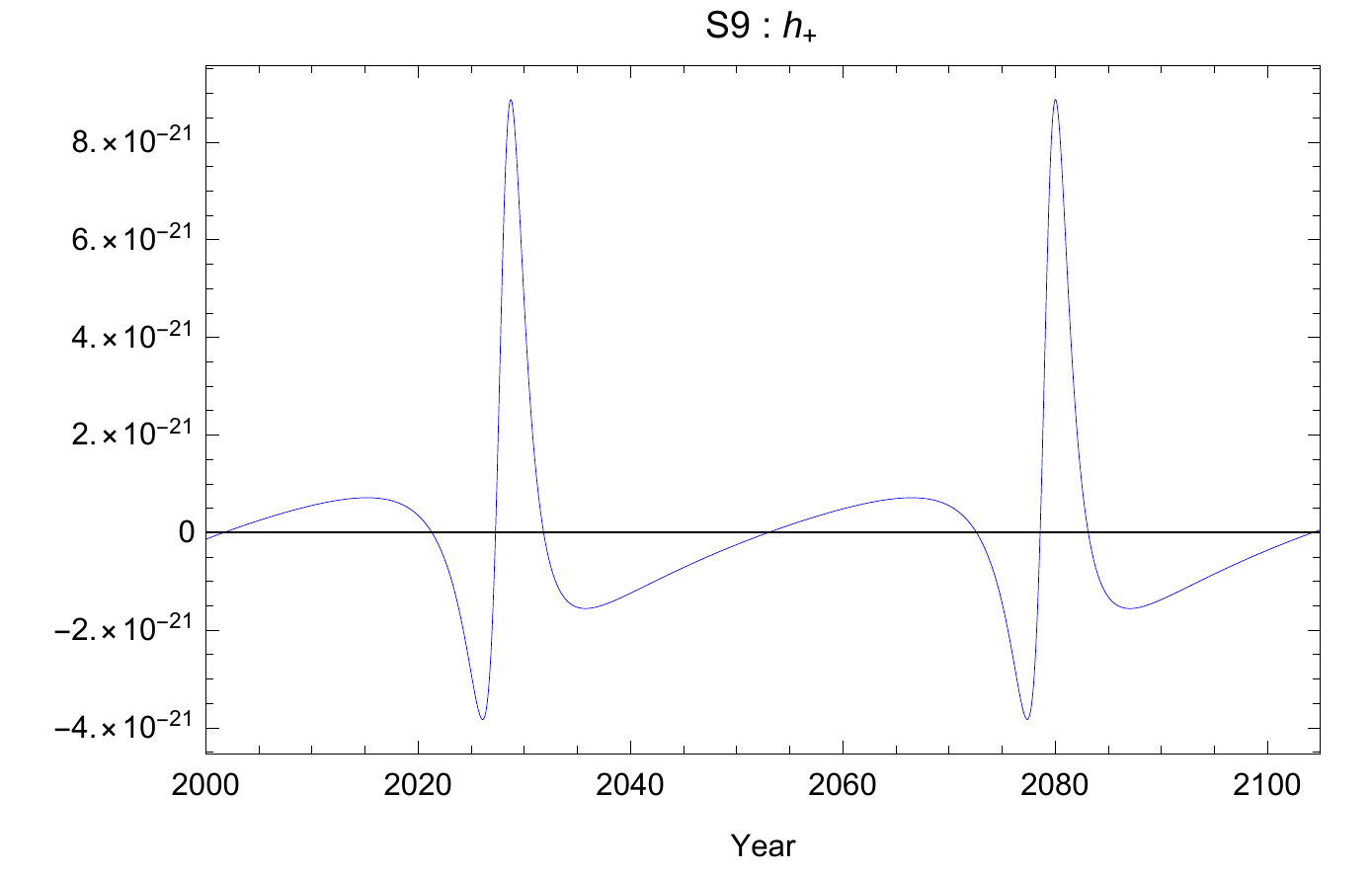}
\includegraphics[width=0.4\textwidth]{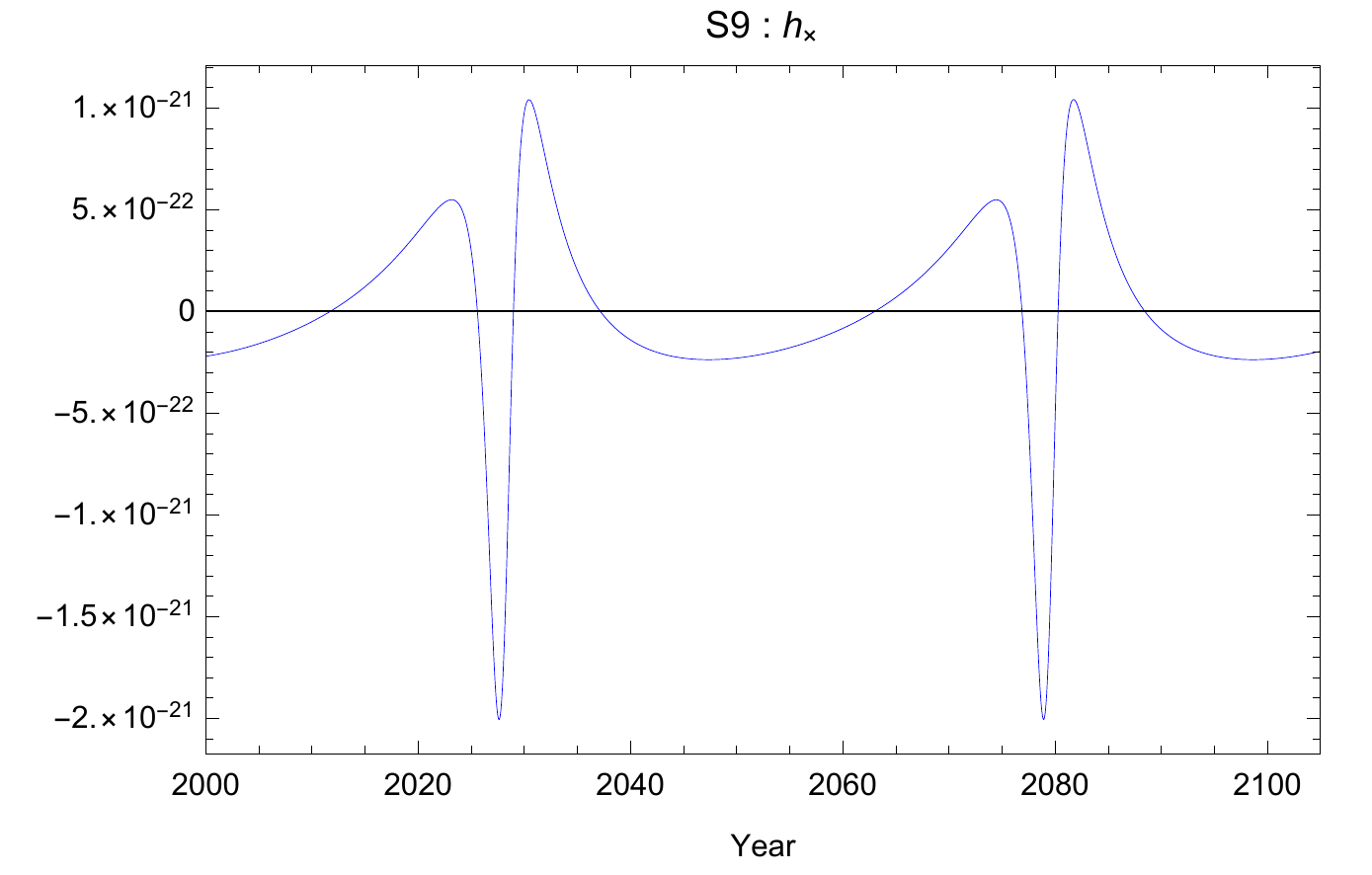}\\
\includegraphics[width=0.4\textwidth]{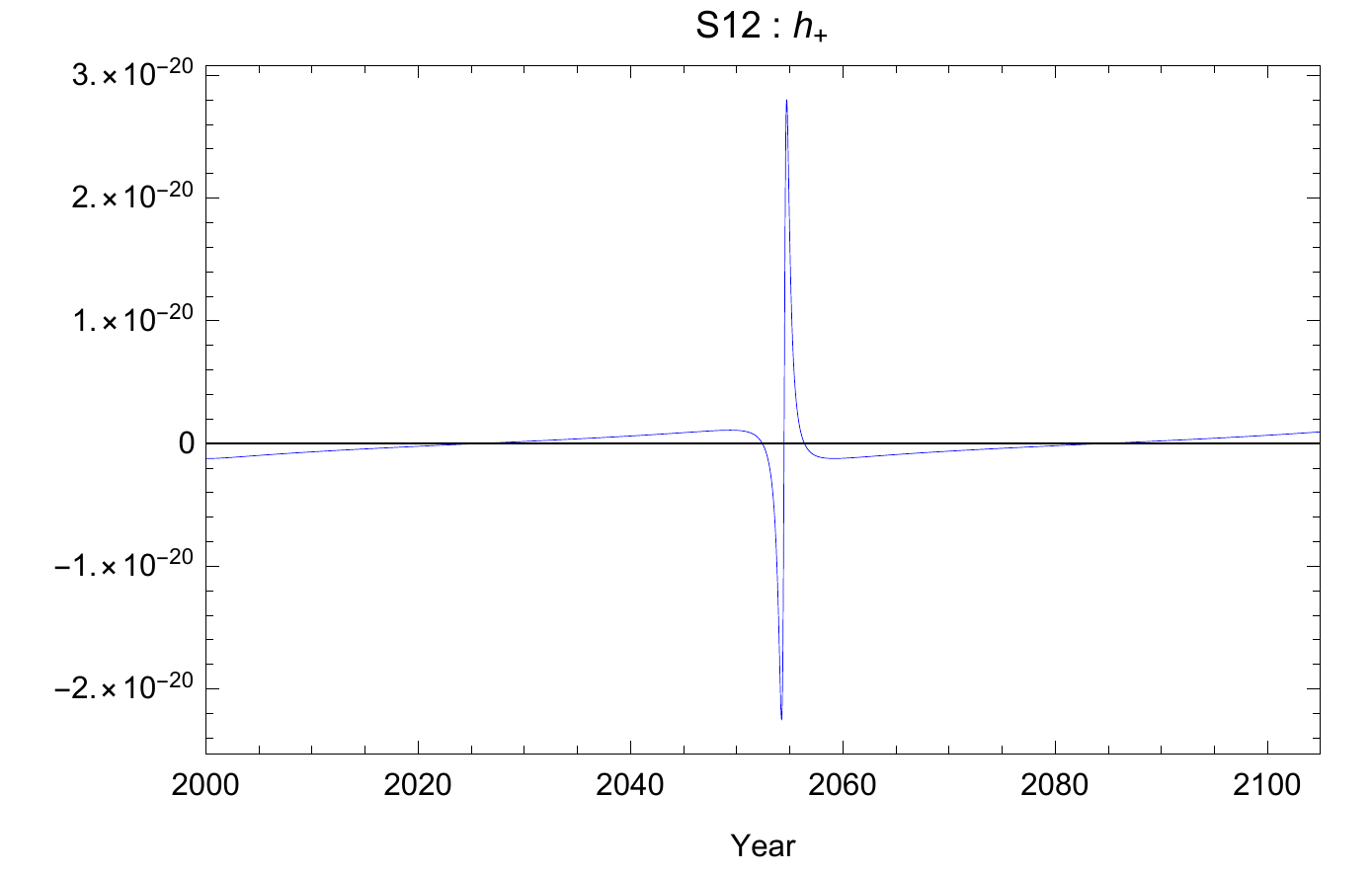}
\includegraphics[width=0.4\textwidth]{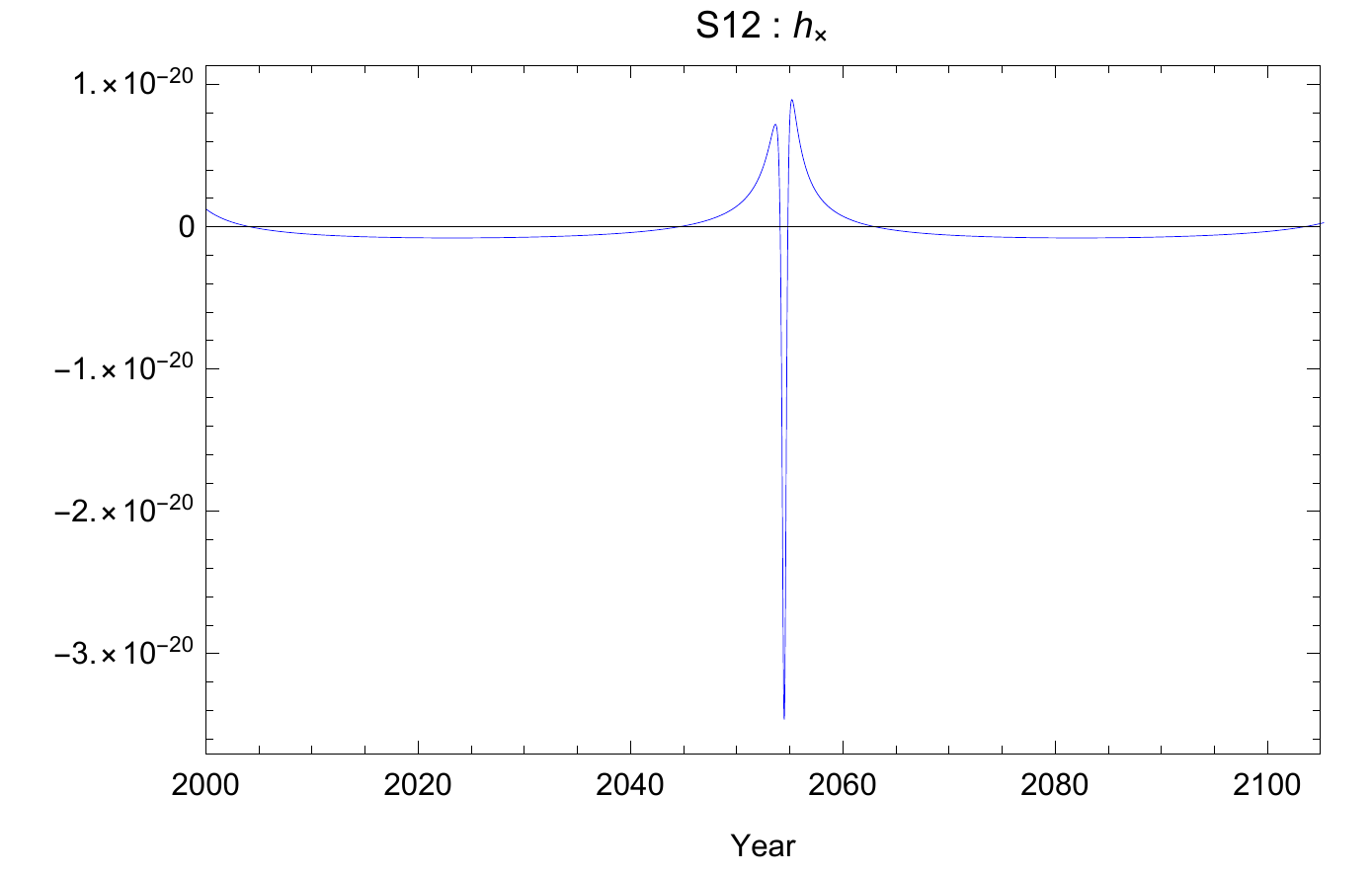}\\
\includegraphics[width=0.4\textwidth]{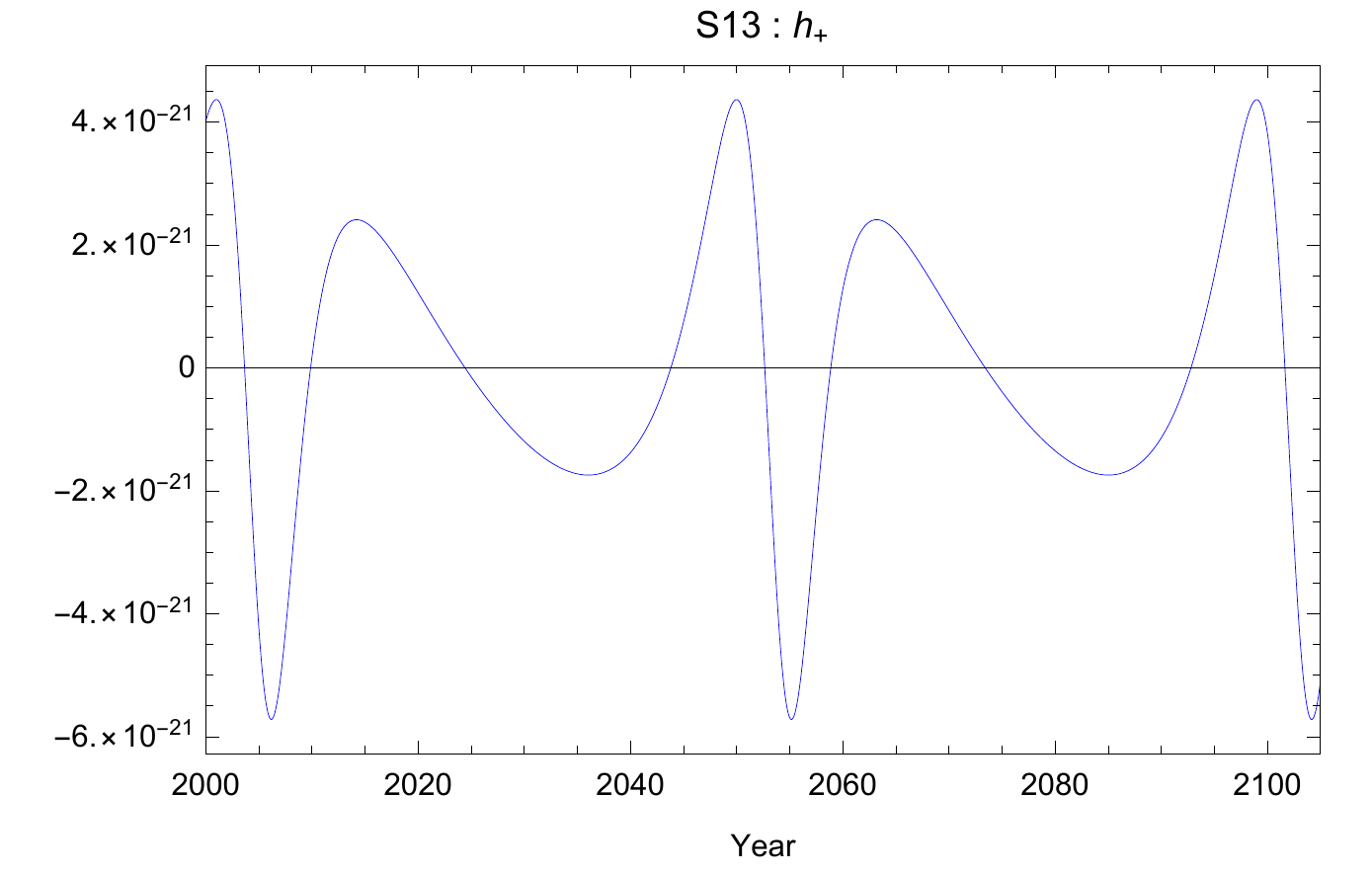}
\includegraphics[width=0.4\textwidth]{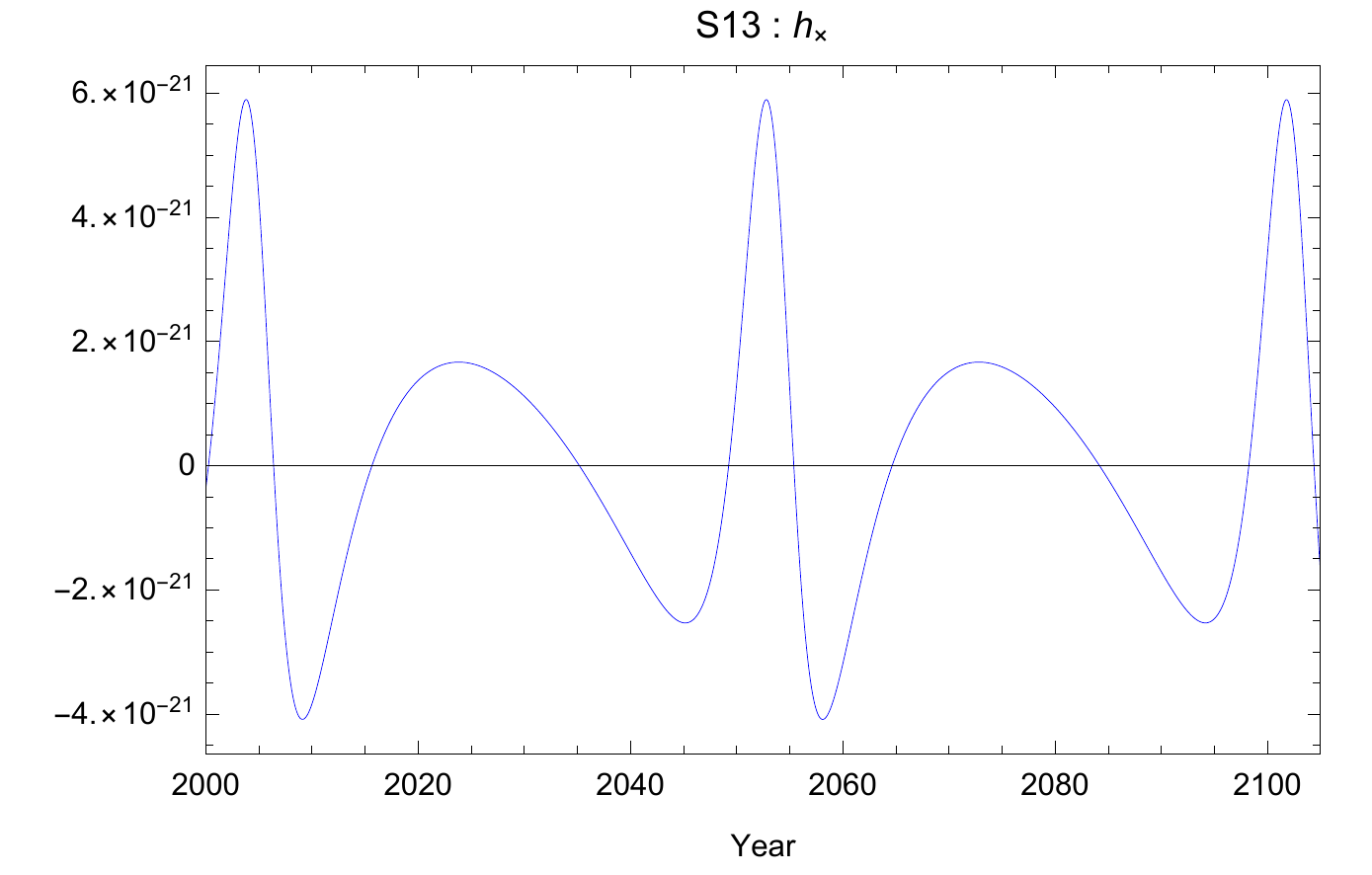}\\
\includegraphics[width=0.4\textwidth]{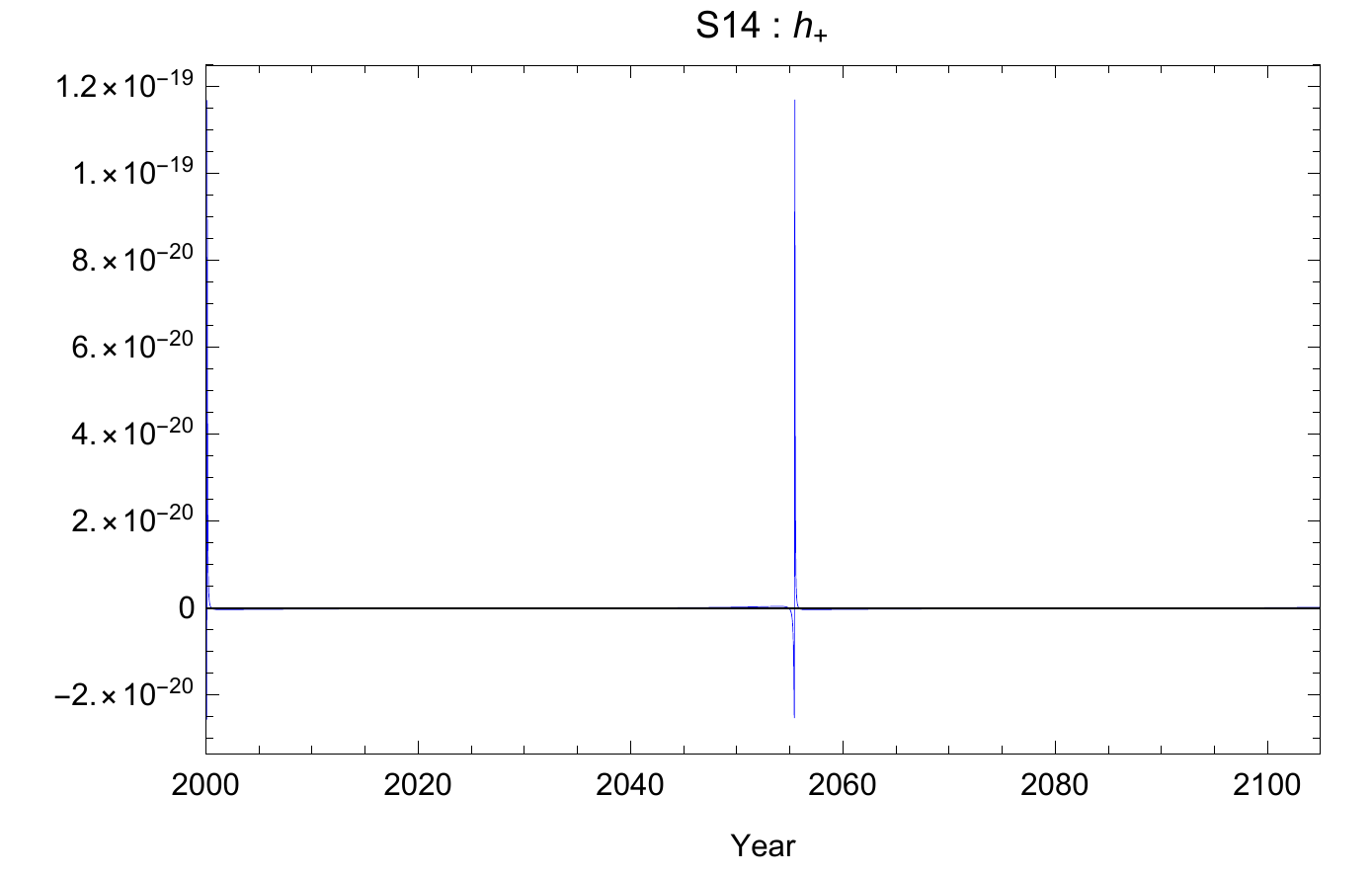}
\includegraphics[width=0.4\textwidth]{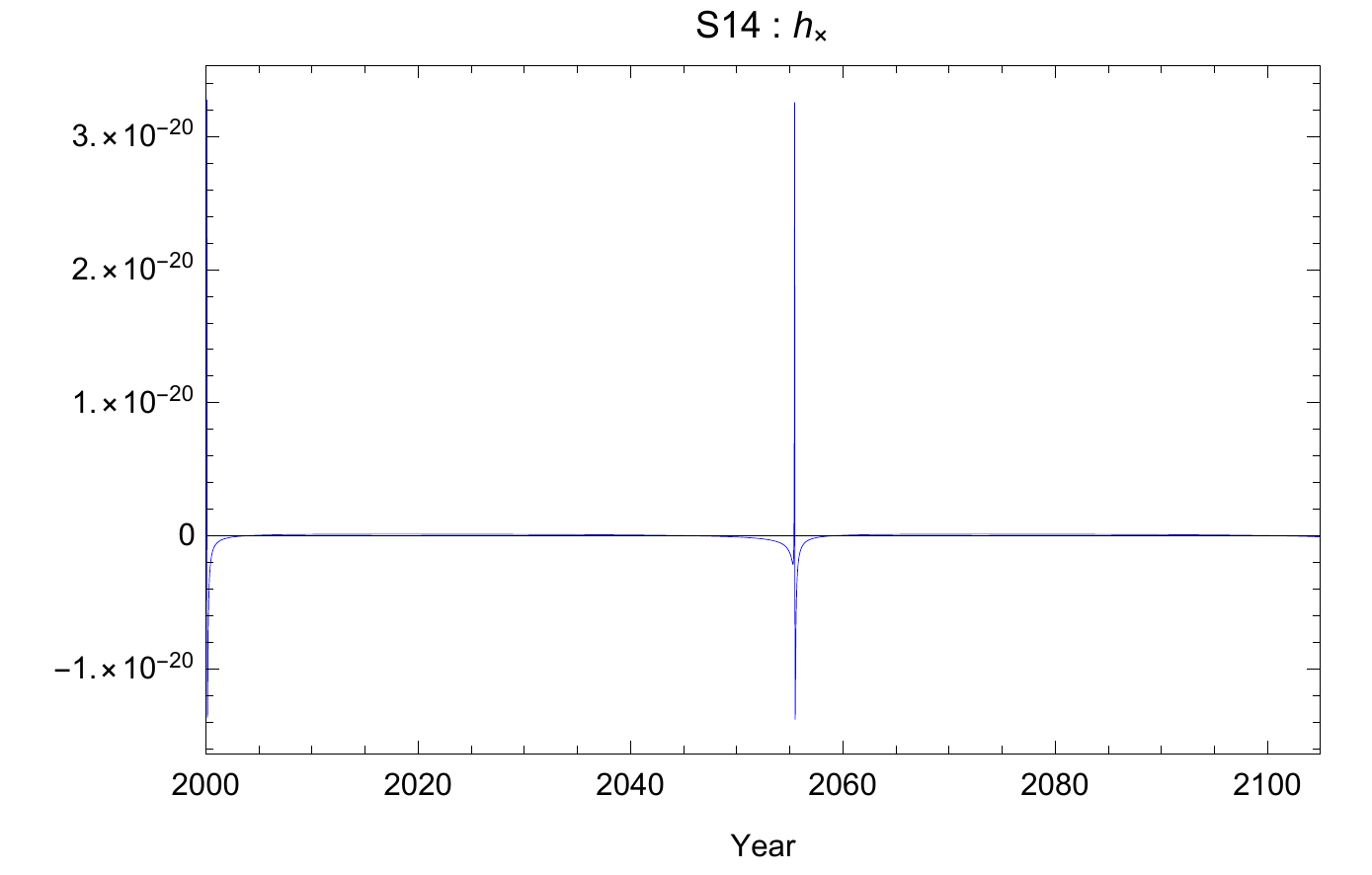}\\
\includegraphics[width=0.4\textwidth]{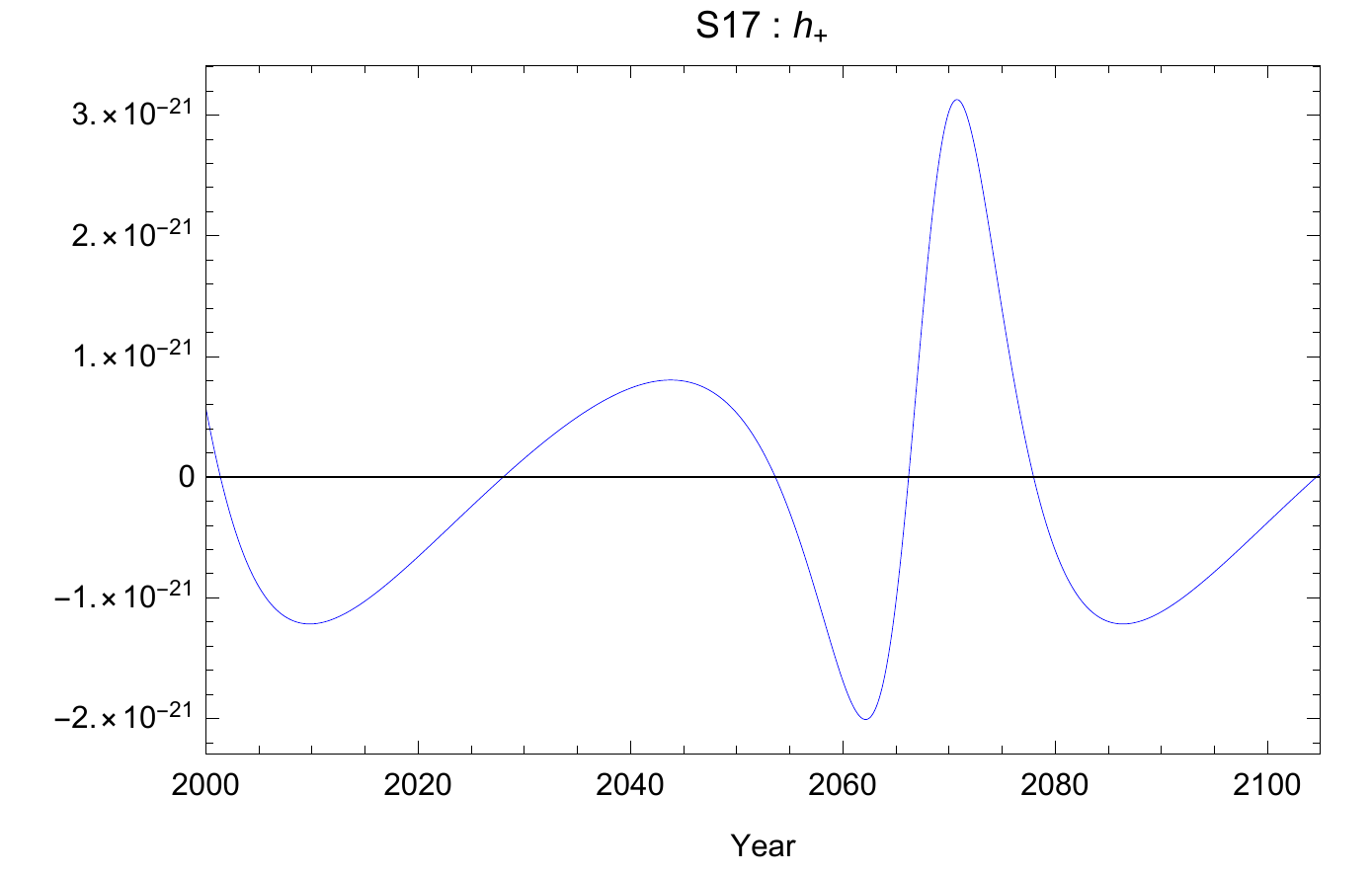}
\includegraphics[width=0.4\textwidth]{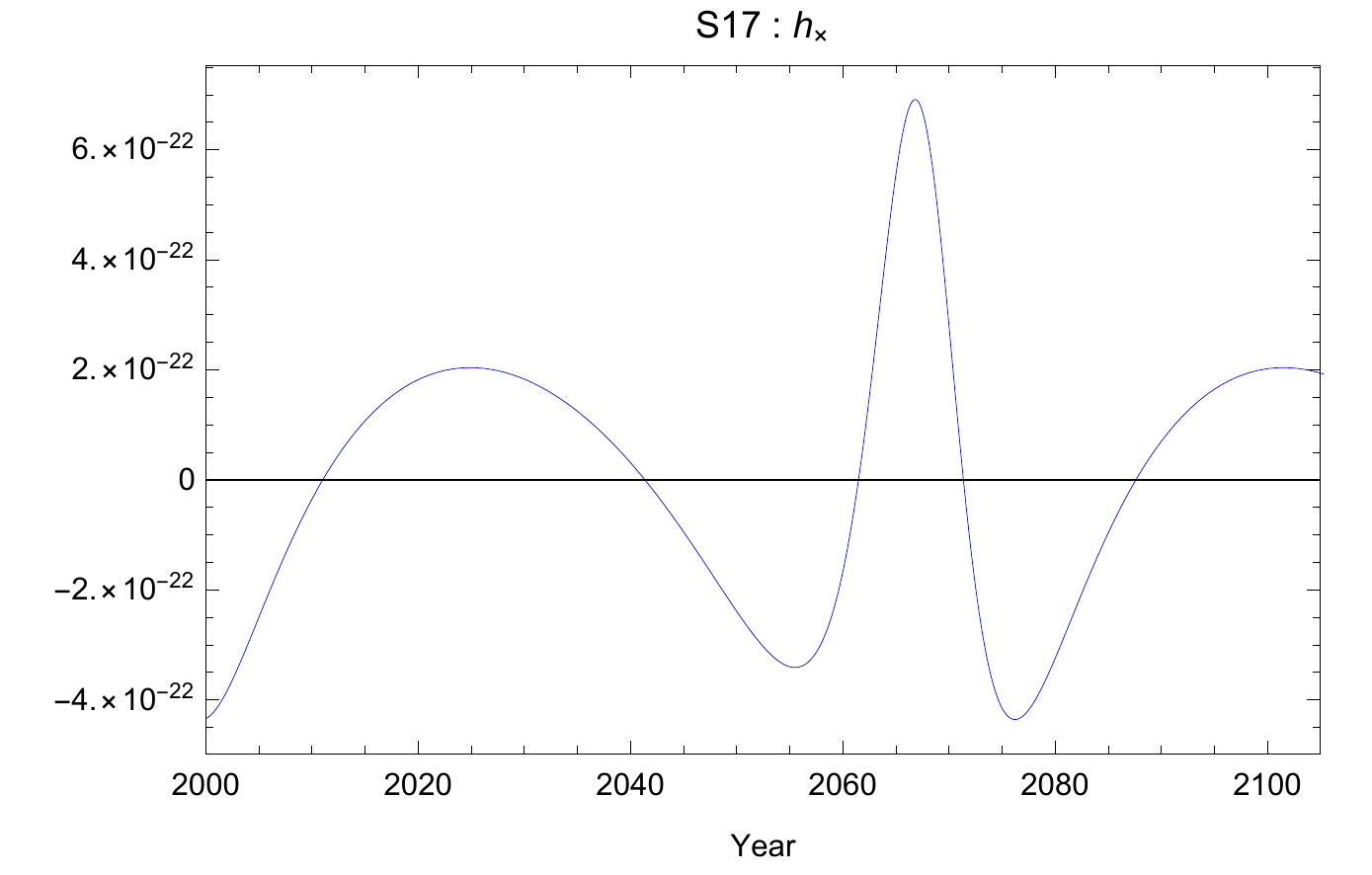}
\end{figure*} 

\begin{figure*}
\includegraphics[width=0.4\textwidth]{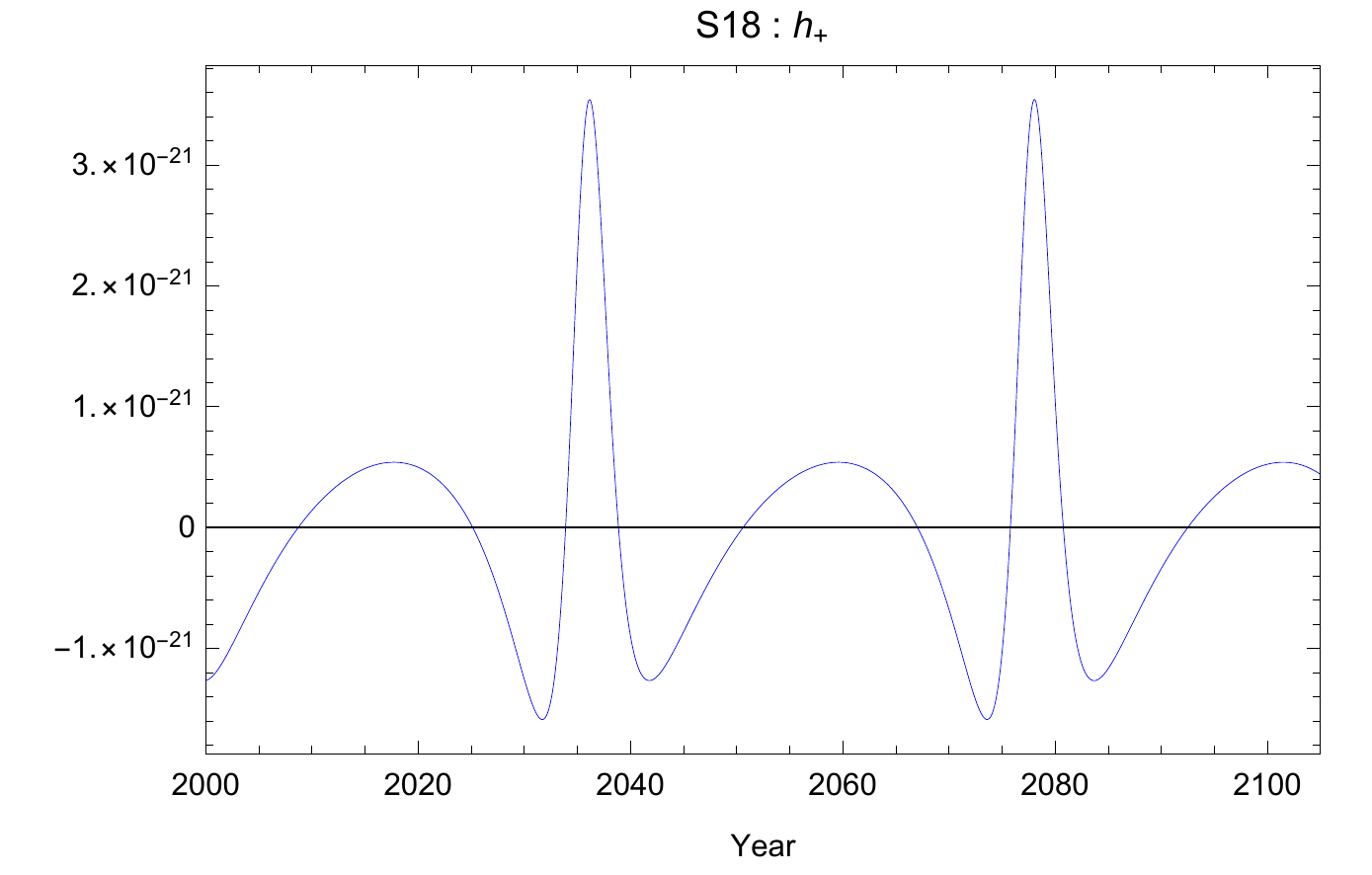}
\includegraphics[width=0.4\textwidth]{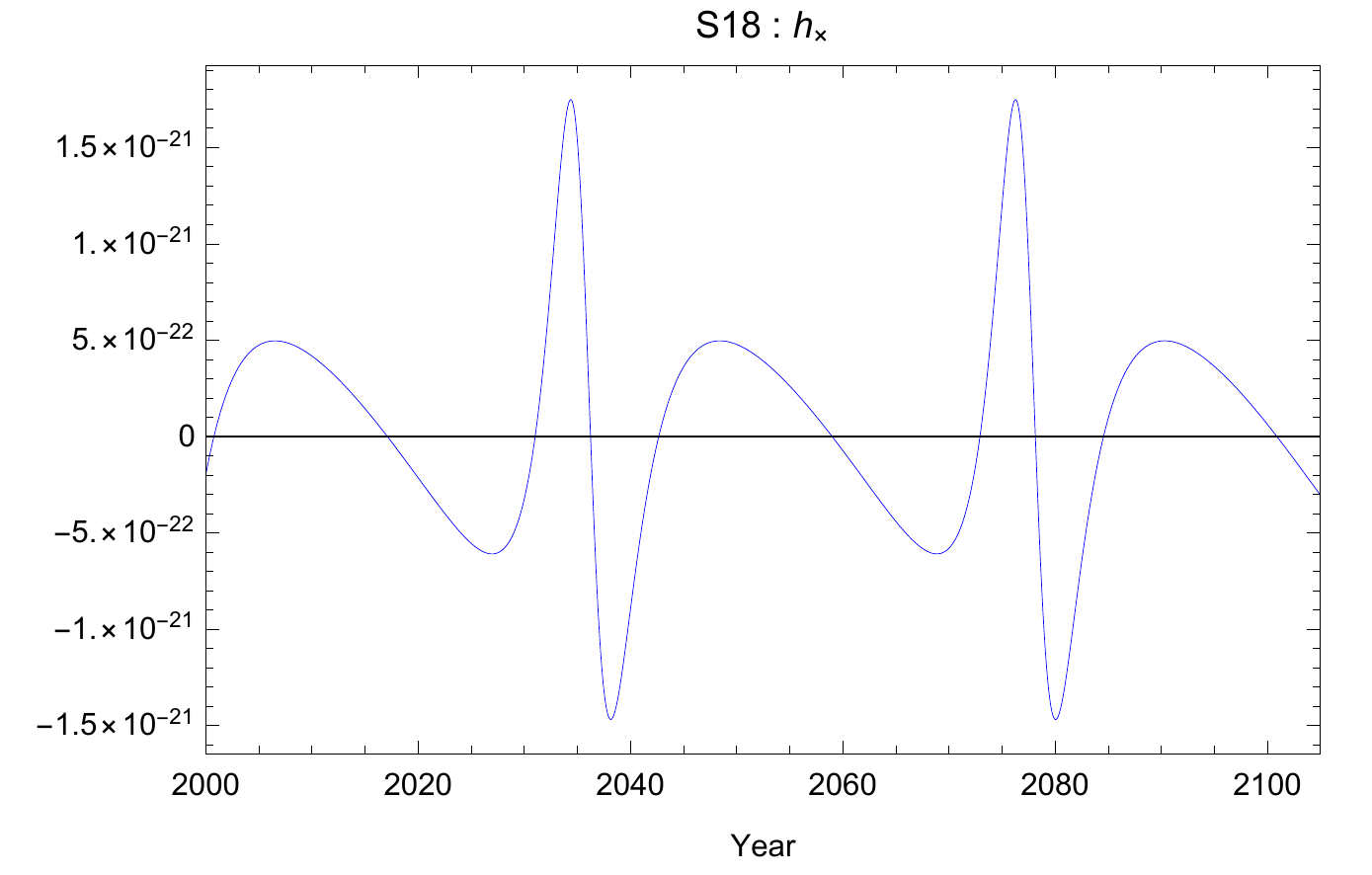}\\
\includegraphics[width=0.4\textwidth]{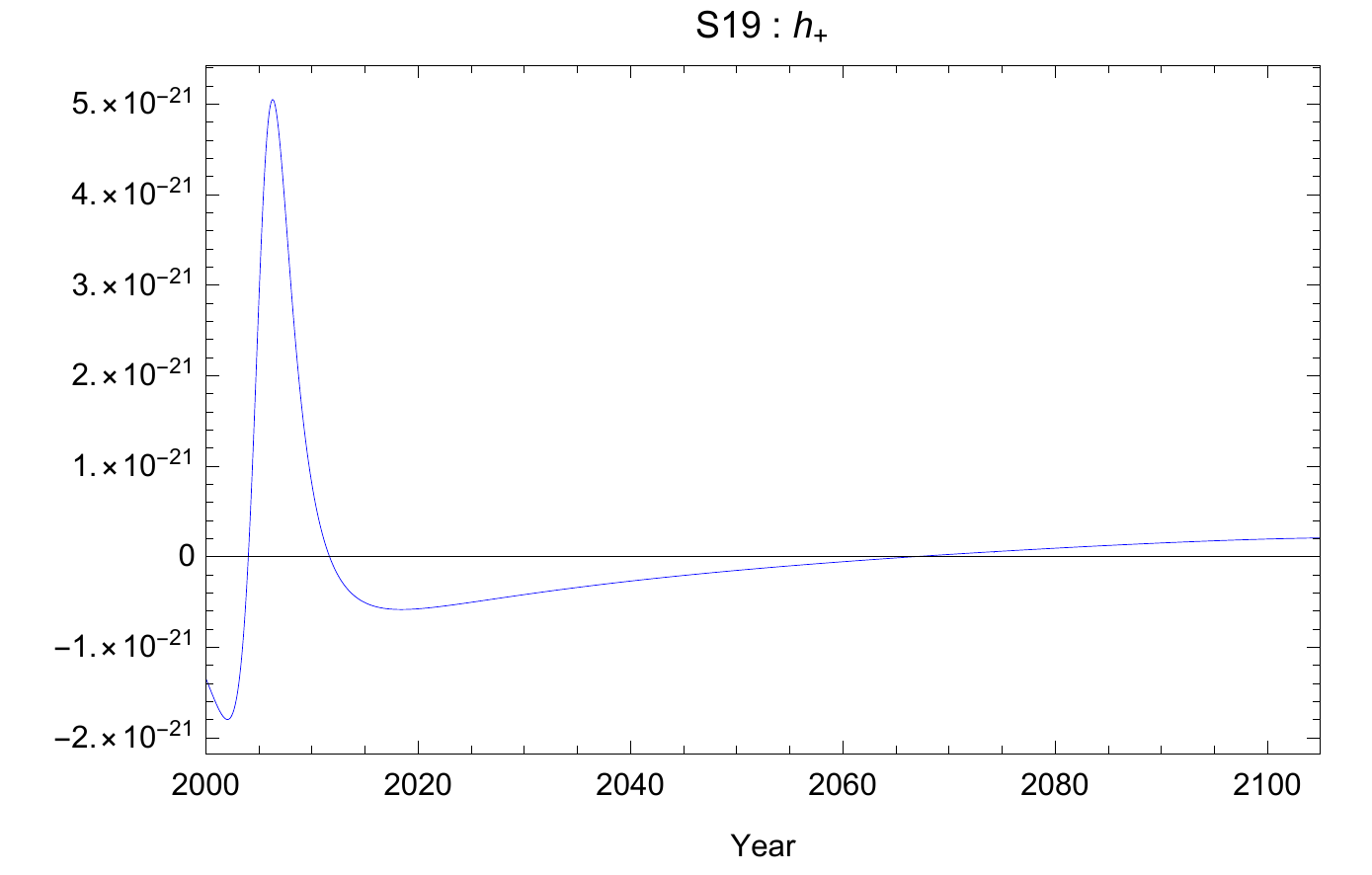}
\includegraphics[width=0.4\textwidth]{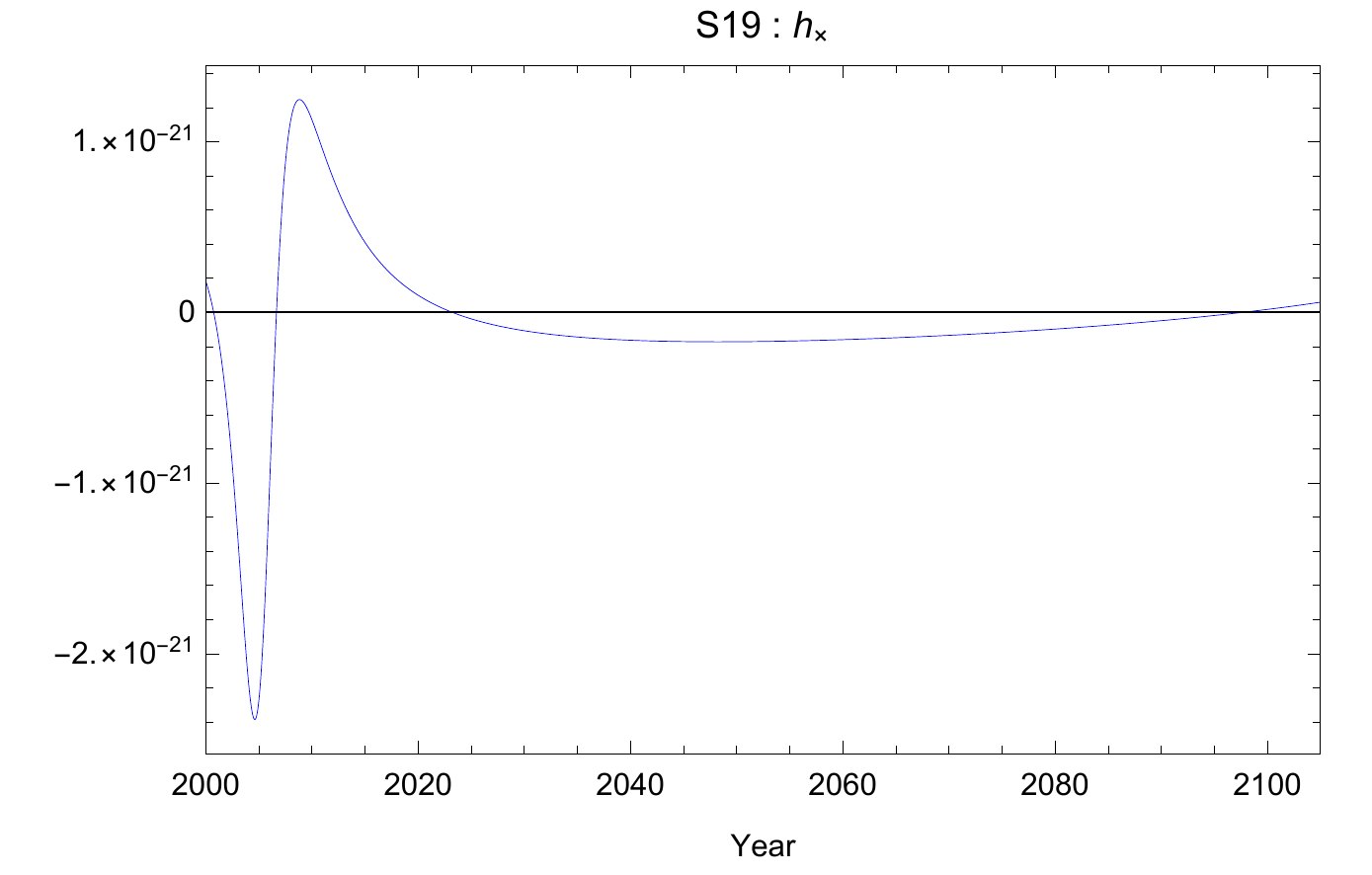}\\
\includegraphics[width=0.4\textwidth]{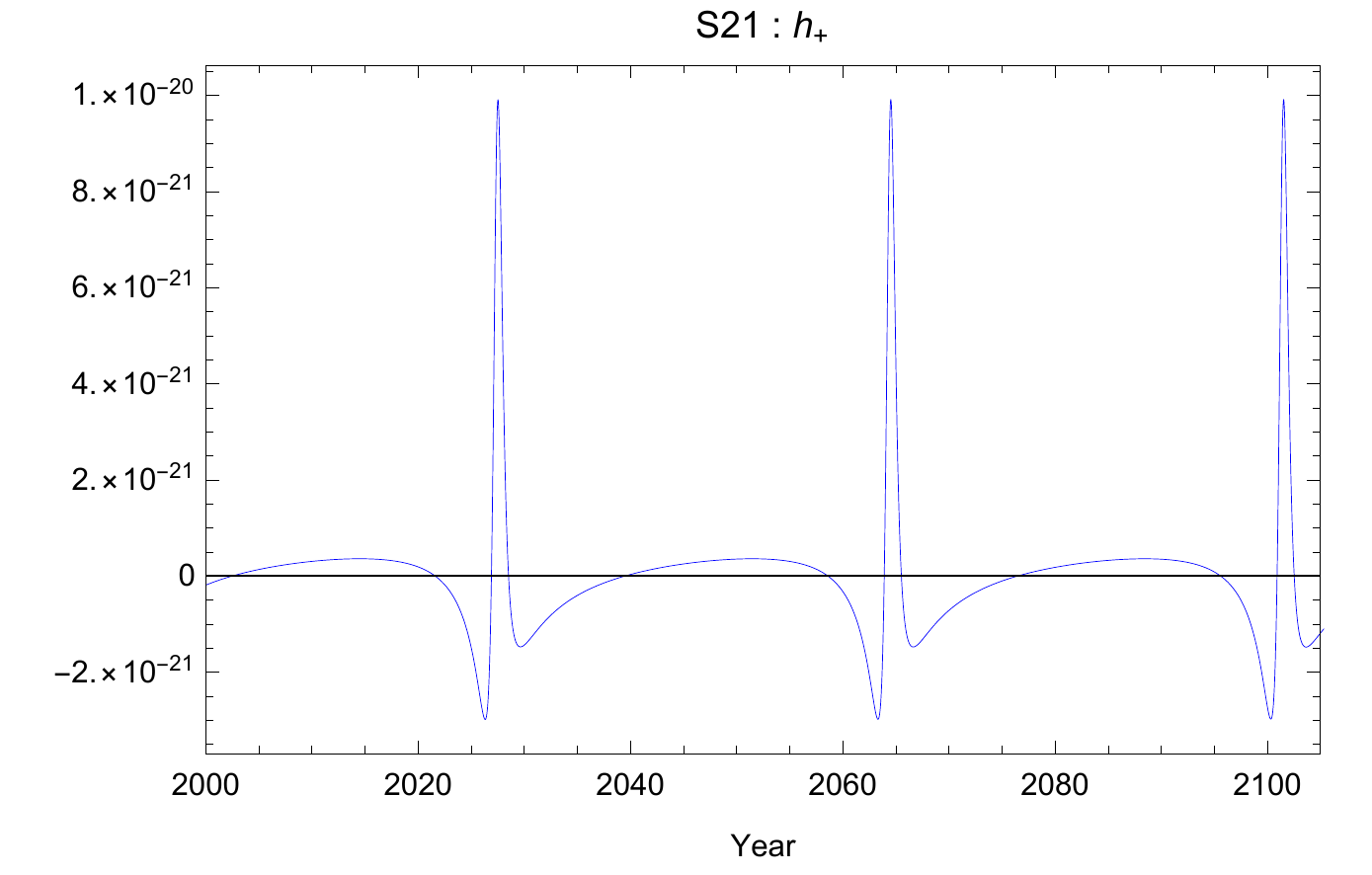}
\includegraphics[width=0.4\textwidth]{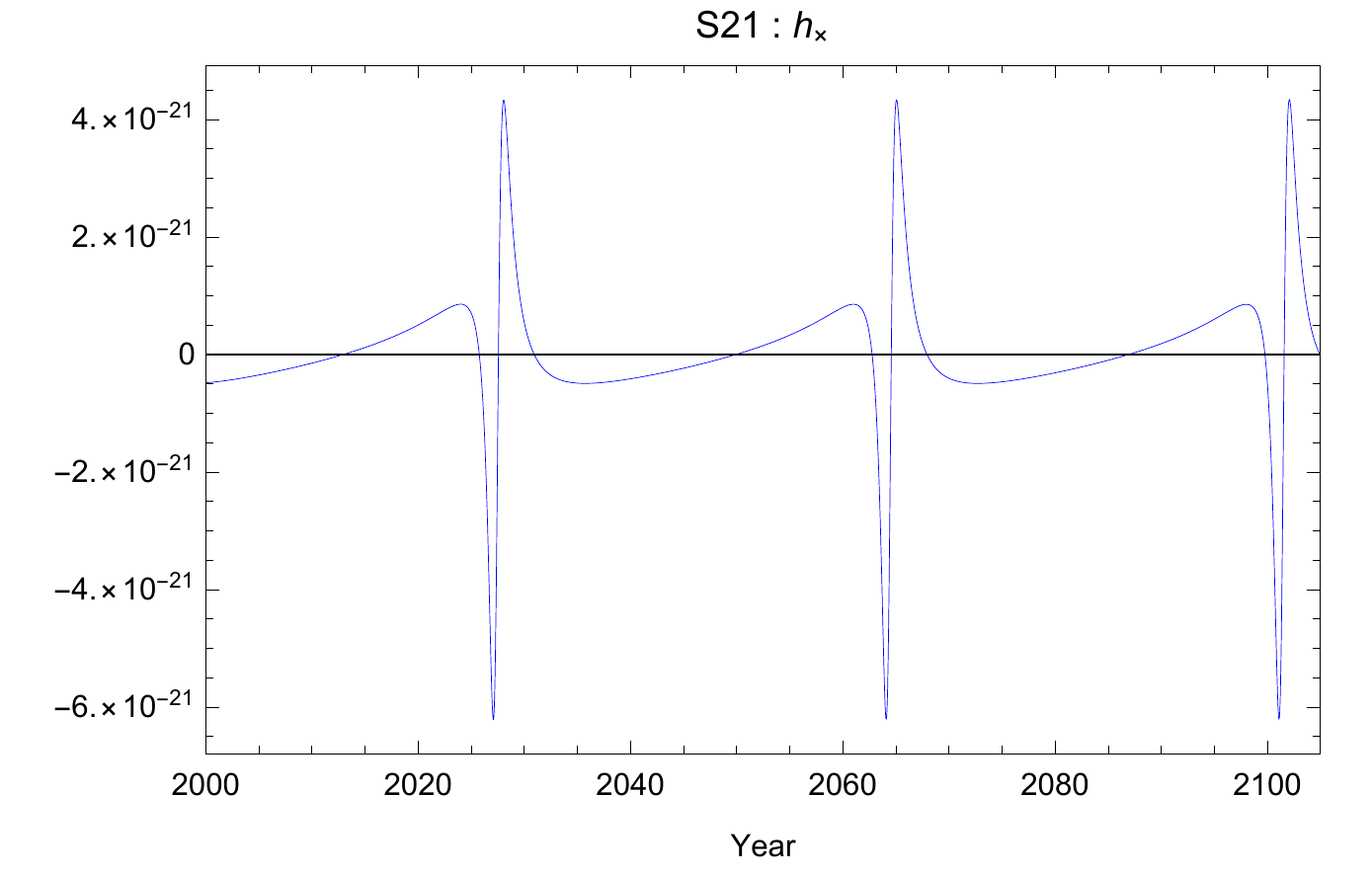}\\
\includegraphics[width=0.4\textwidth]{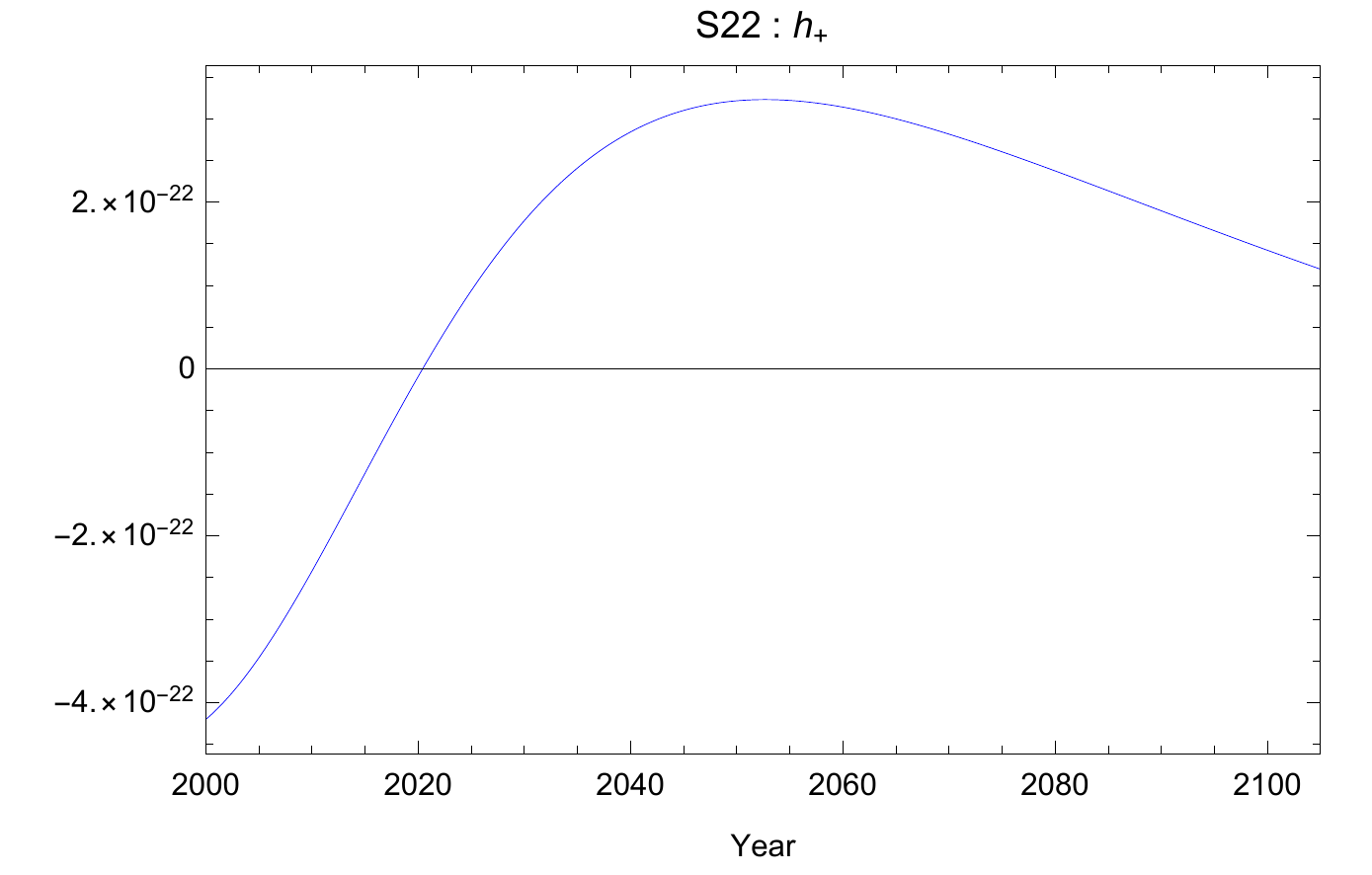}
\includegraphics[width=0.4\textwidth]{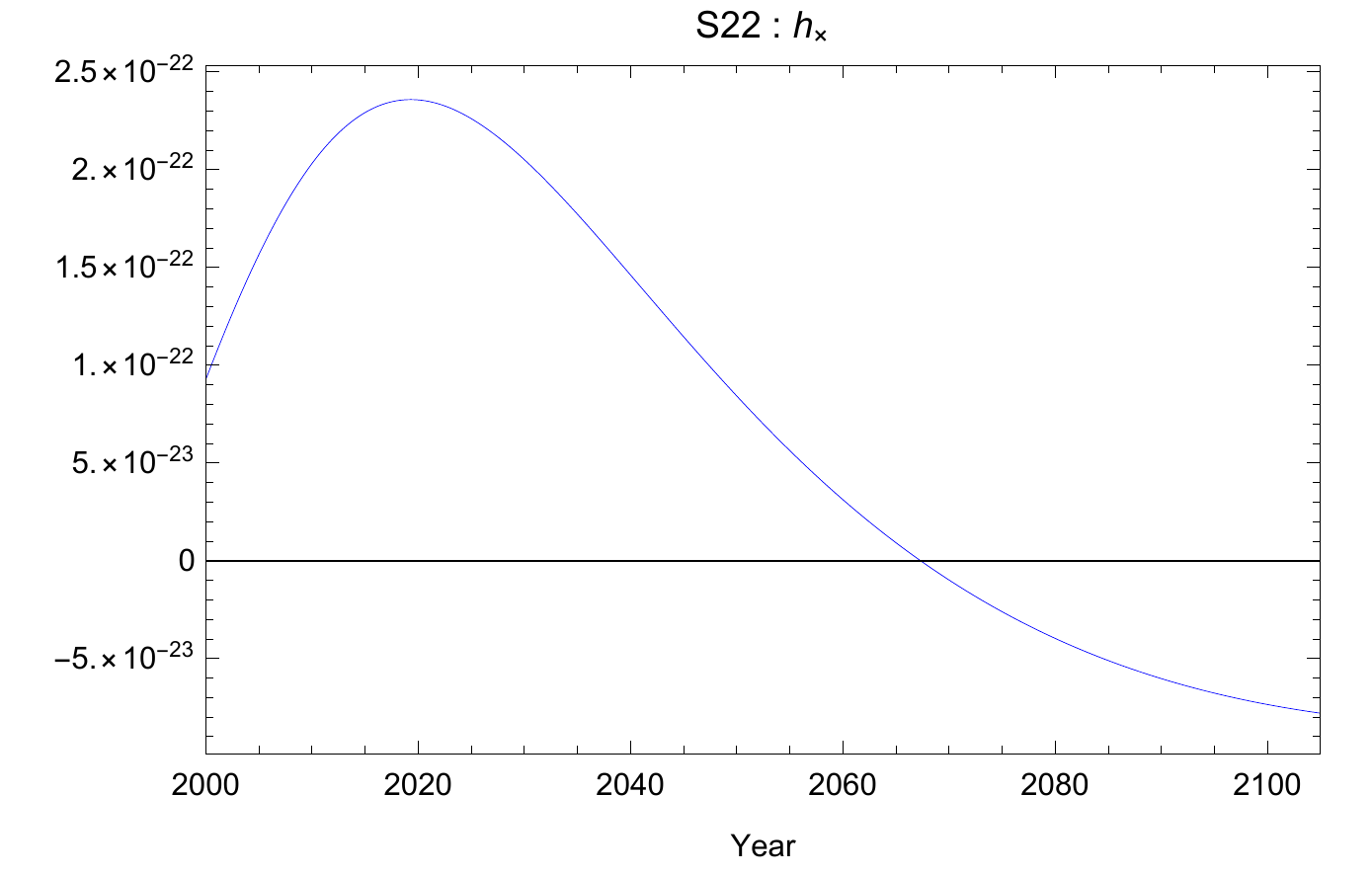}\\
\includegraphics[width=0.4\textwidth]{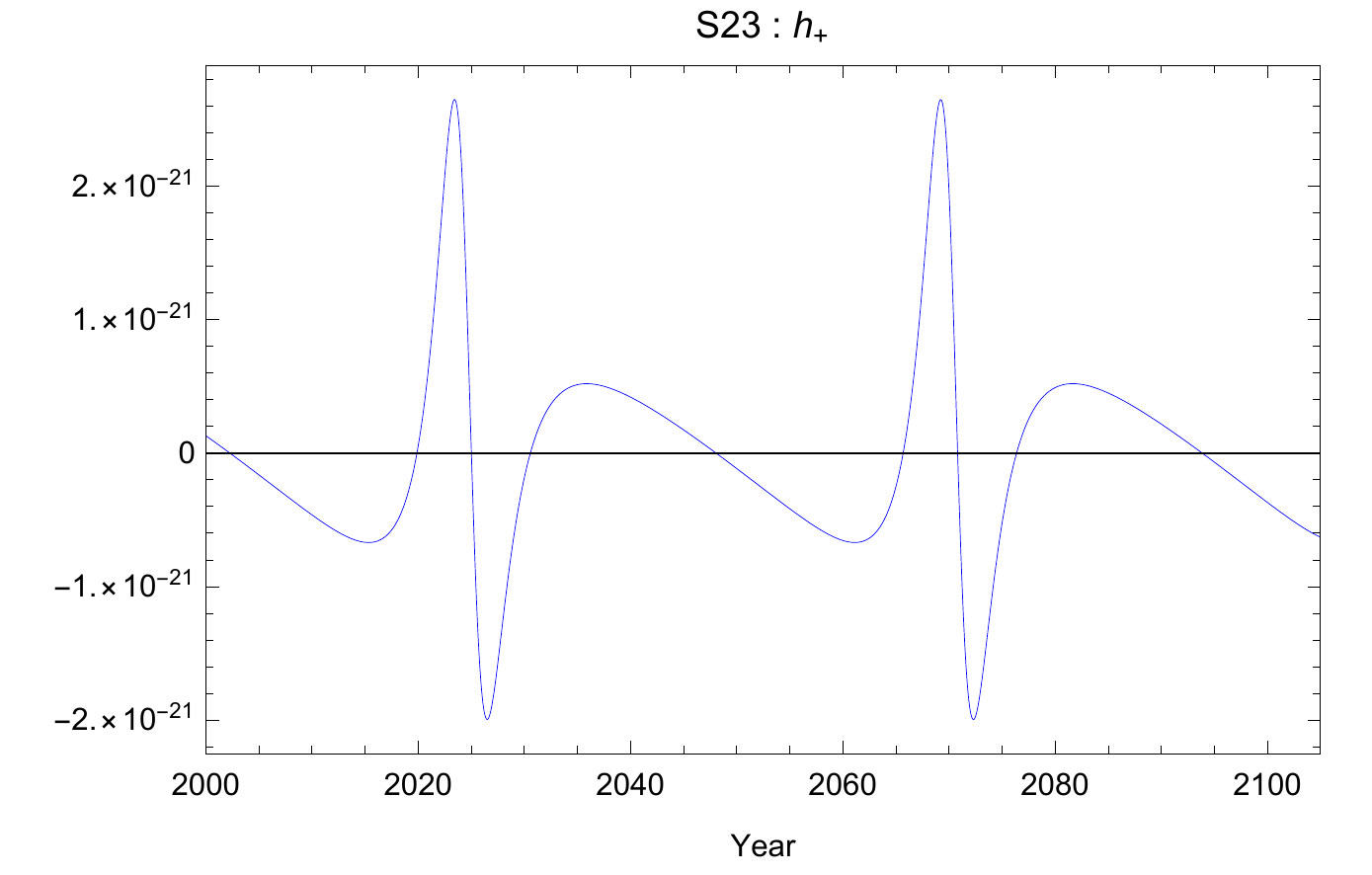}
\includegraphics[width=0.4\textwidth]{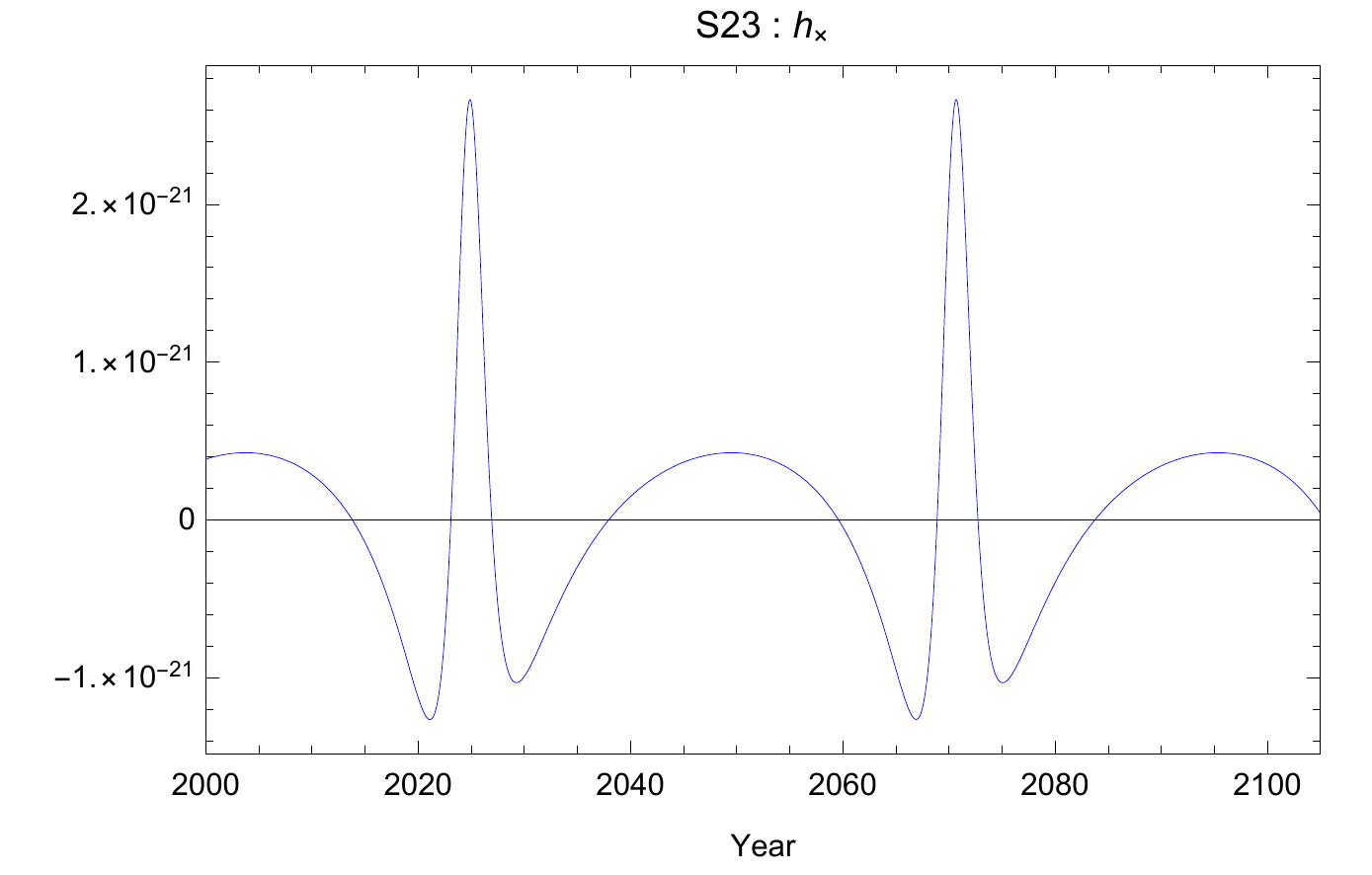}
\end{figure*} 

\begin{figure*}
\includegraphics[width=0.4\textwidth]{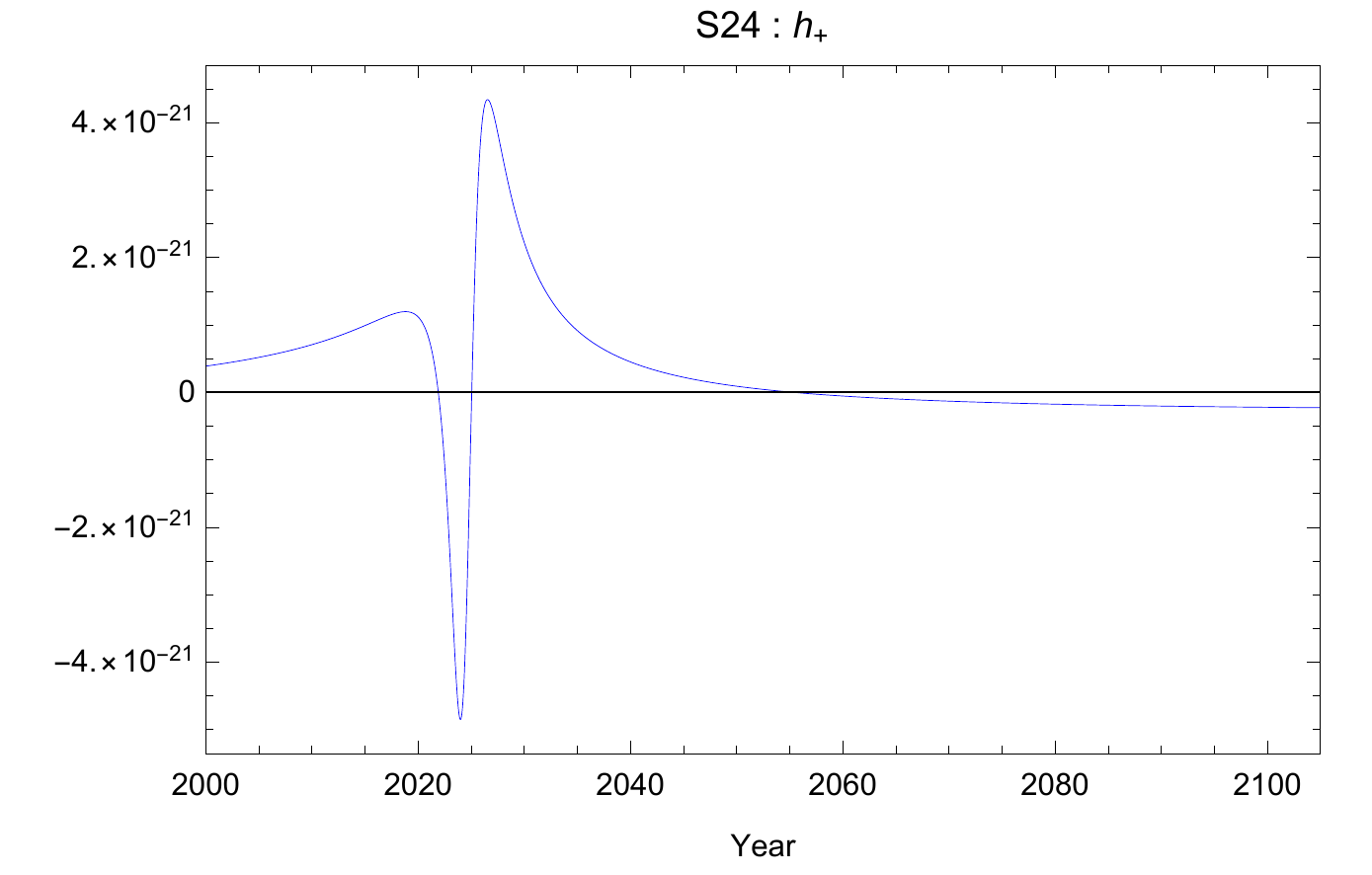}
\includegraphics[width=0.4\textwidth]{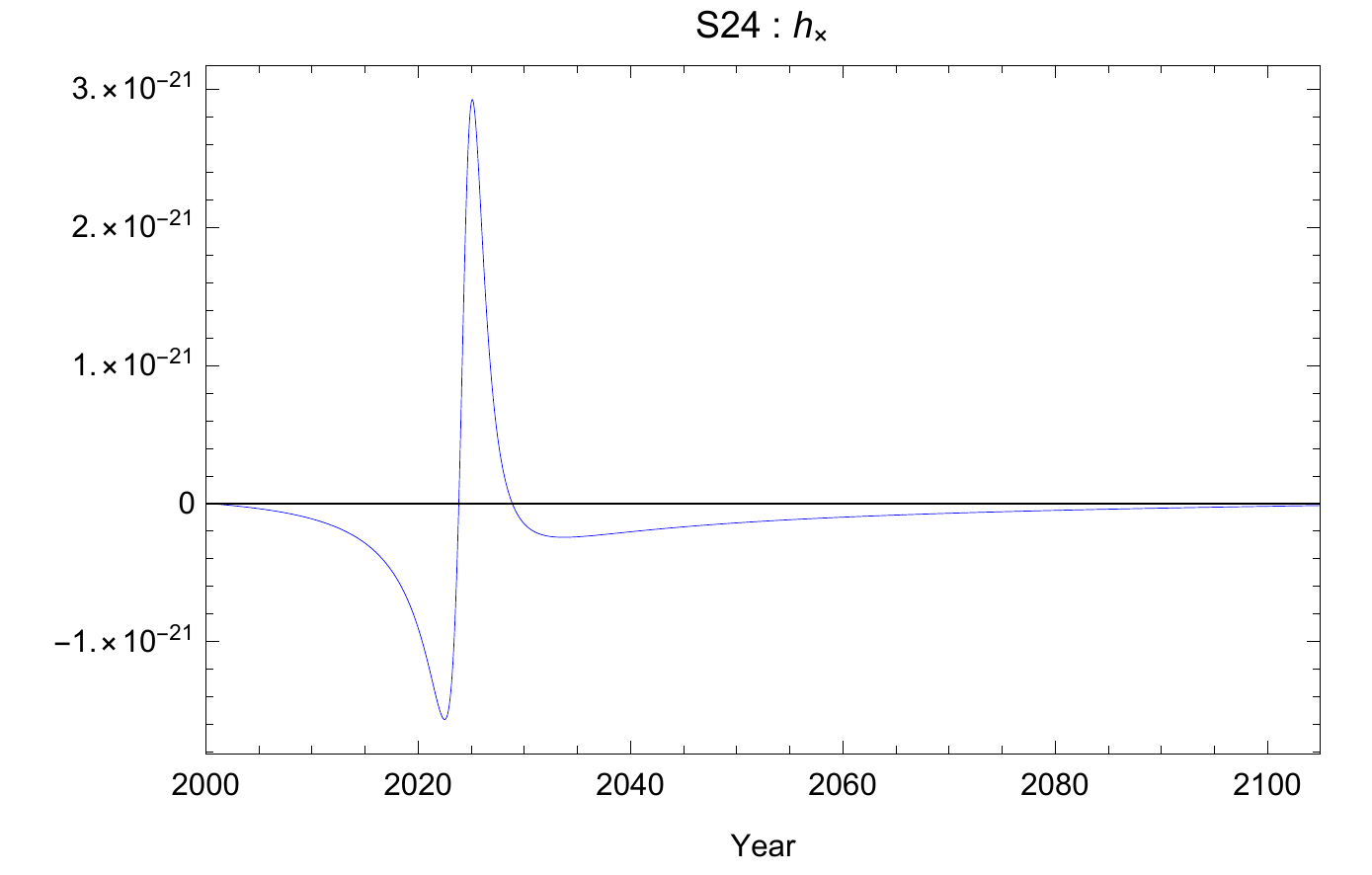}\\
\includegraphics[width=0.4\textwidth]{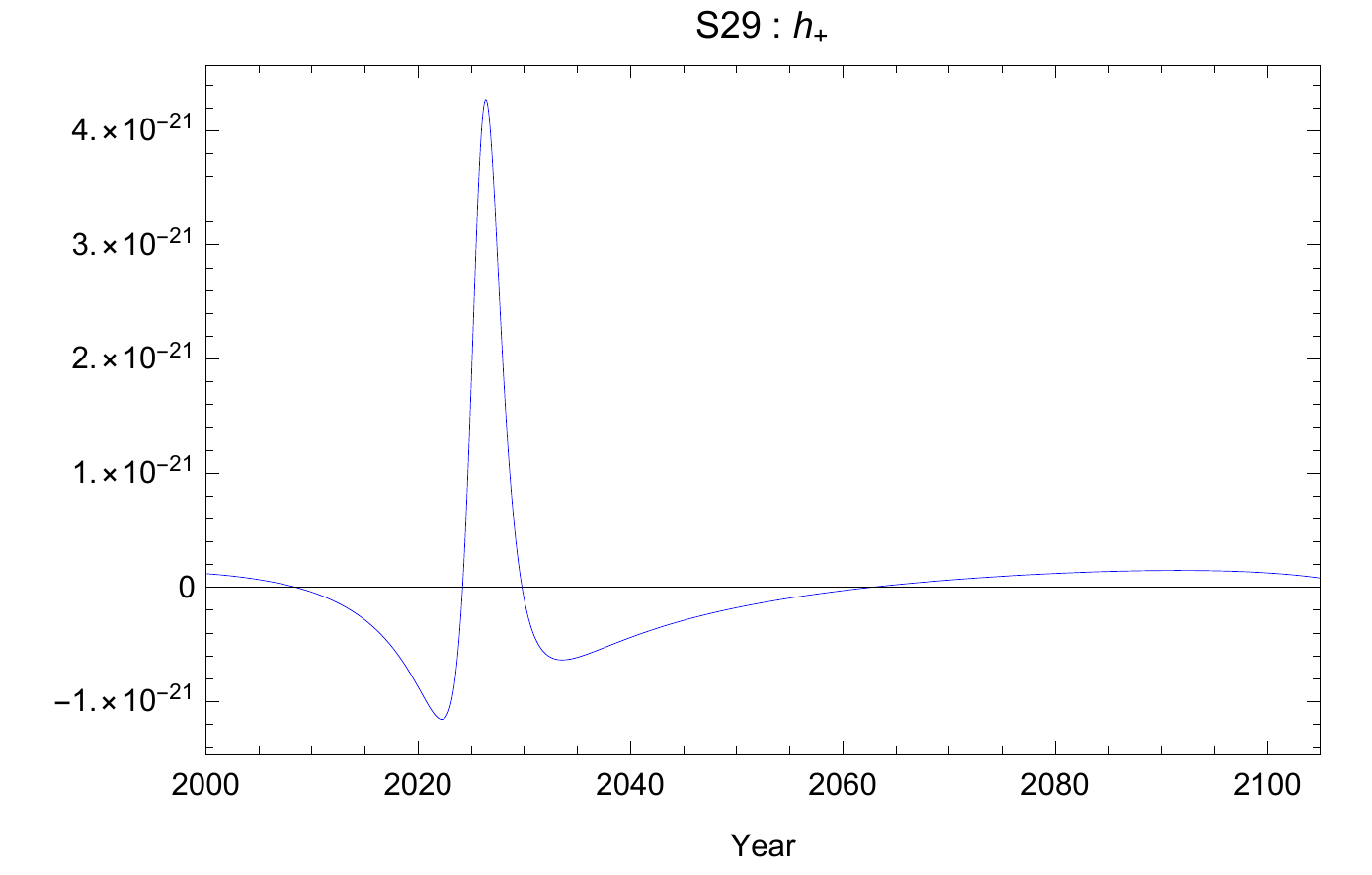}
\includegraphics[width=0.4\textwidth]{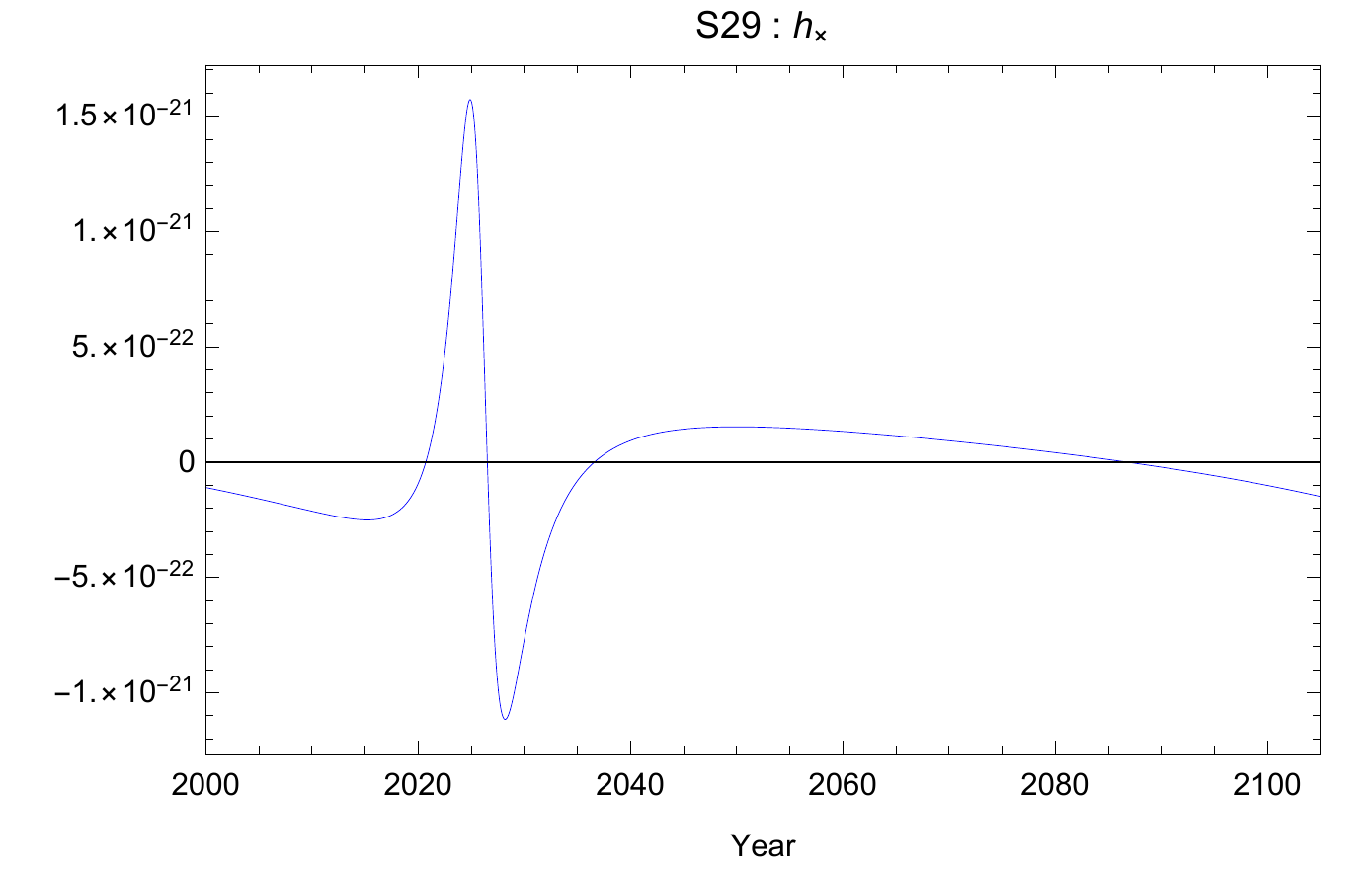}\\
\includegraphics[width=0.4\textwidth]{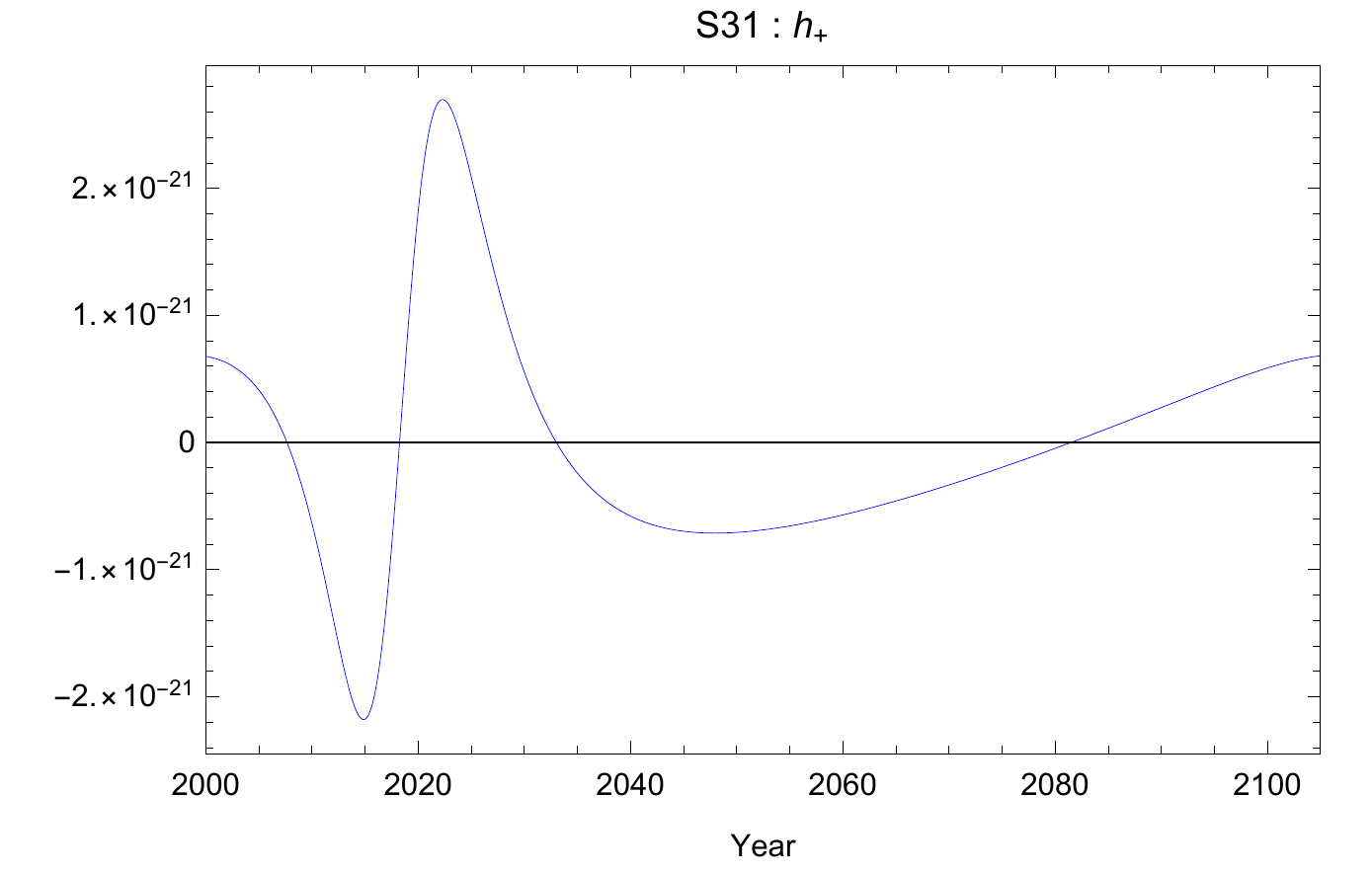}
\includegraphics[width=0.4\textwidth]{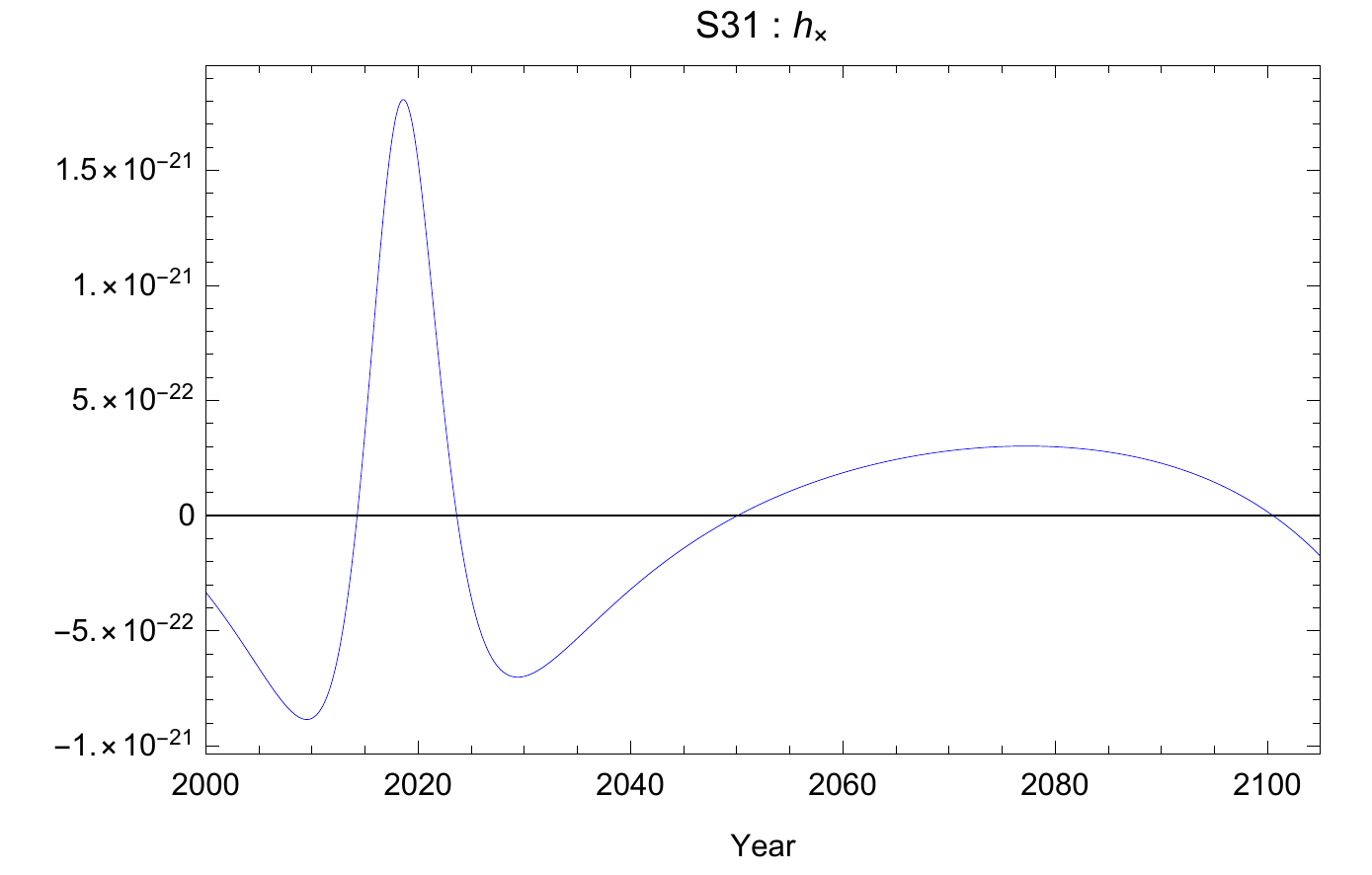}\\
\includegraphics[width=0.4\textwidth]{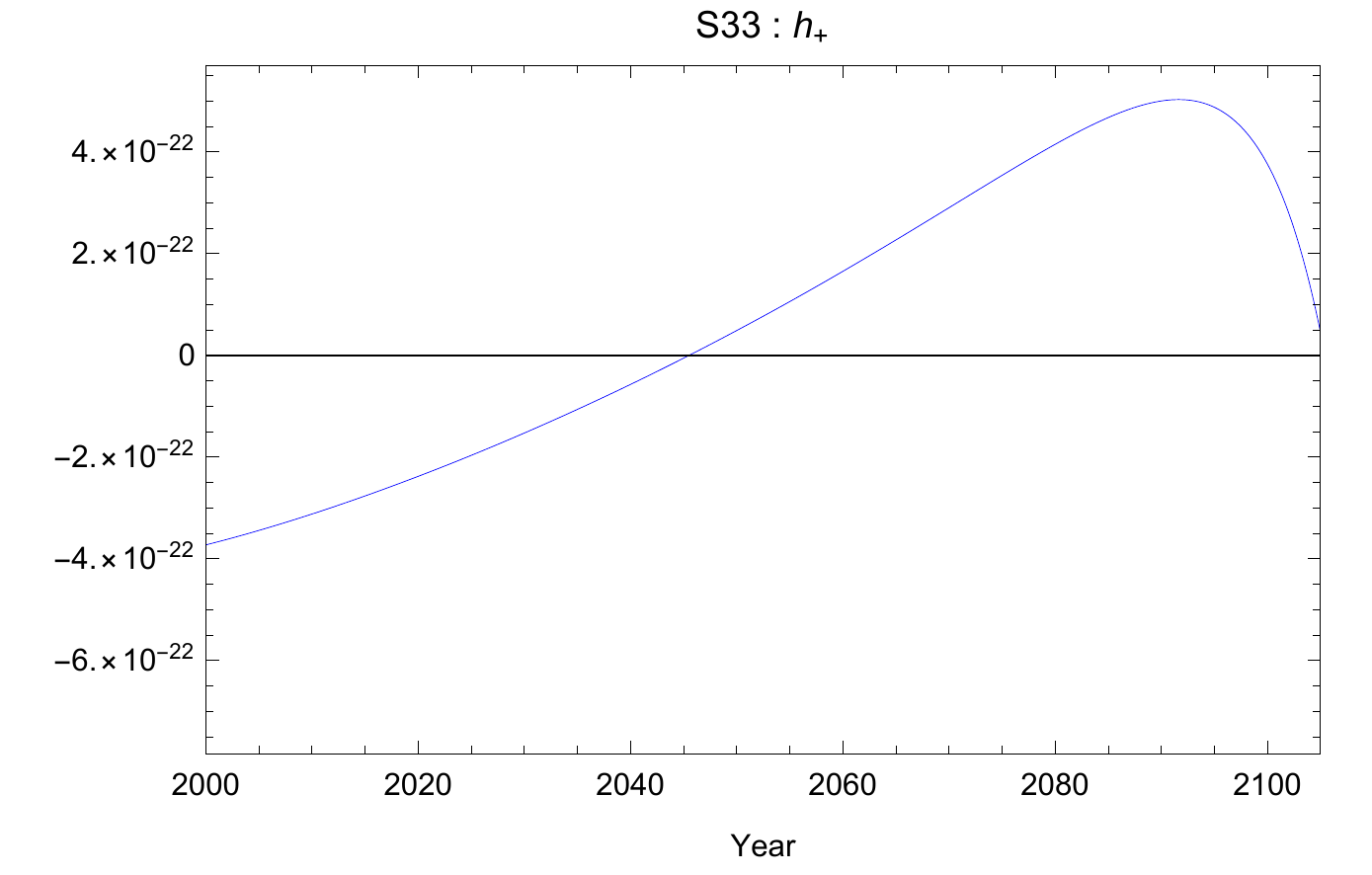}
\includegraphics[width=0.4\textwidth]{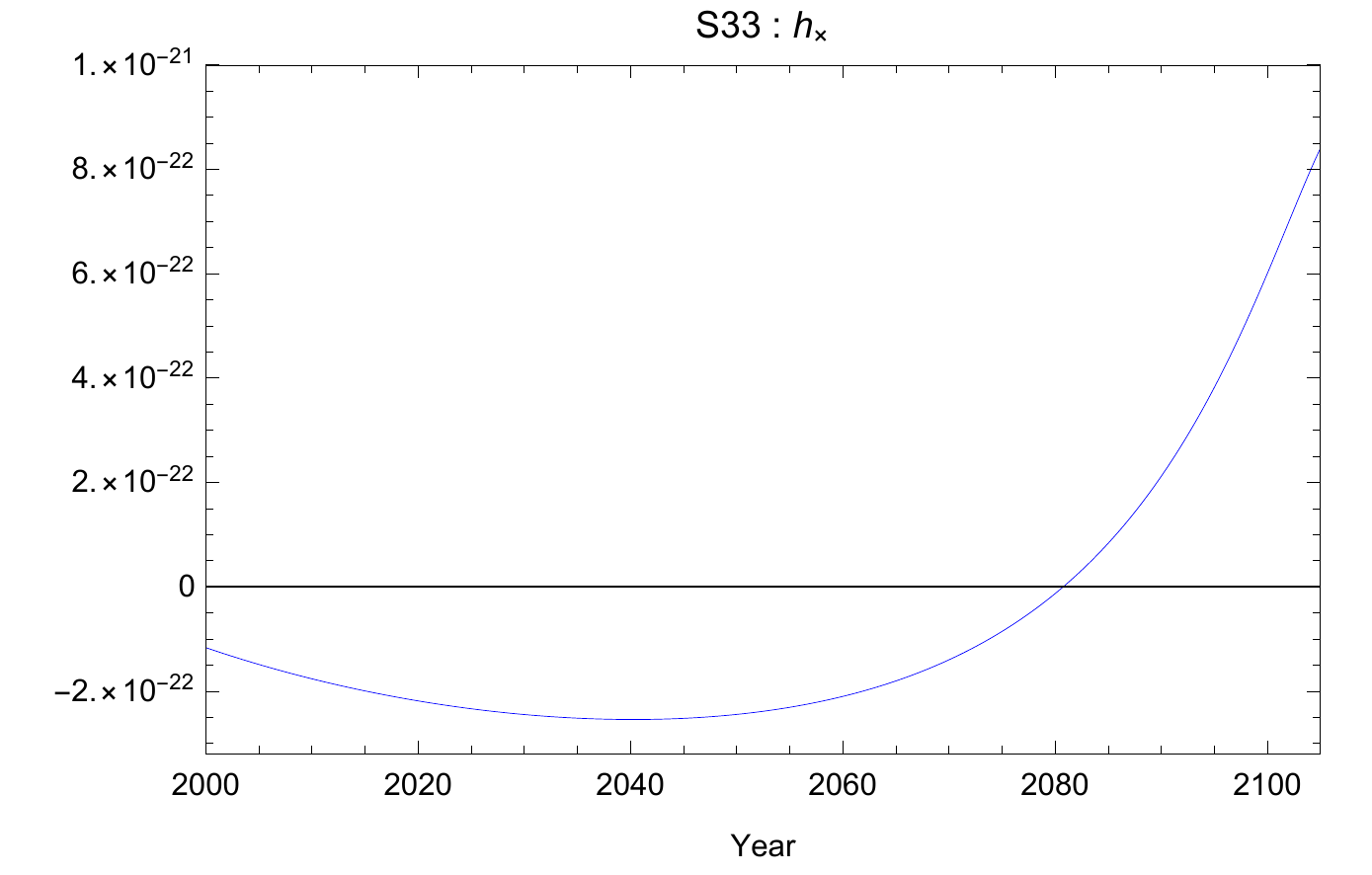}\\
\includegraphics[width=0.4\textwidth]{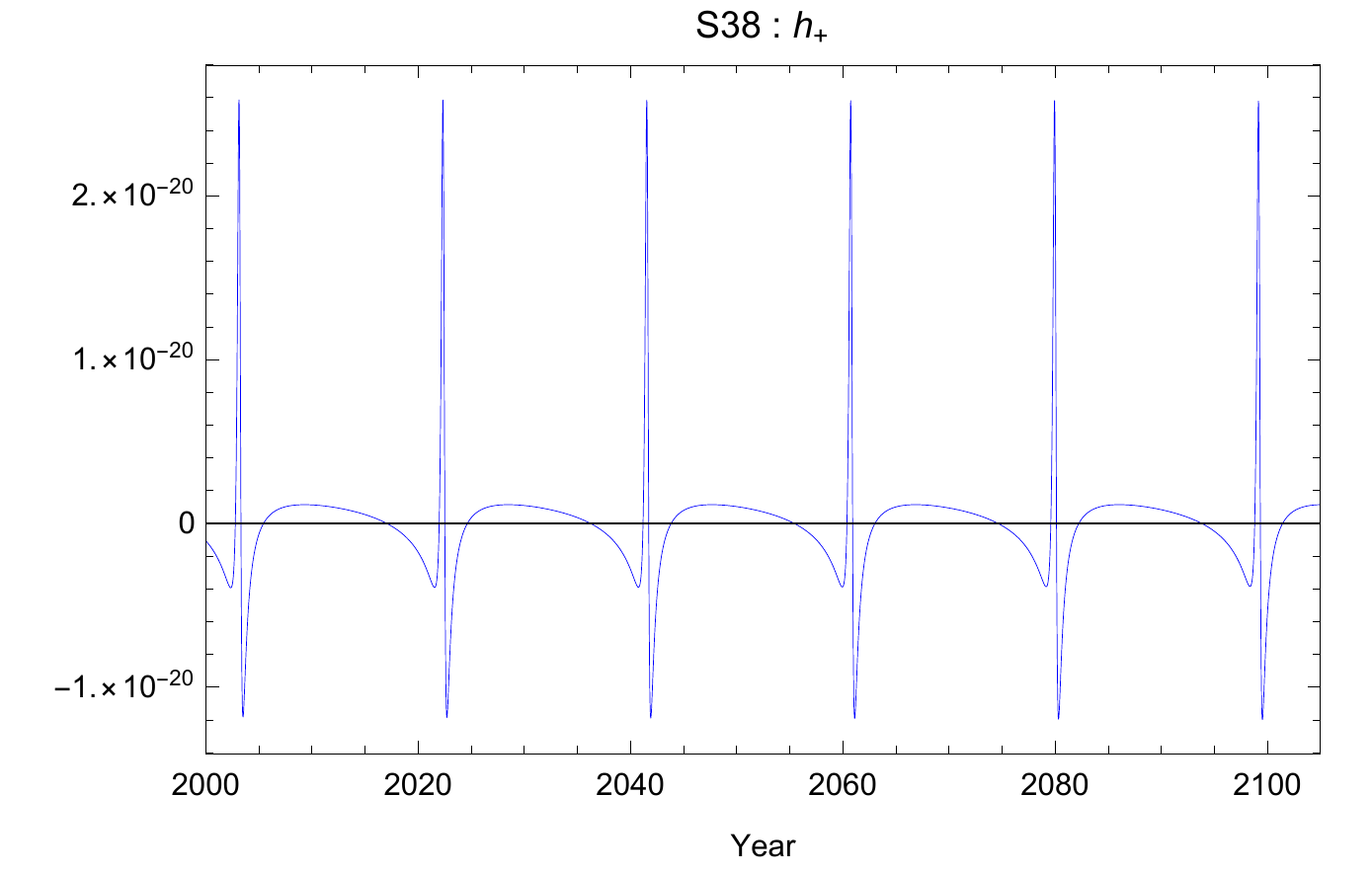}
\includegraphics[width=0.4\textwidth]{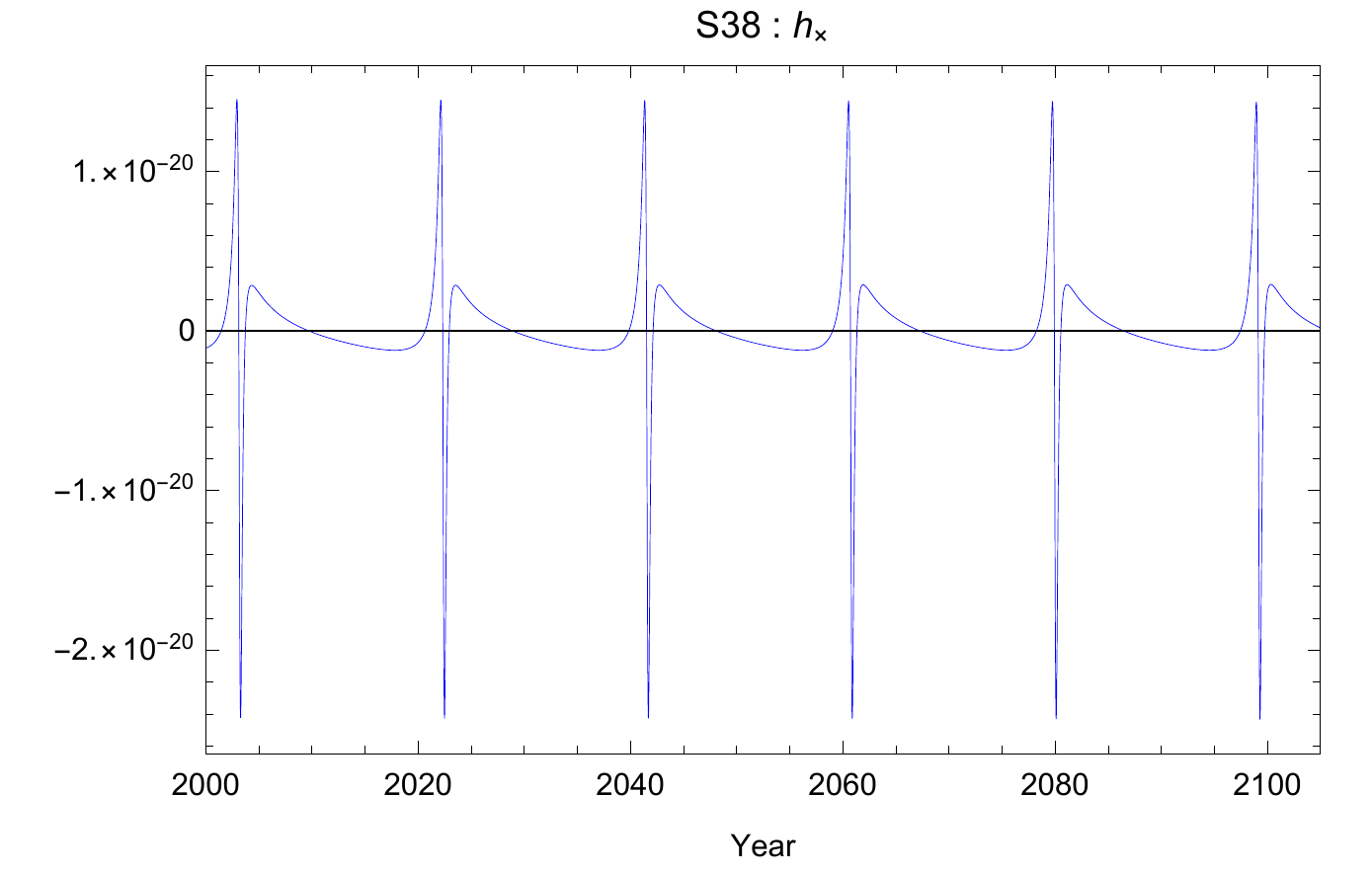}
\end{figure*} 

\begin{figure*}
\includegraphics[width=0.4\textwidth]{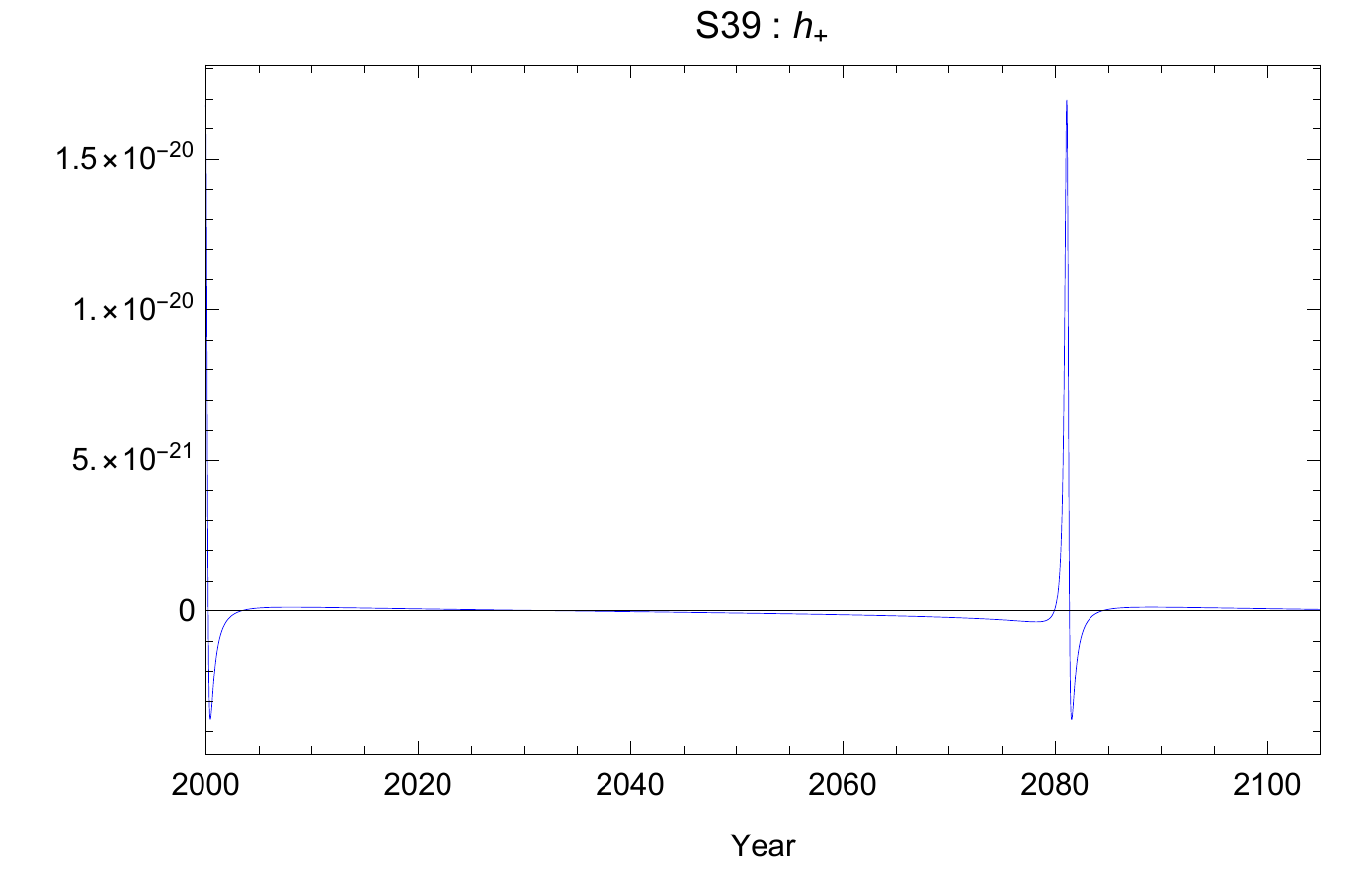}
\includegraphics[width=0.4\textwidth]{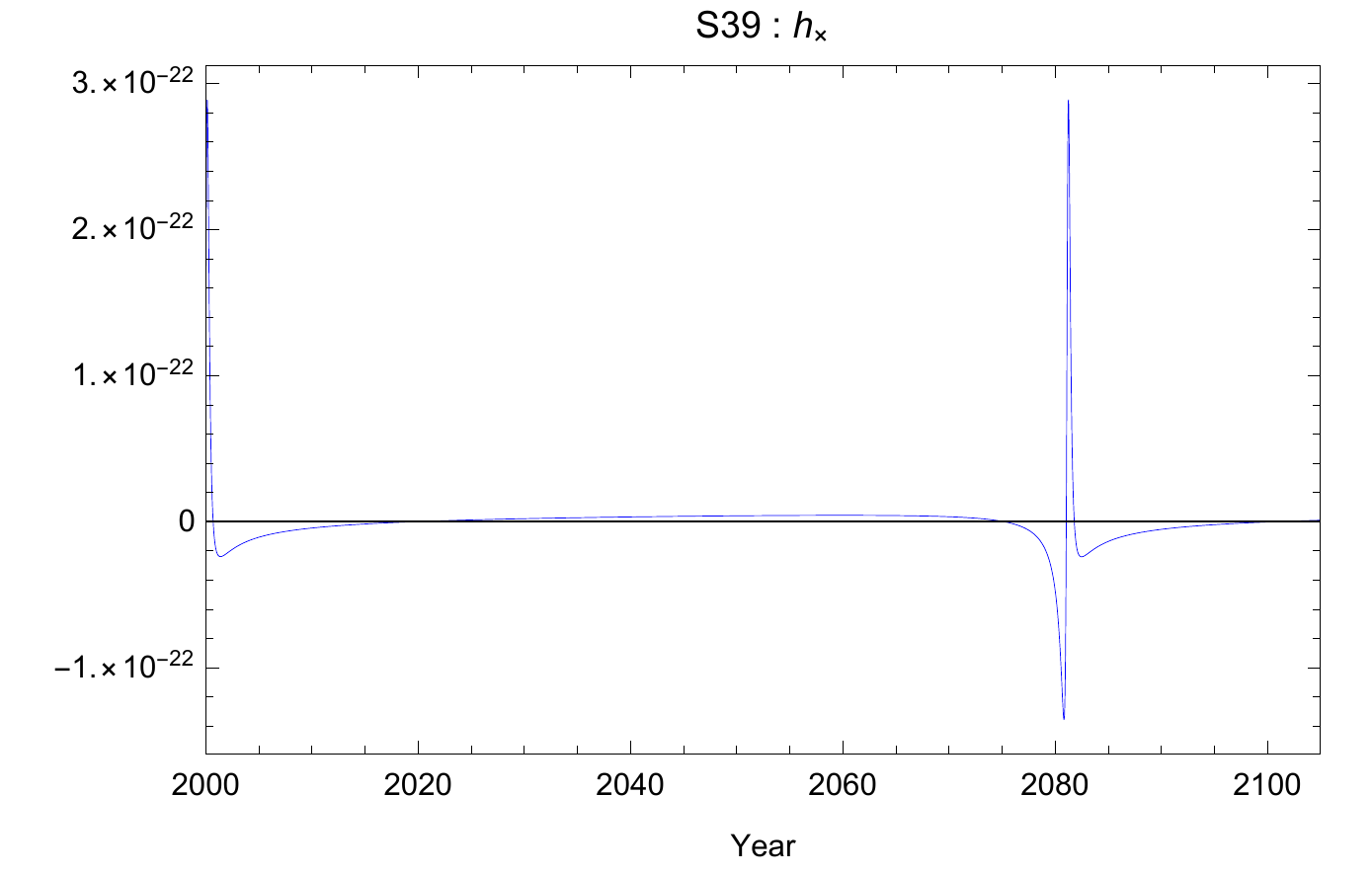}\\
\includegraphics[width=0.4\textwidth]{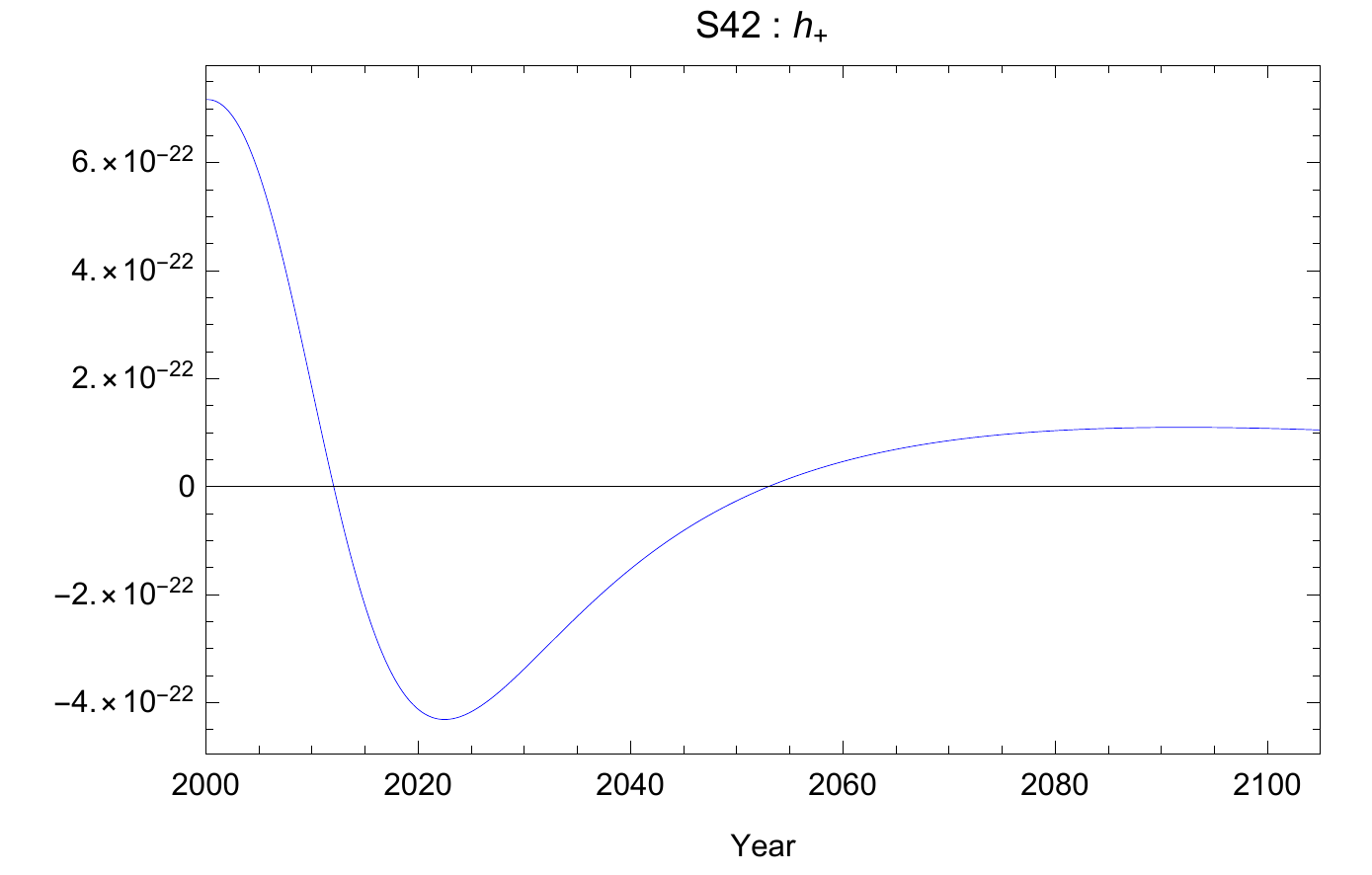}
\includegraphics[width=0.4\textwidth]{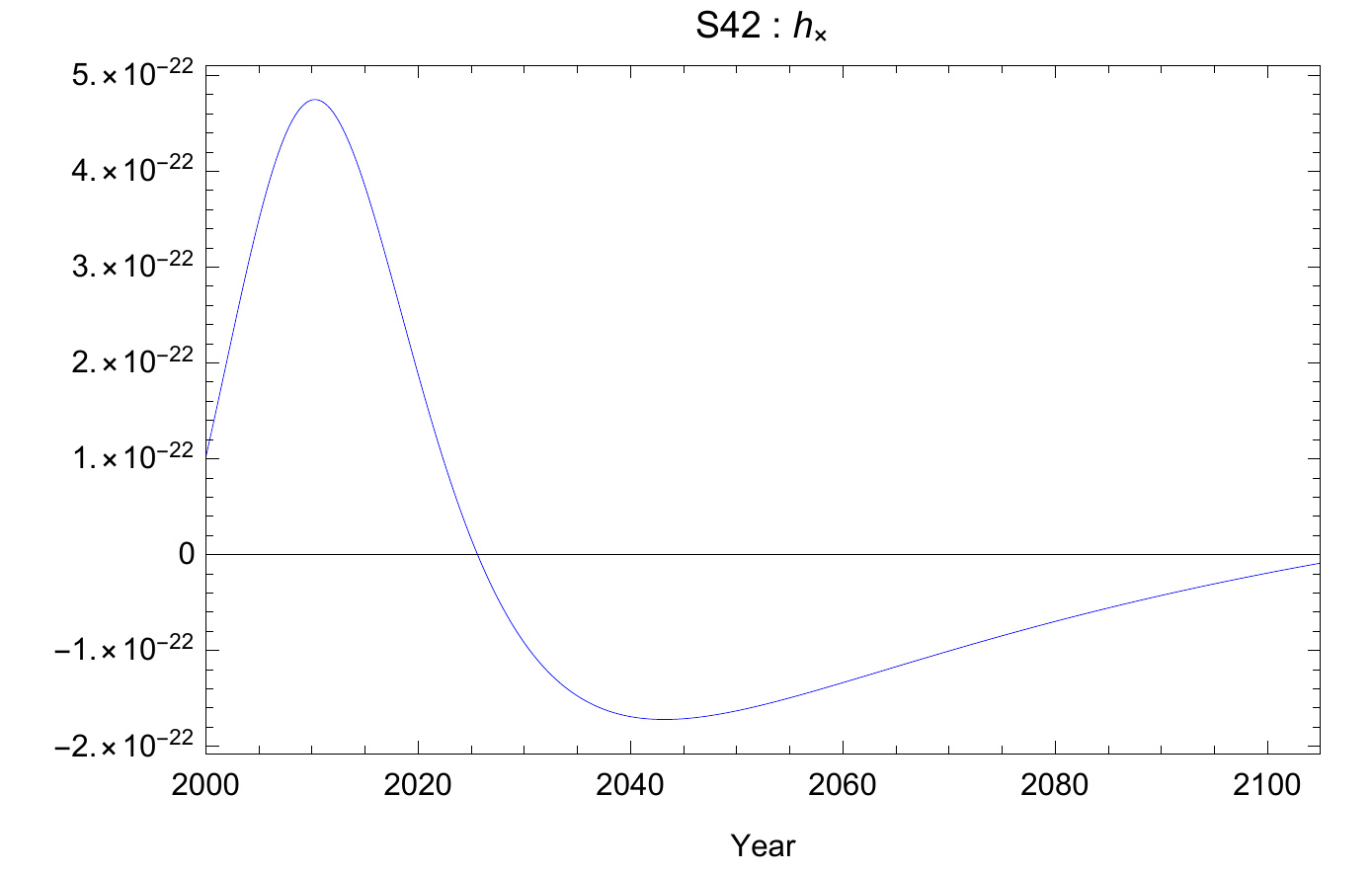}\\
\includegraphics[width=0.4\textwidth]{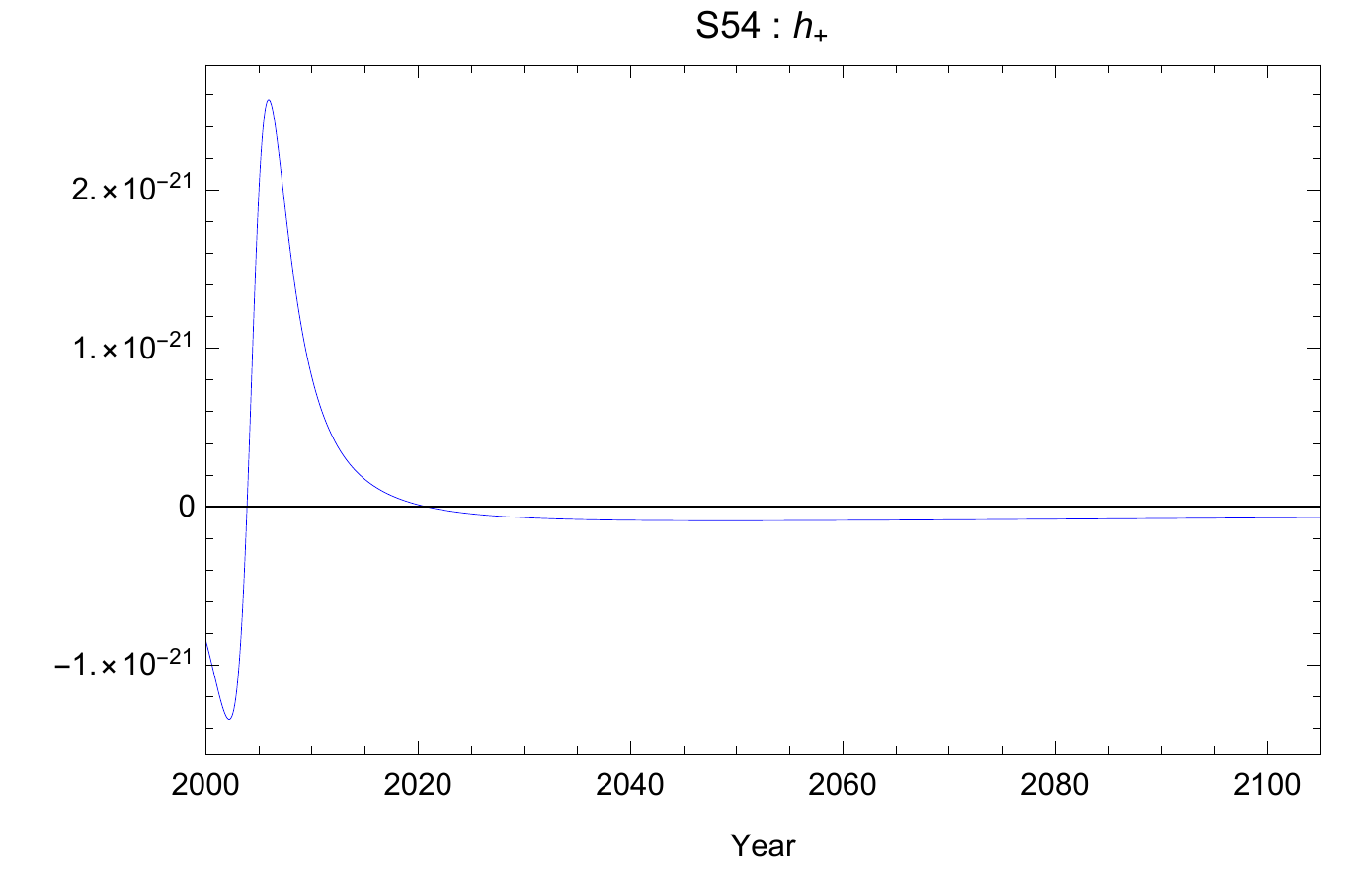}
\includegraphics[width=0.4\textwidth]{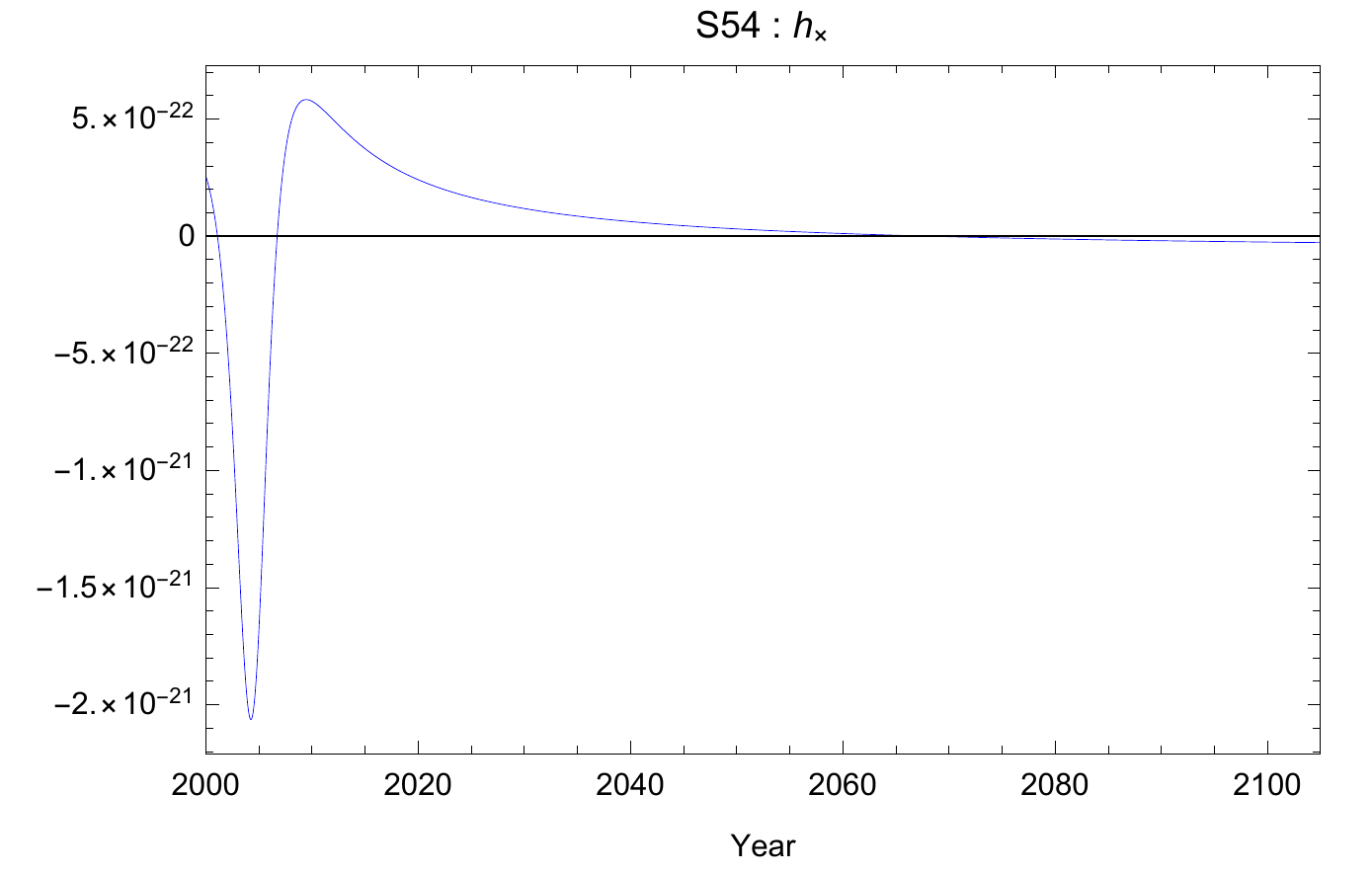}\\
\includegraphics[width=0.4\textwidth]{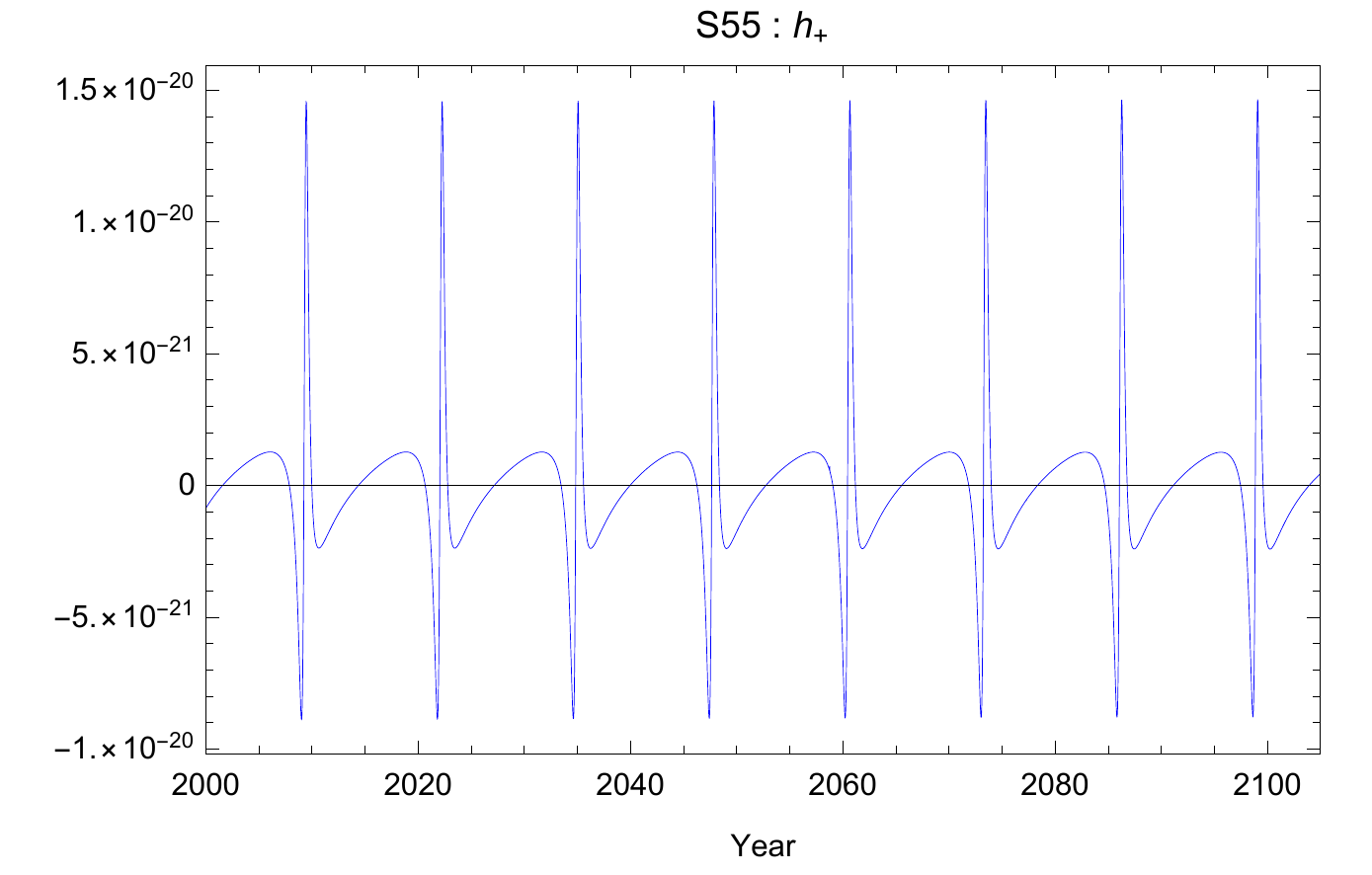}
\includegraphics[width=0.4\textwidth]{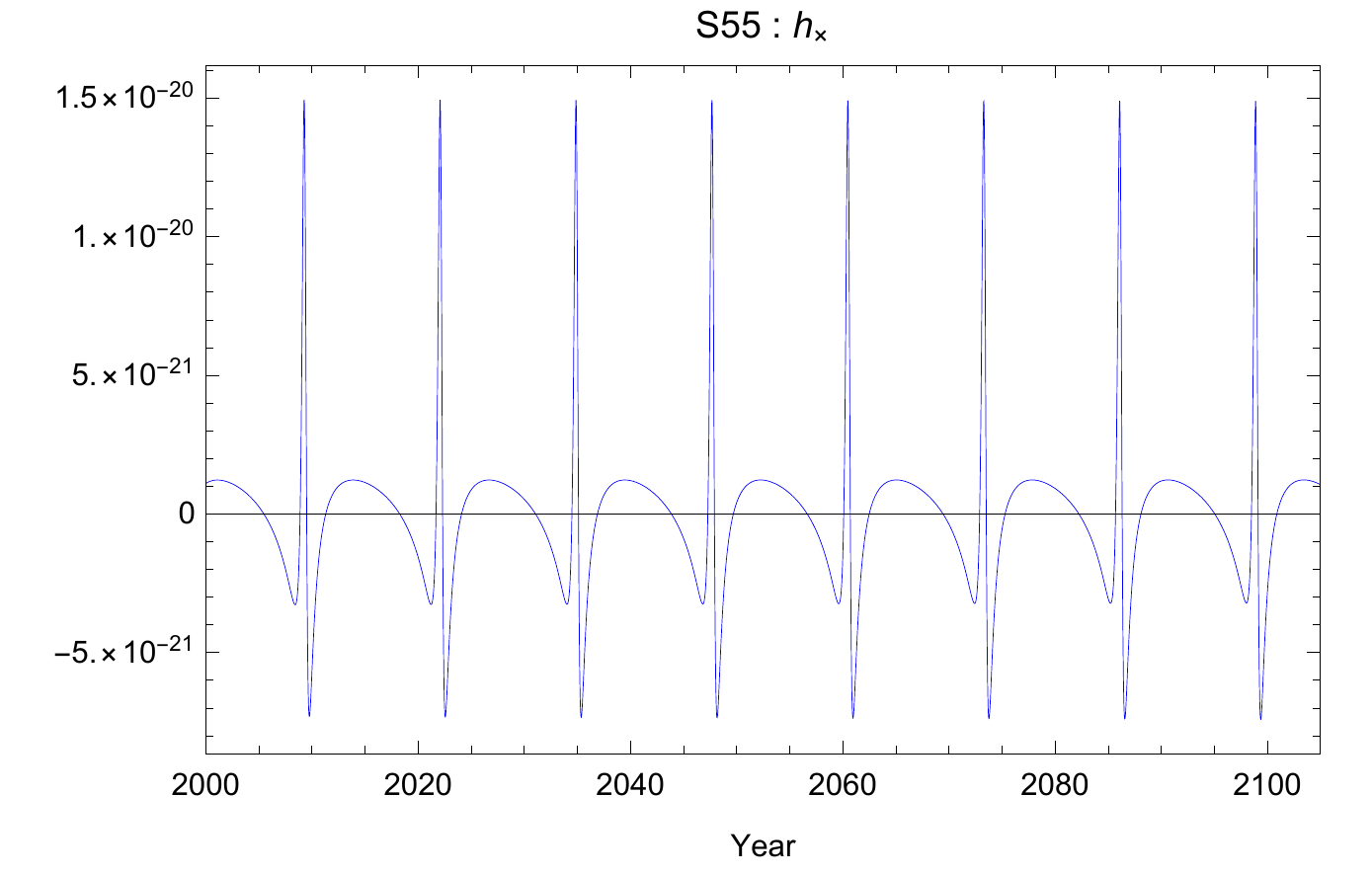}\\
\includegraphics[width=0.4\textwidth]{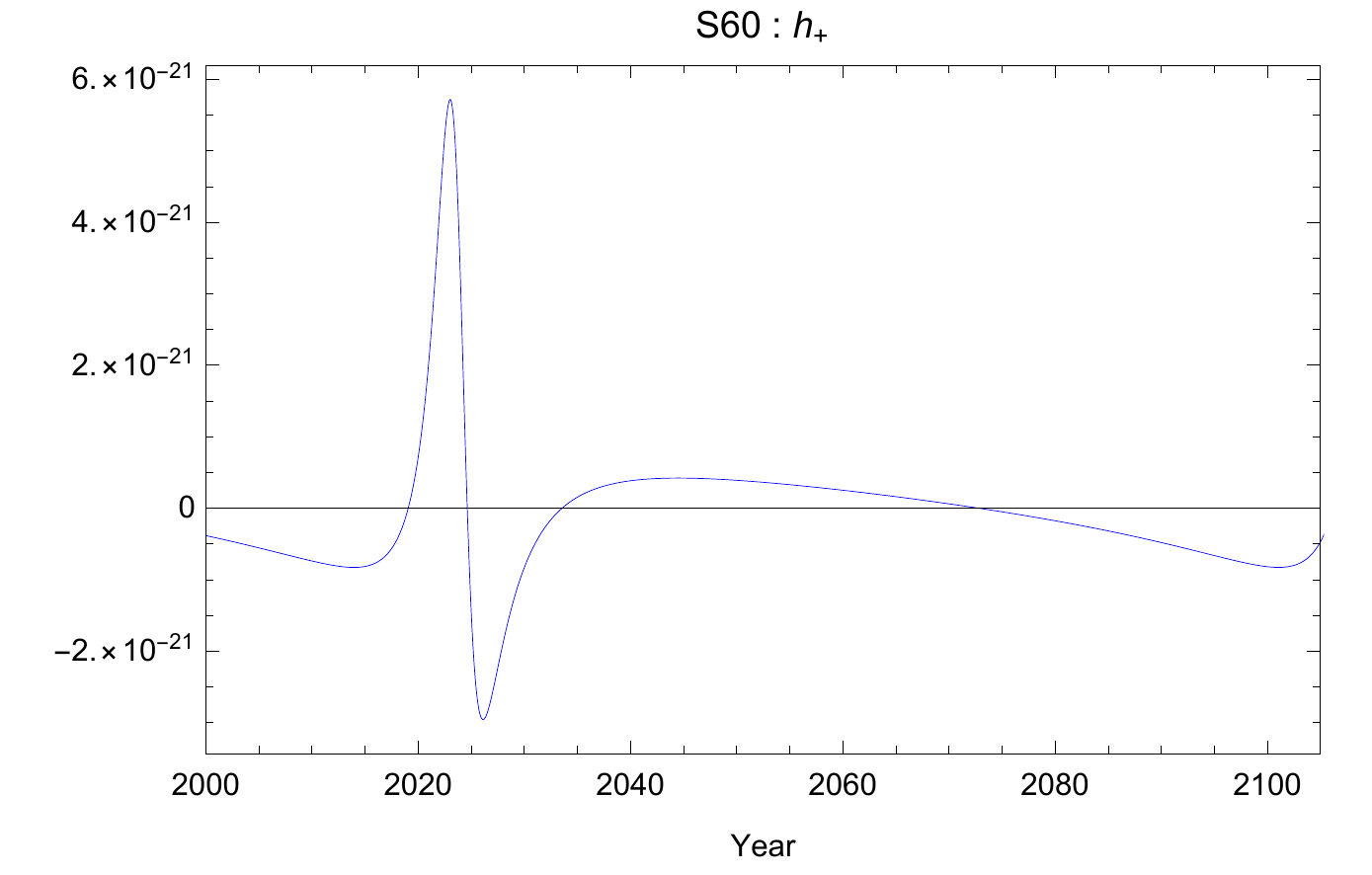}
\includegraphics[width=0.4\textwidth]{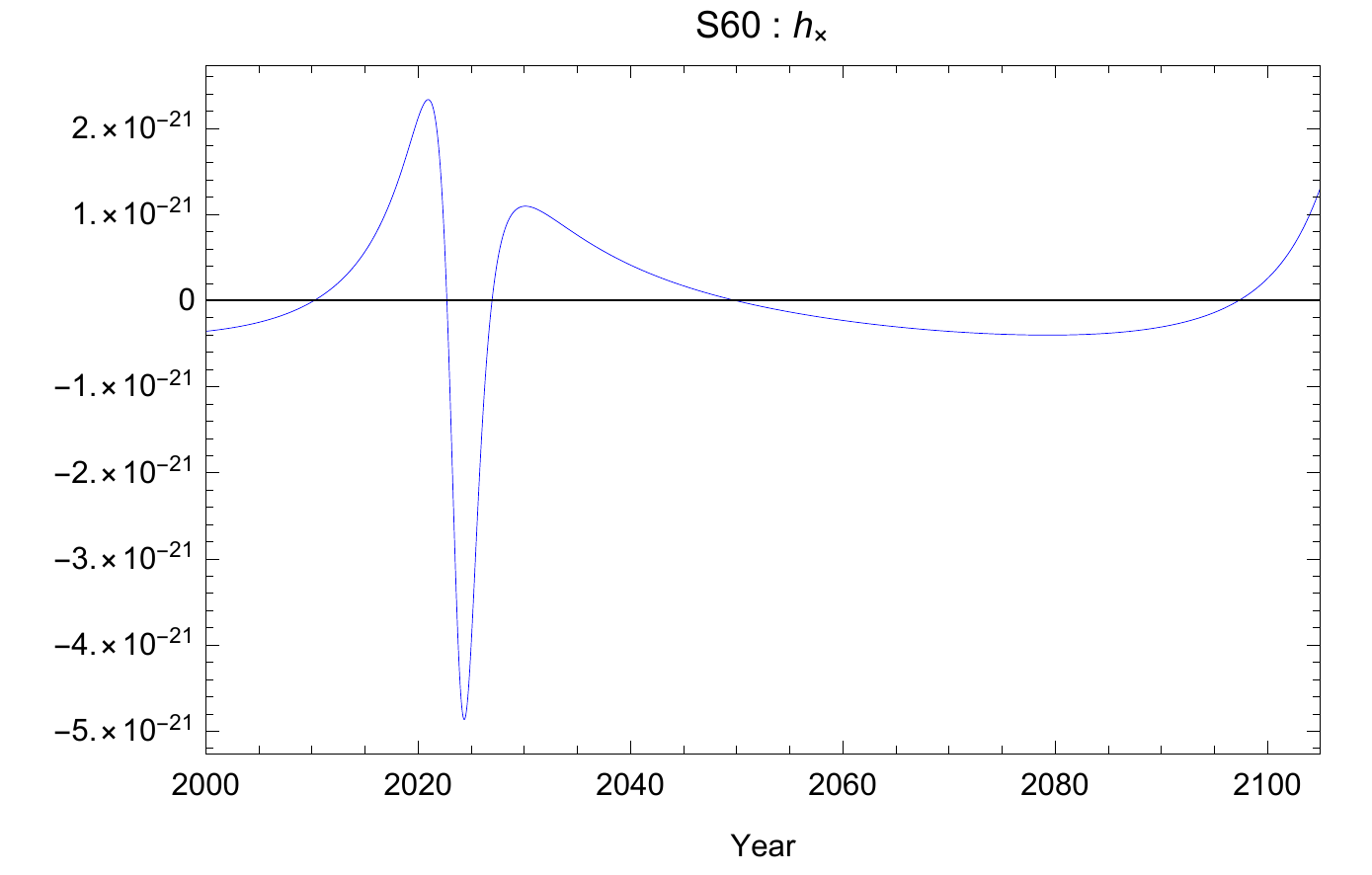}
\end{figure*} 

\begin{figure*}
\includegraphics[width=0.4\textwidth]{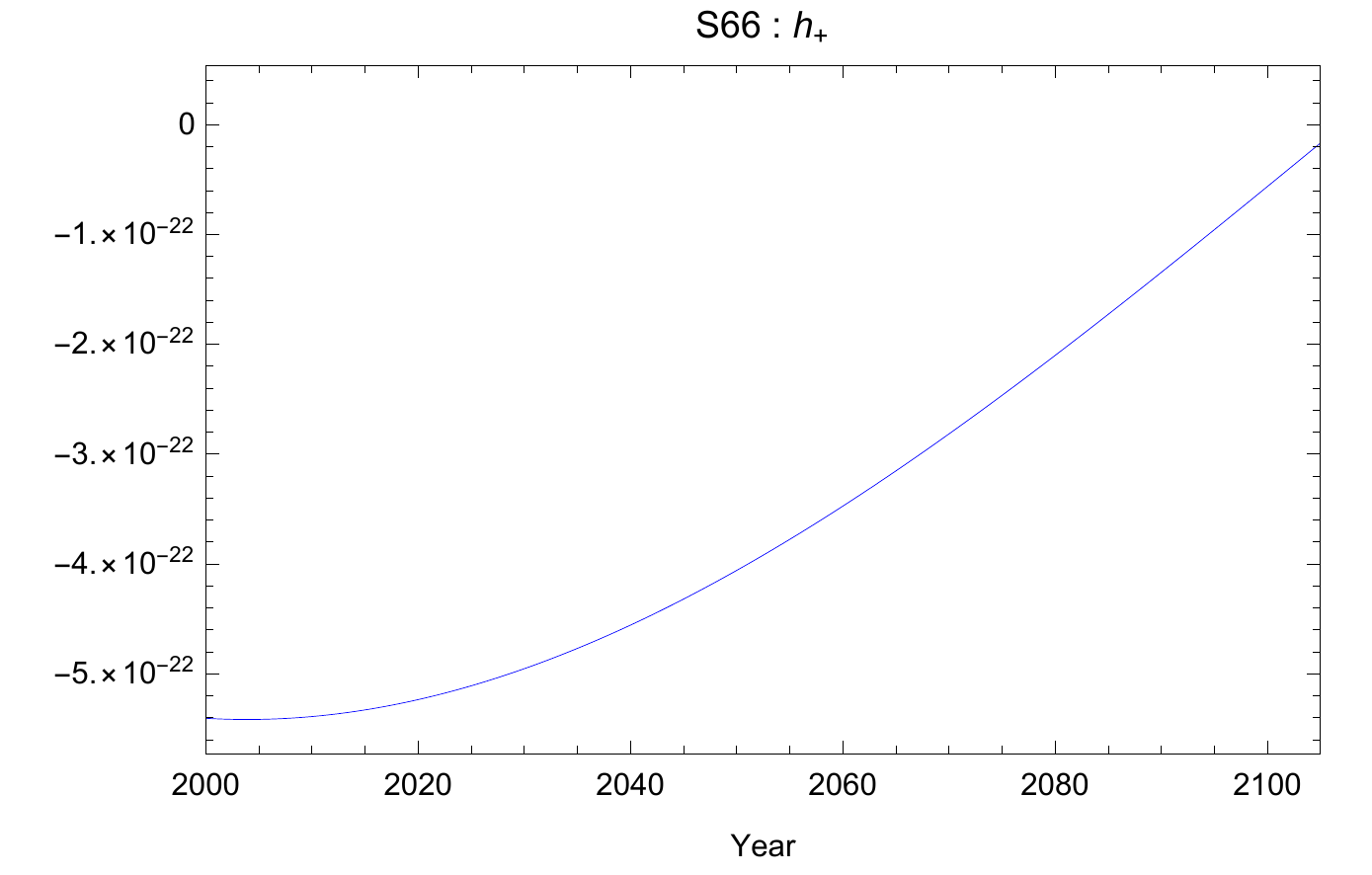}
\includegraphics[width=0.4\textwidth]{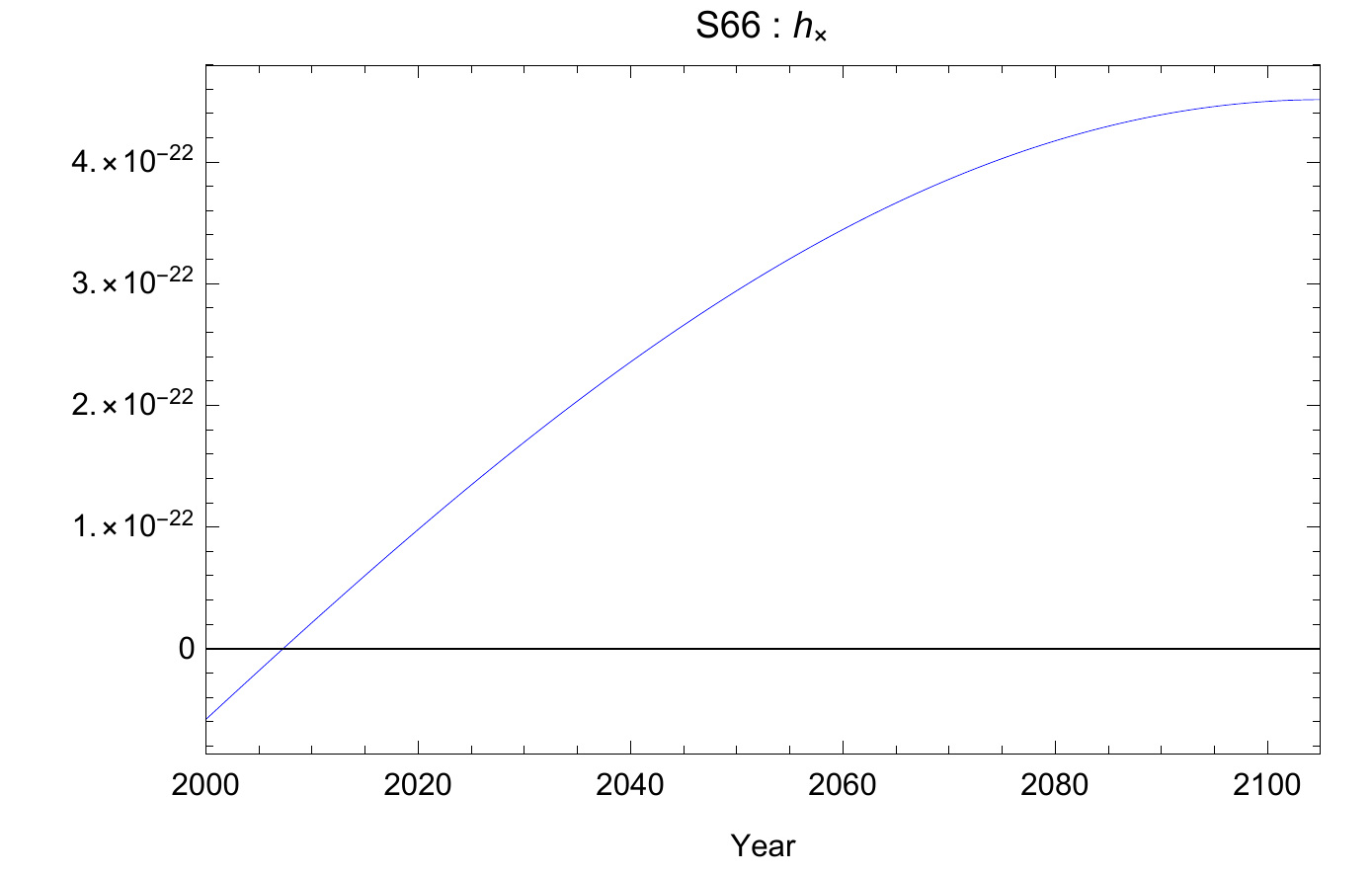}\\
\includegraphics[width=0.4\textwidth]{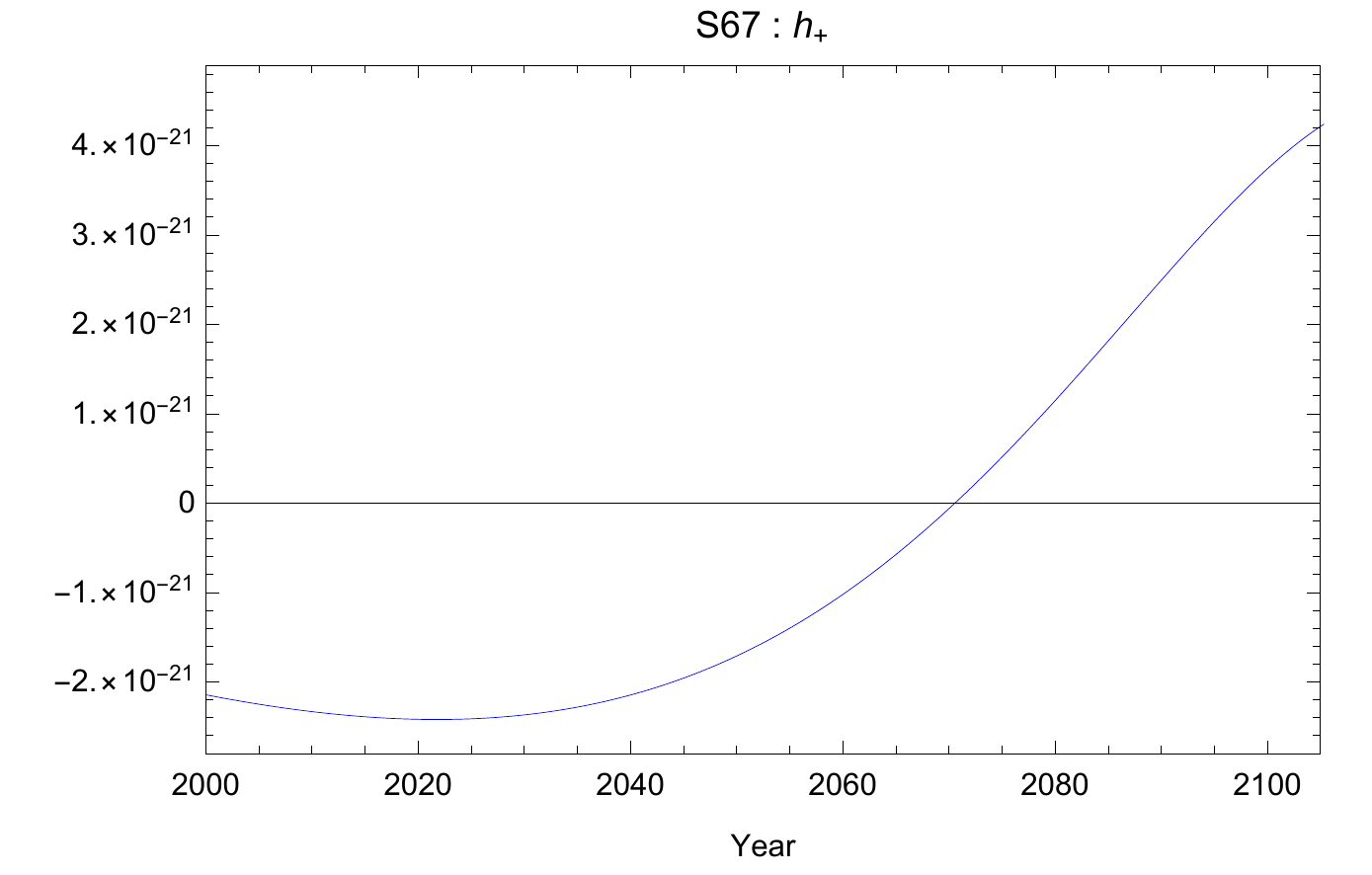}
\includegraphics[width=0.4\textwidth]{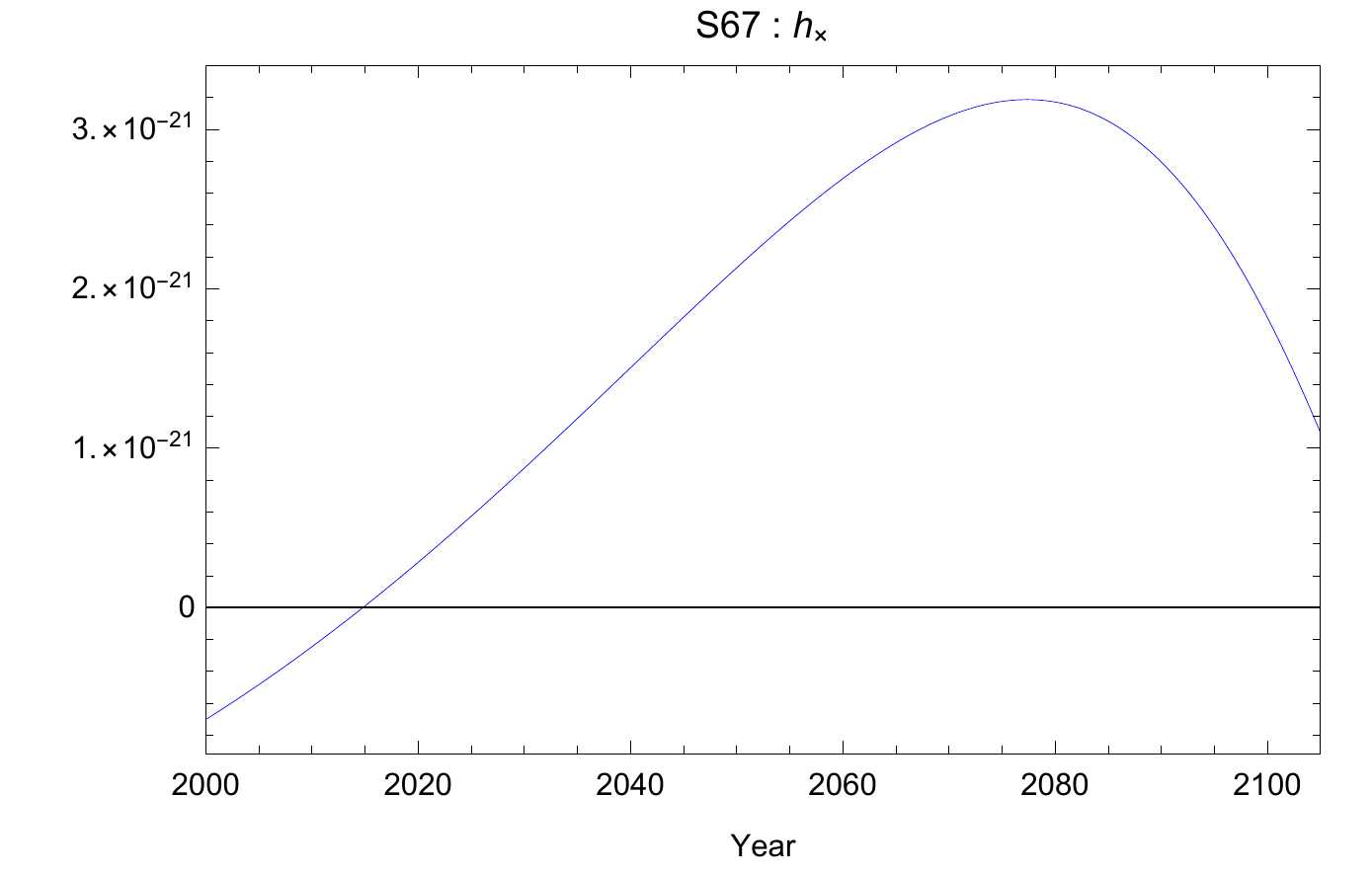}\\
\includegraphics[width=0.4\textwidth]{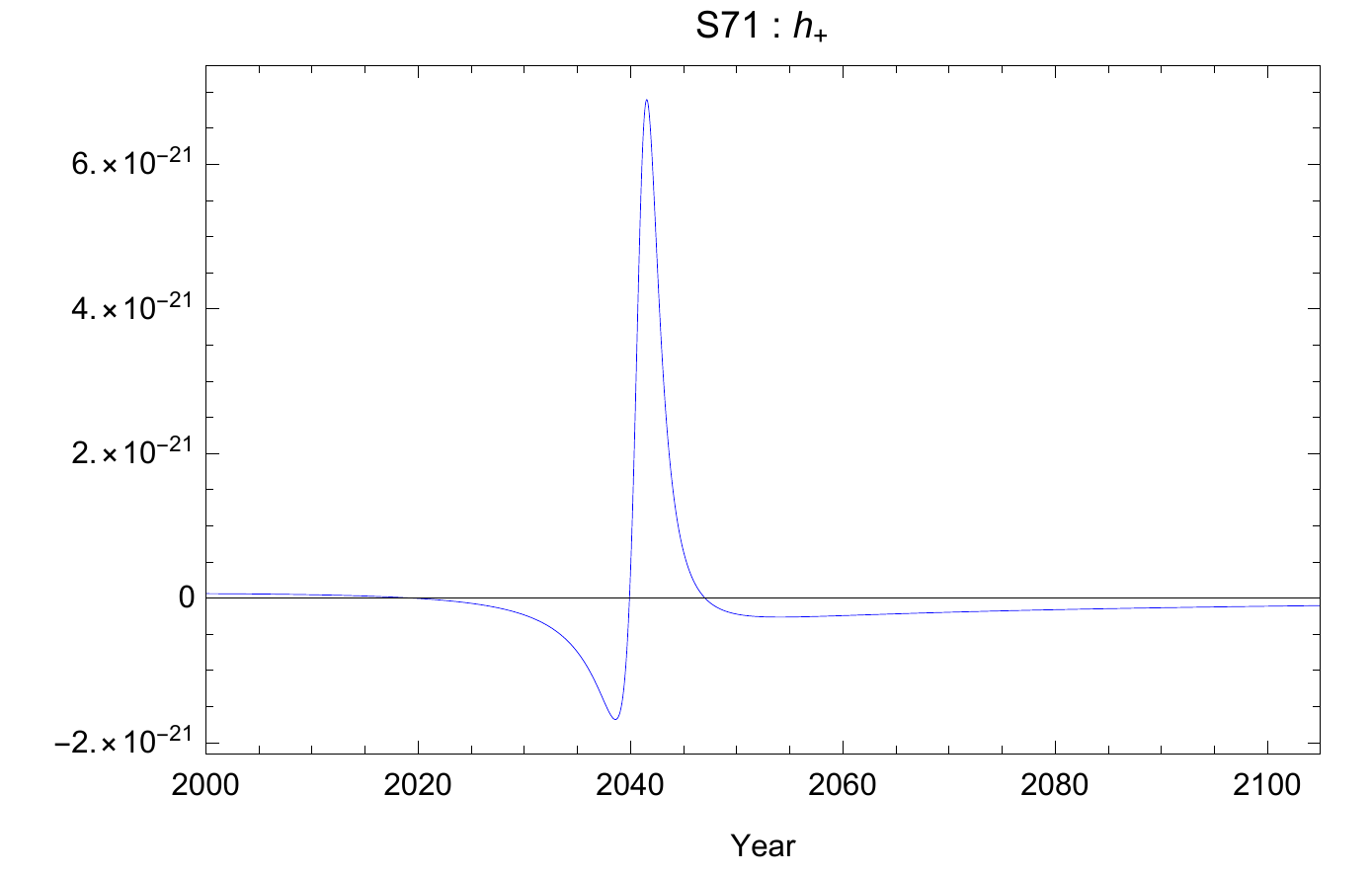}
\includegraphics[width=0.4\textwidth]{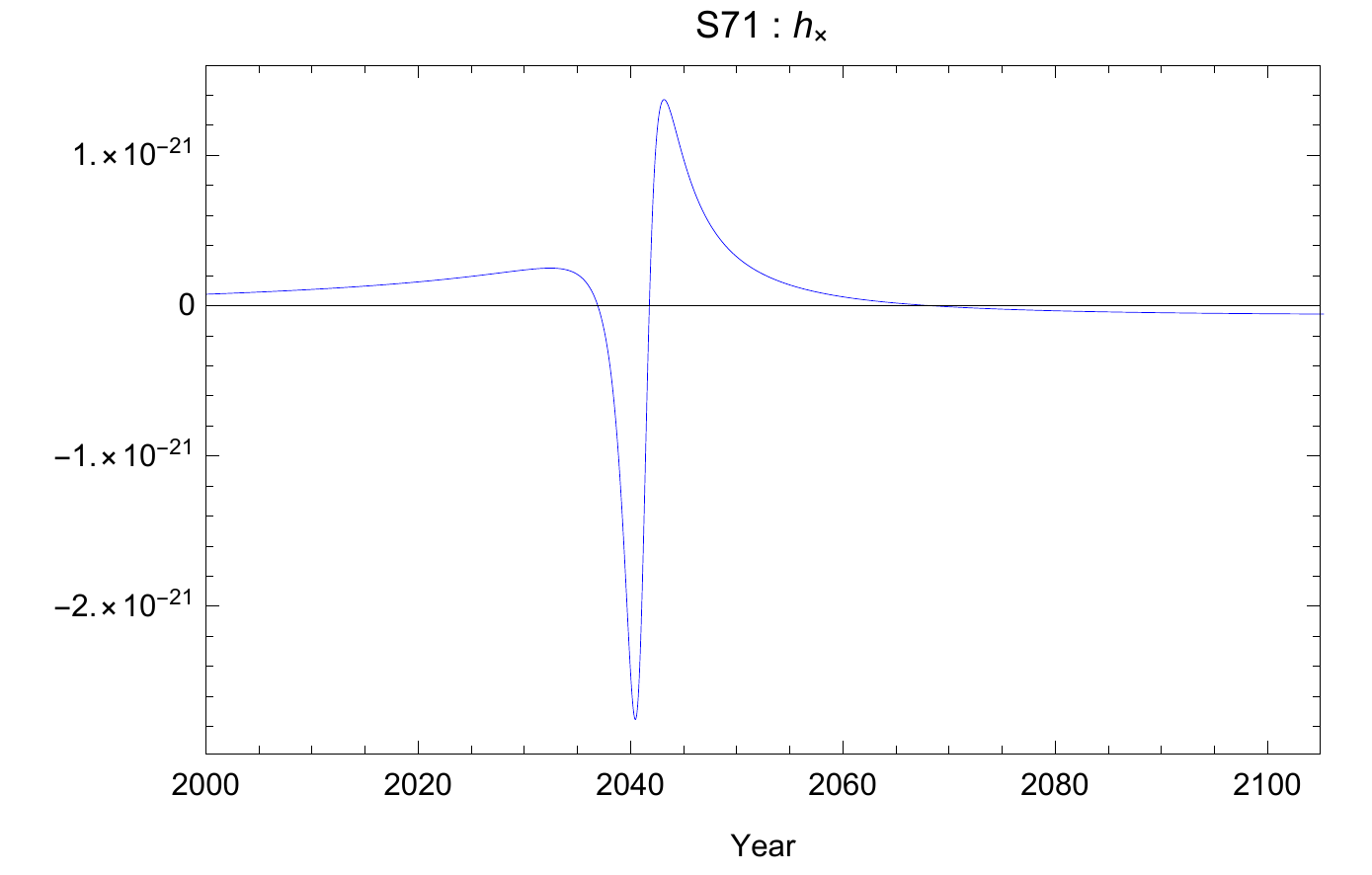}\\
\includegraphics[width=0.4\textwidth]{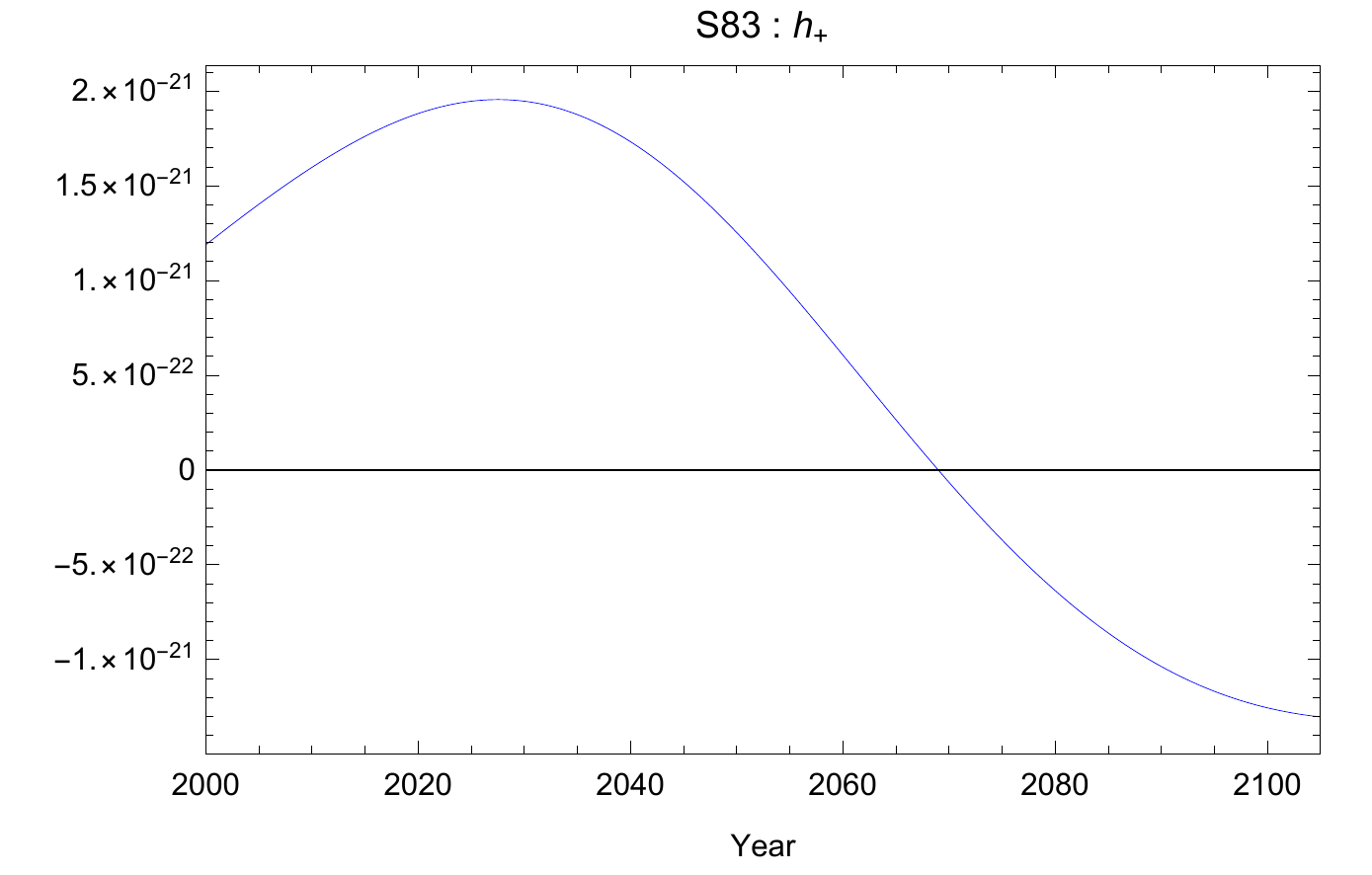}
\includegraphics[width=0.4\textwidth]{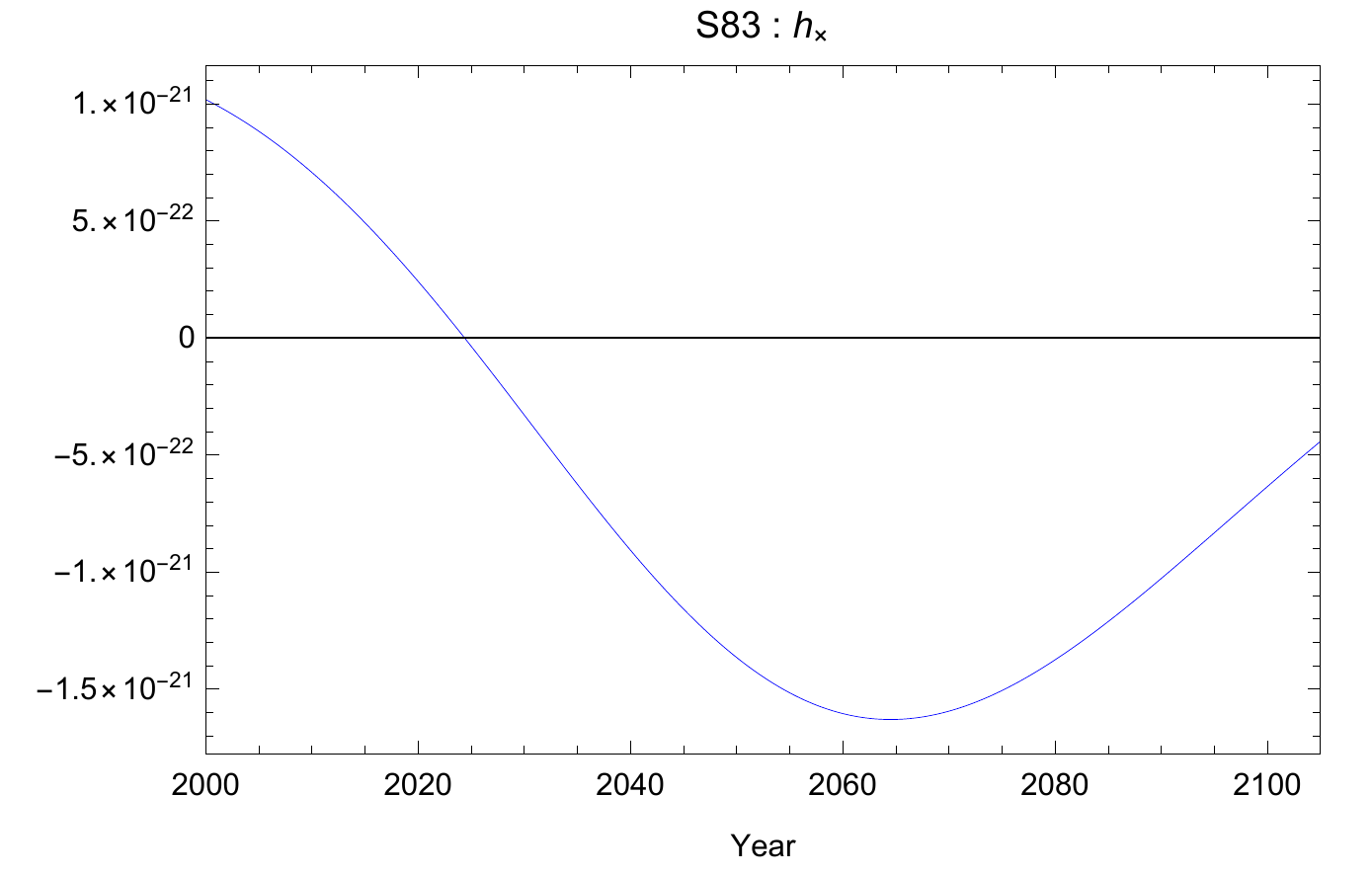}\\
\includegraphics[width=0.4\textwidth]{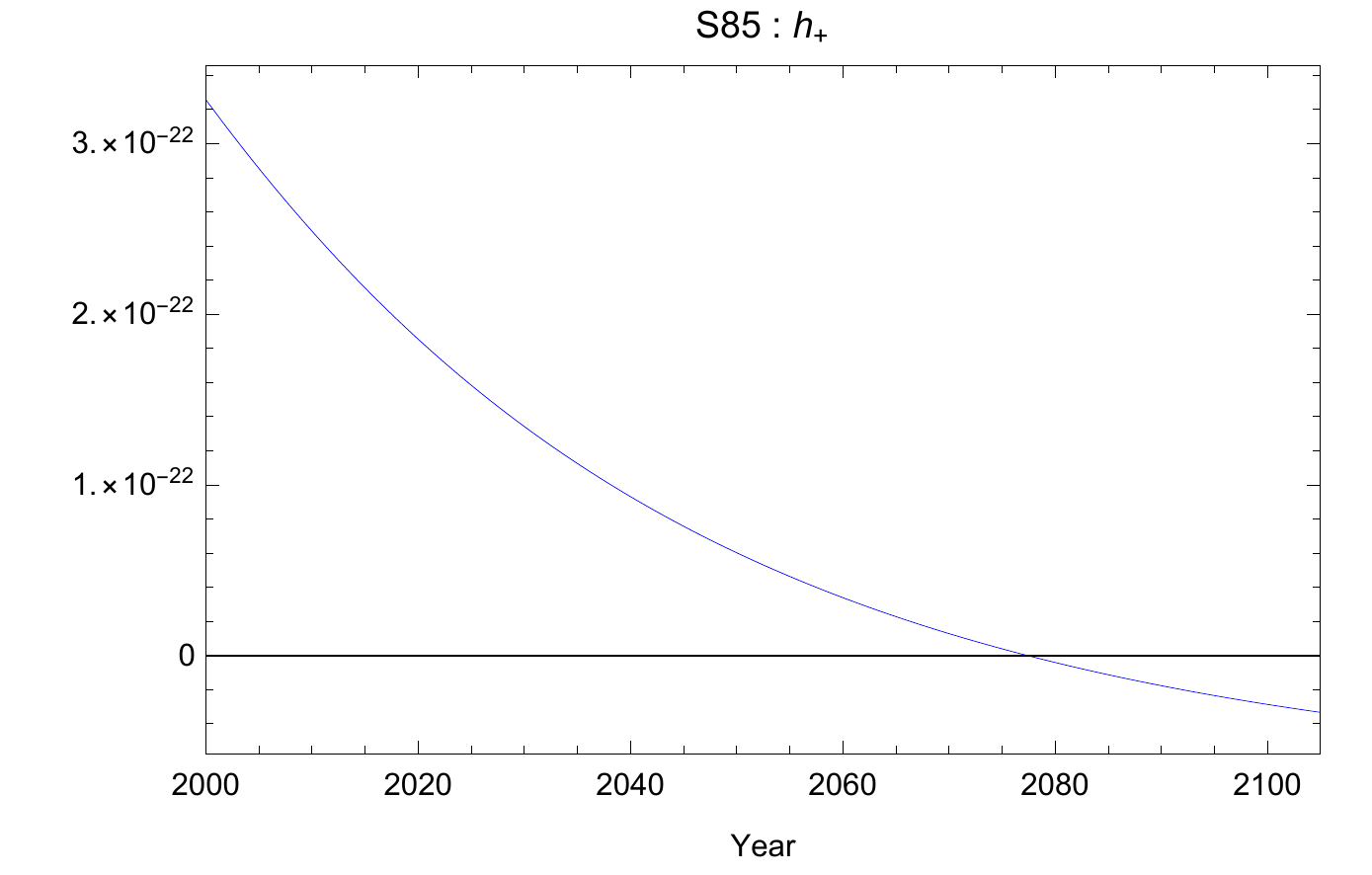}
\includegraphics[width=0.4\textwidth]{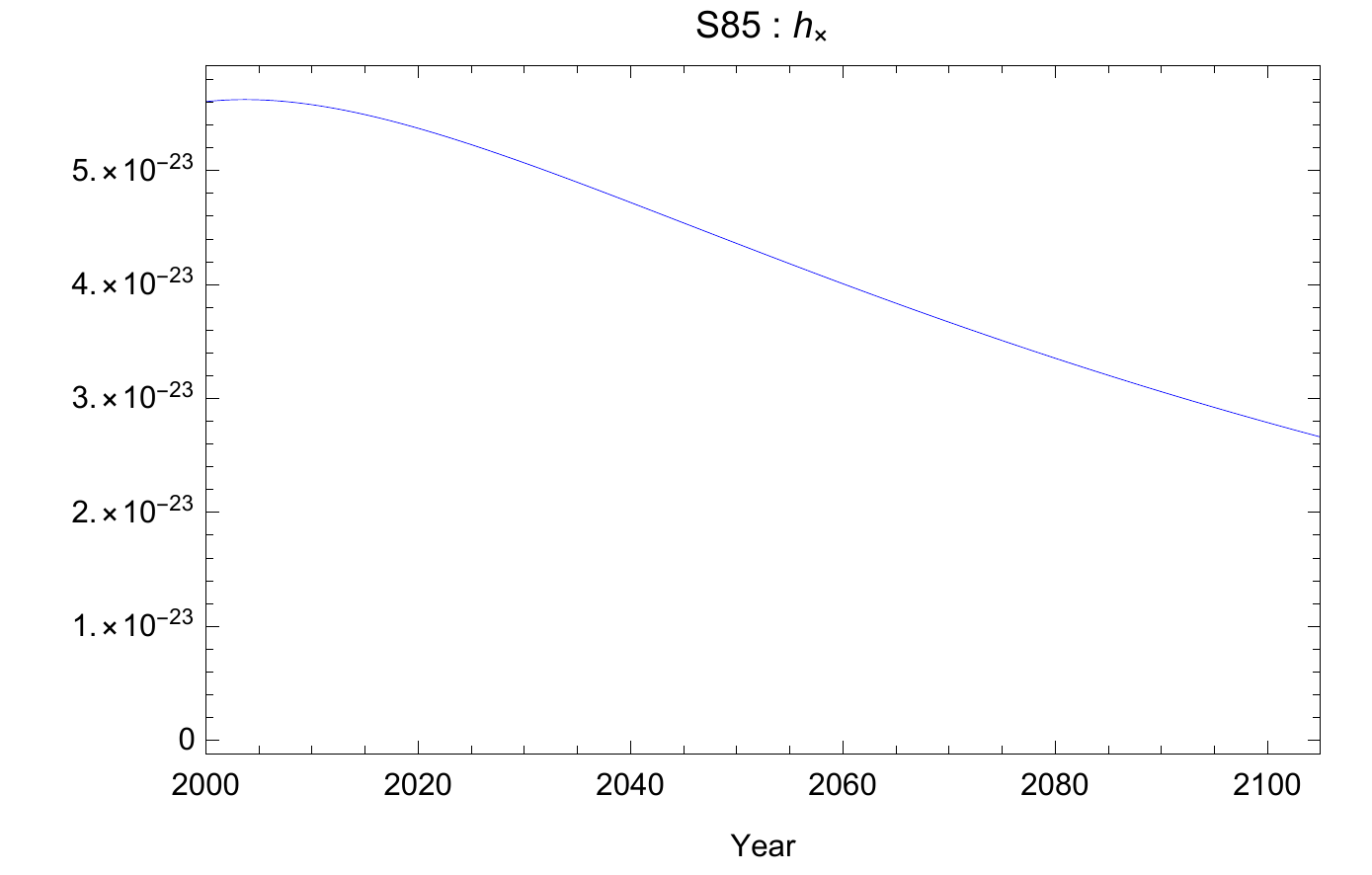}
\end{figure*} 

\begin{figure*}
\includegraphics[width=0.4\textwidth]{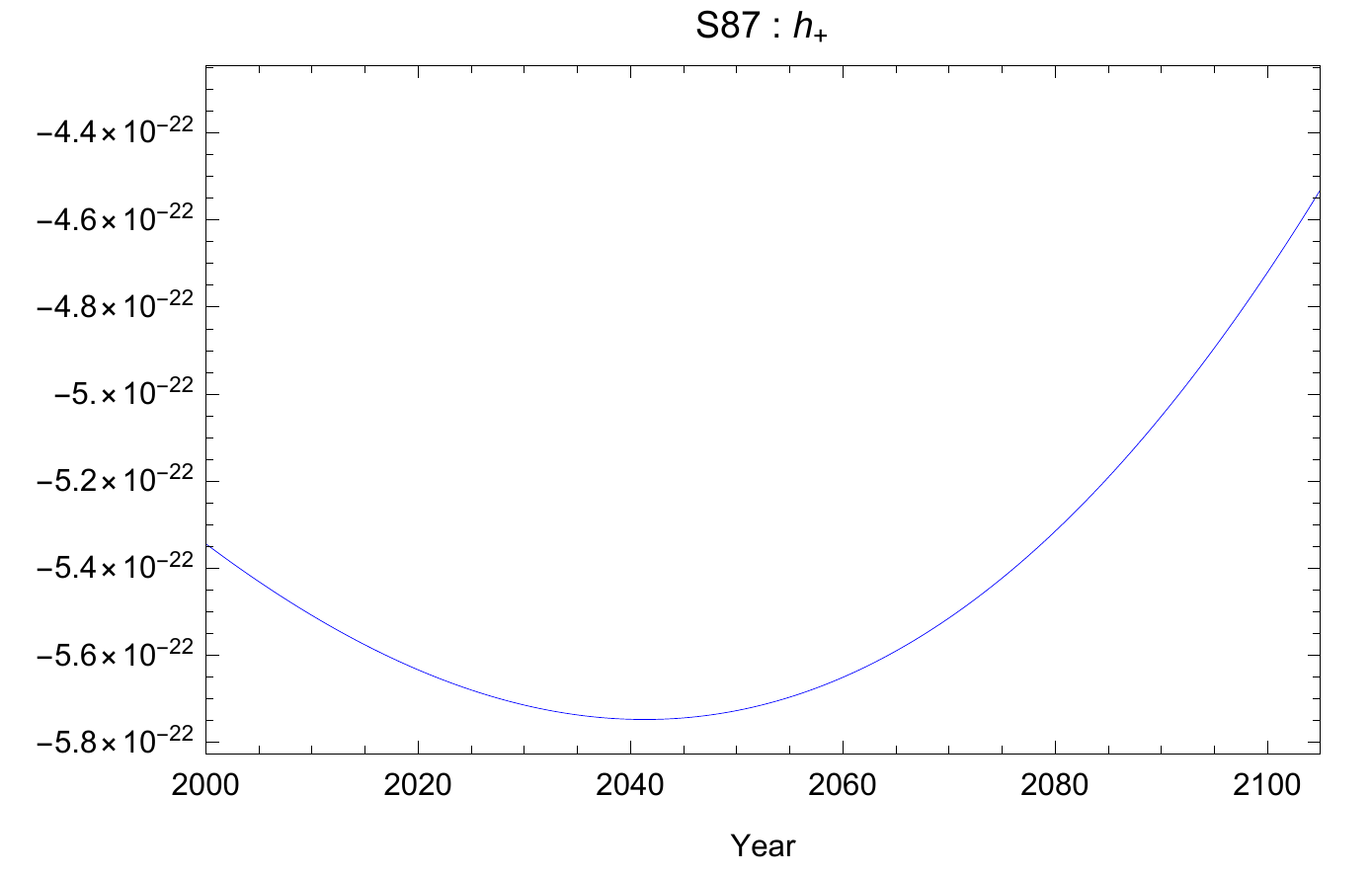}
\includegraphics[width=0.4\textwidth]{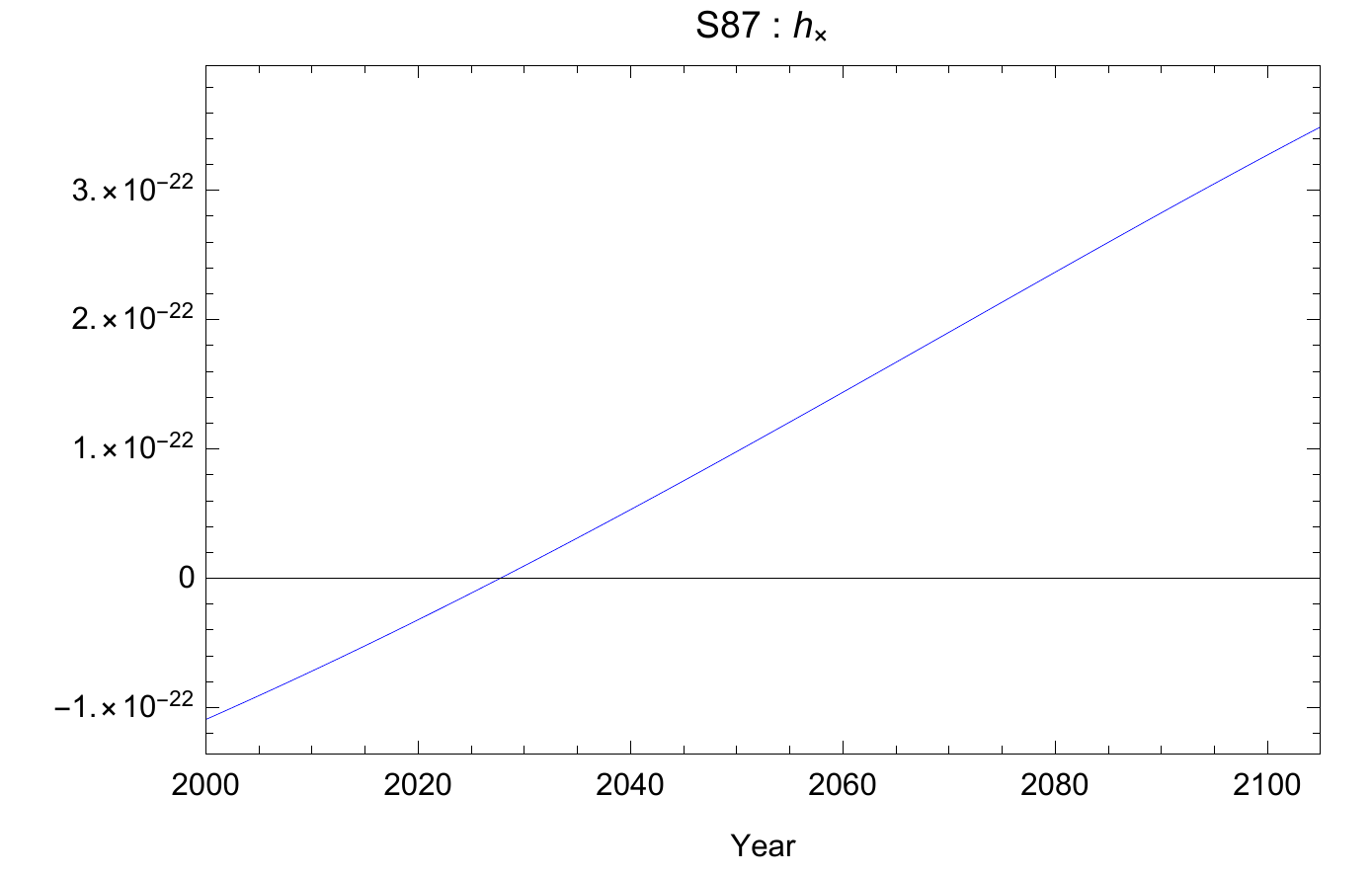}\\
\includegraphics[width=0.4\textwidth]{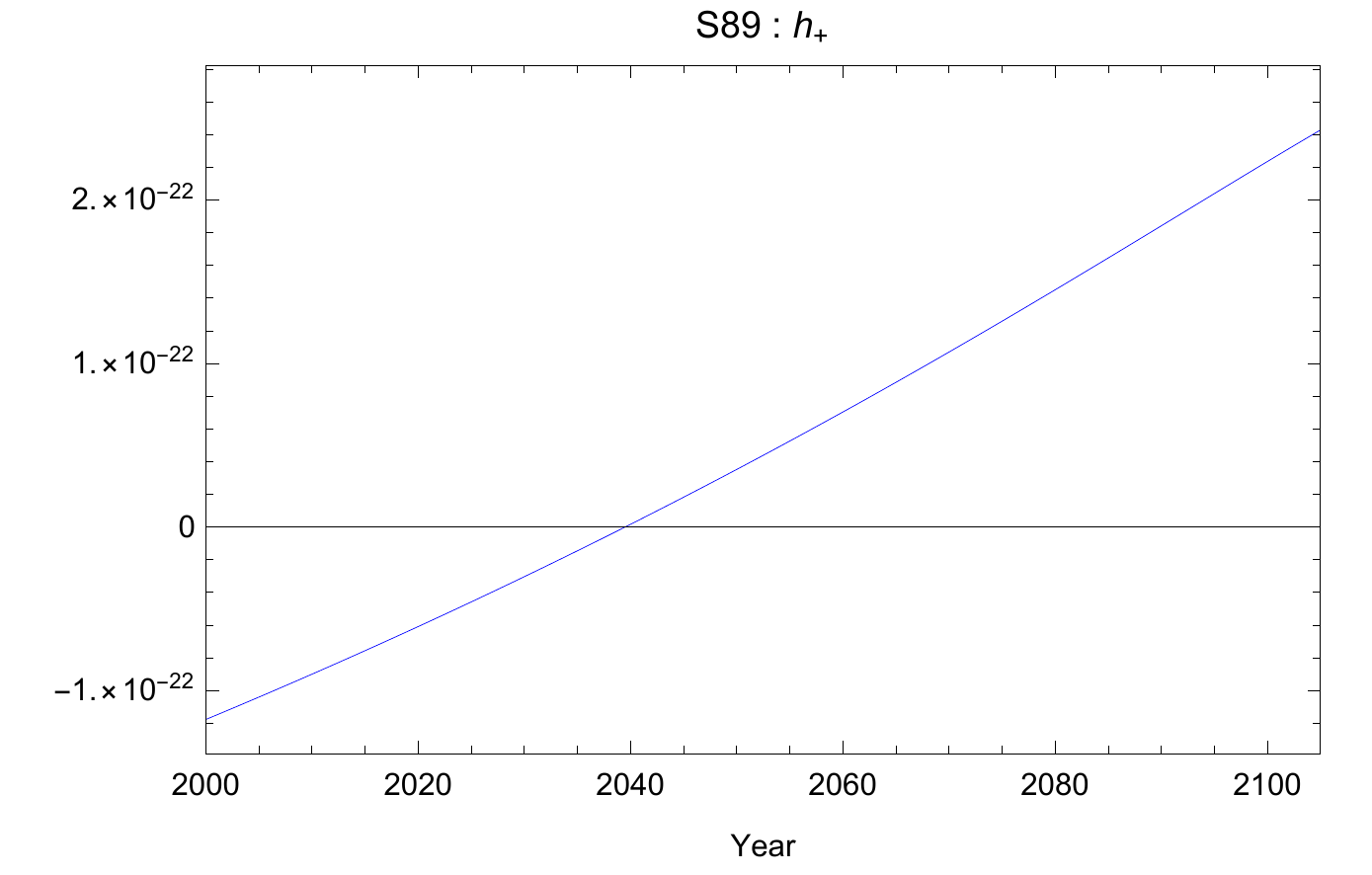}
\includegraphics[width=0.4\textwidth]{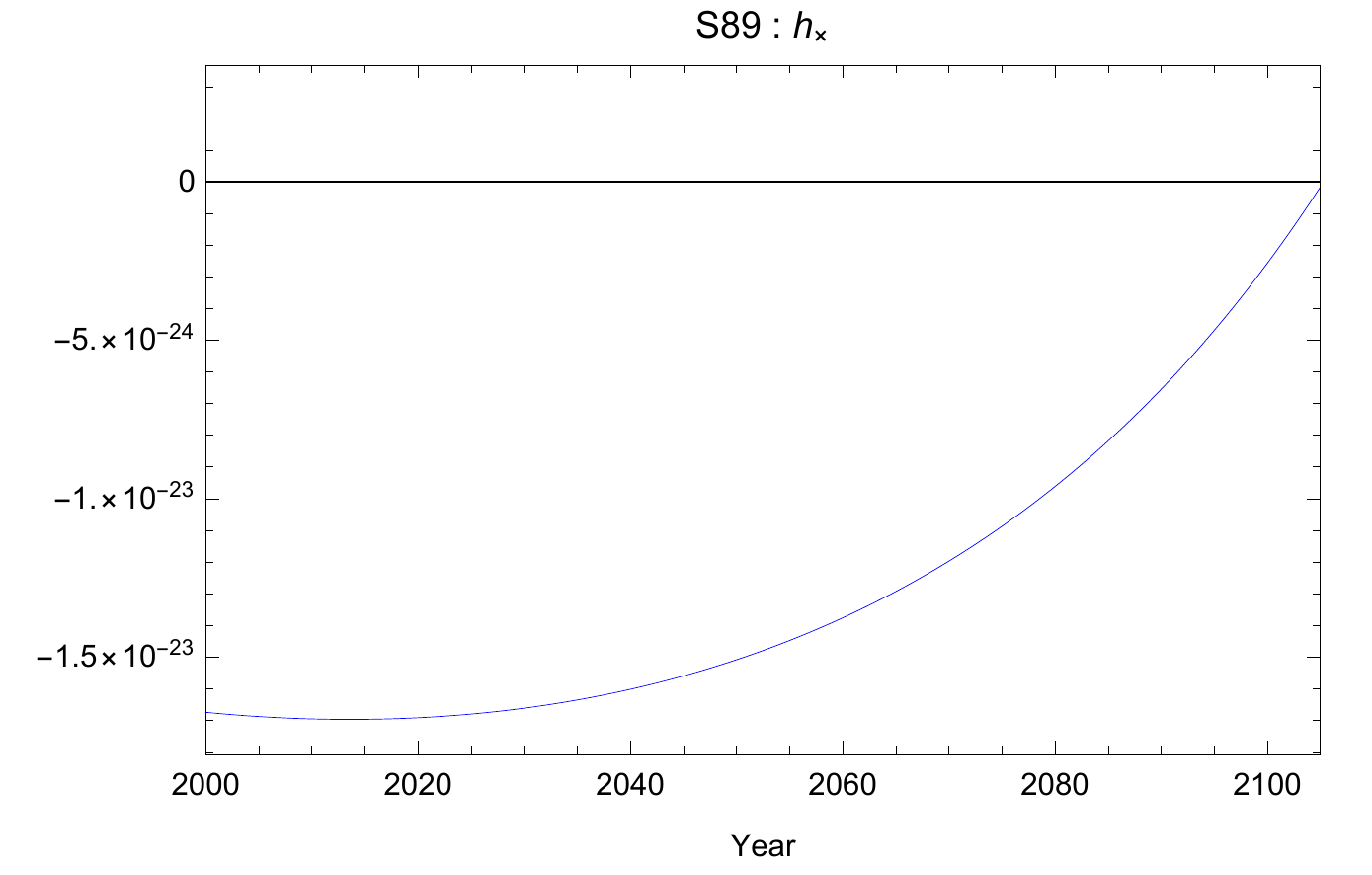}\\
\includegraphics[width=0.4\textwidth]{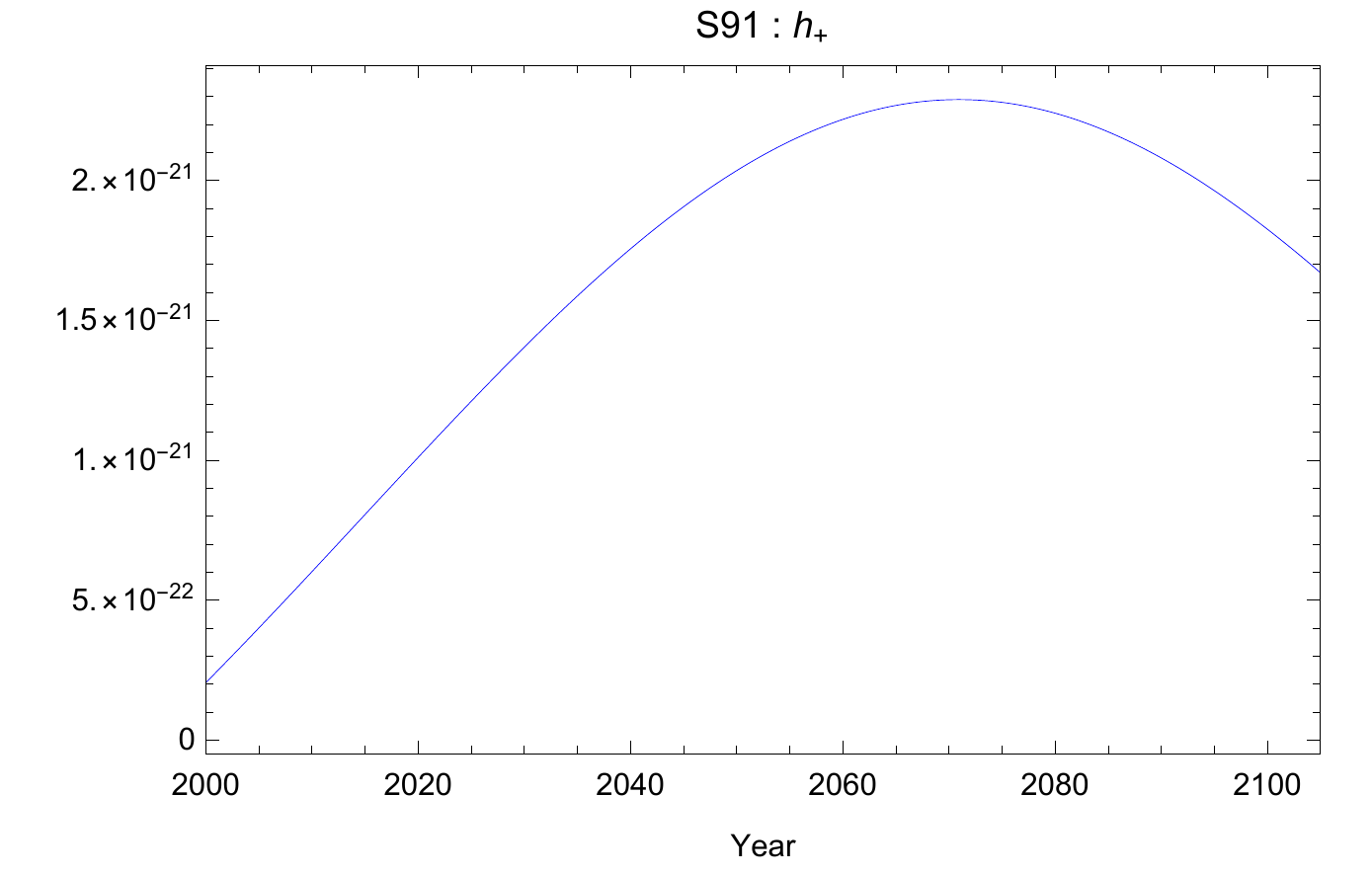}
\includegraphics[width=0.4\textwidth]{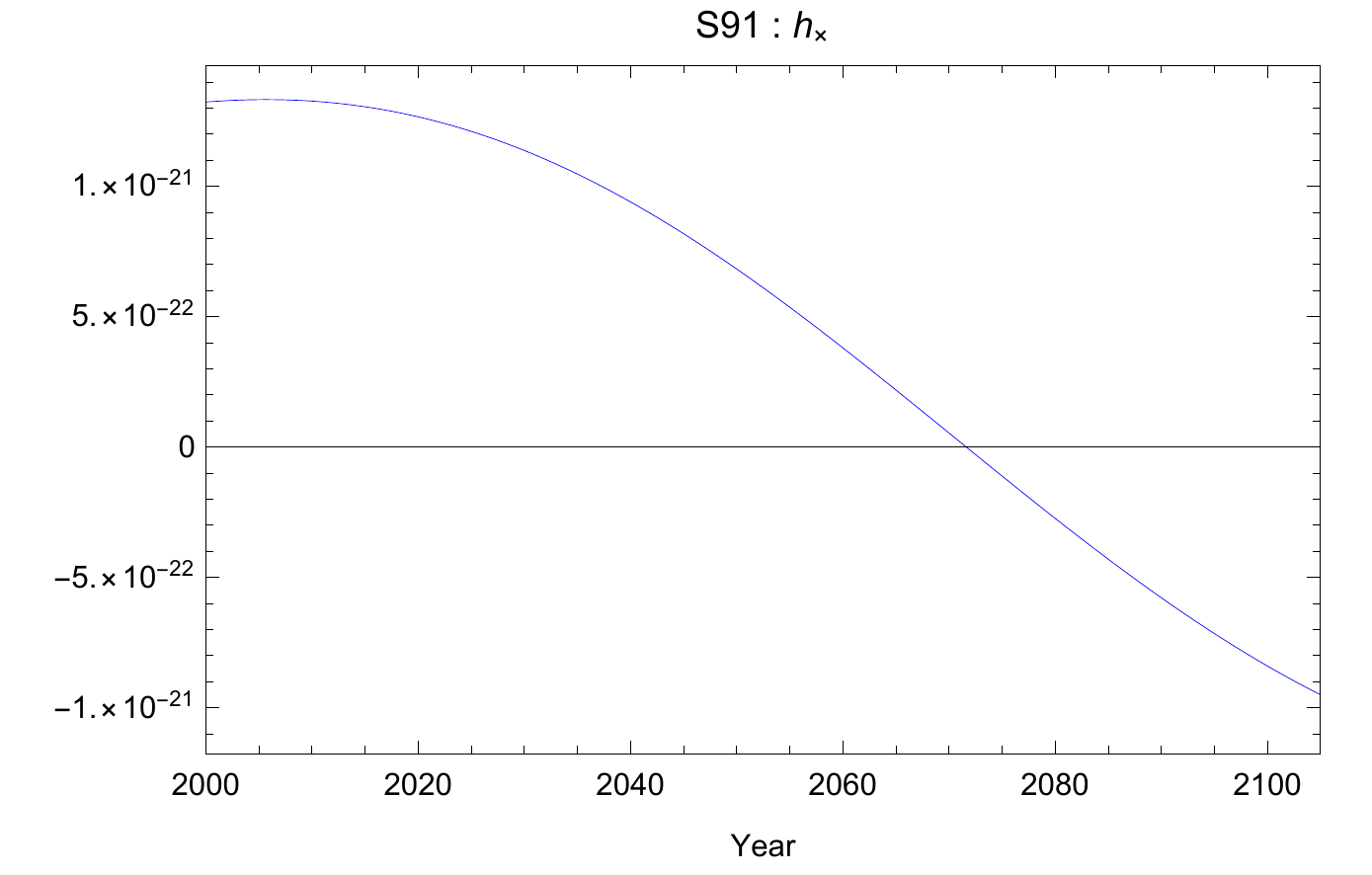}\\
\includegraphics[width=0.4\textwidth]{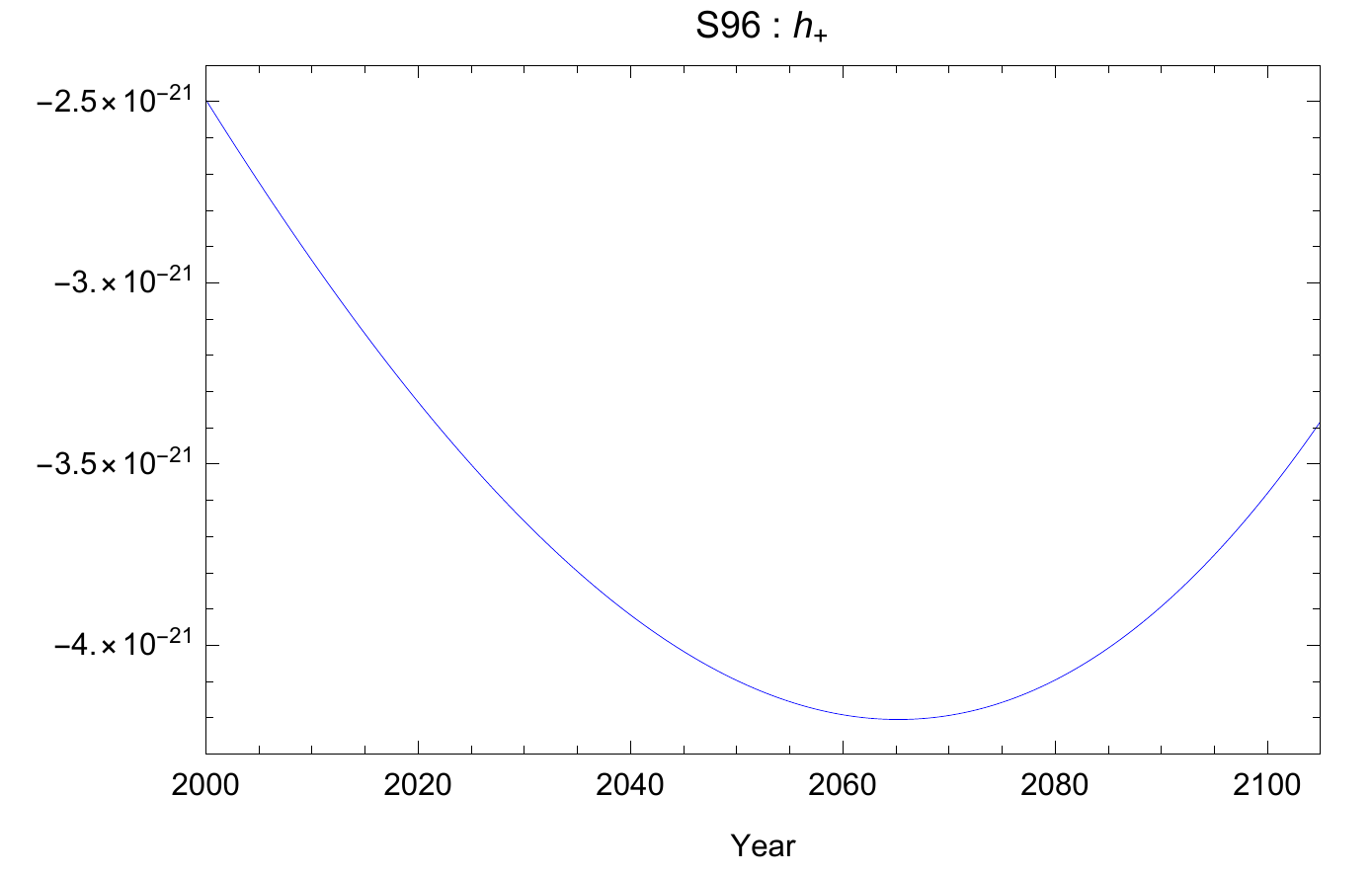}
\includegraphics[width=0.4\textwidth]{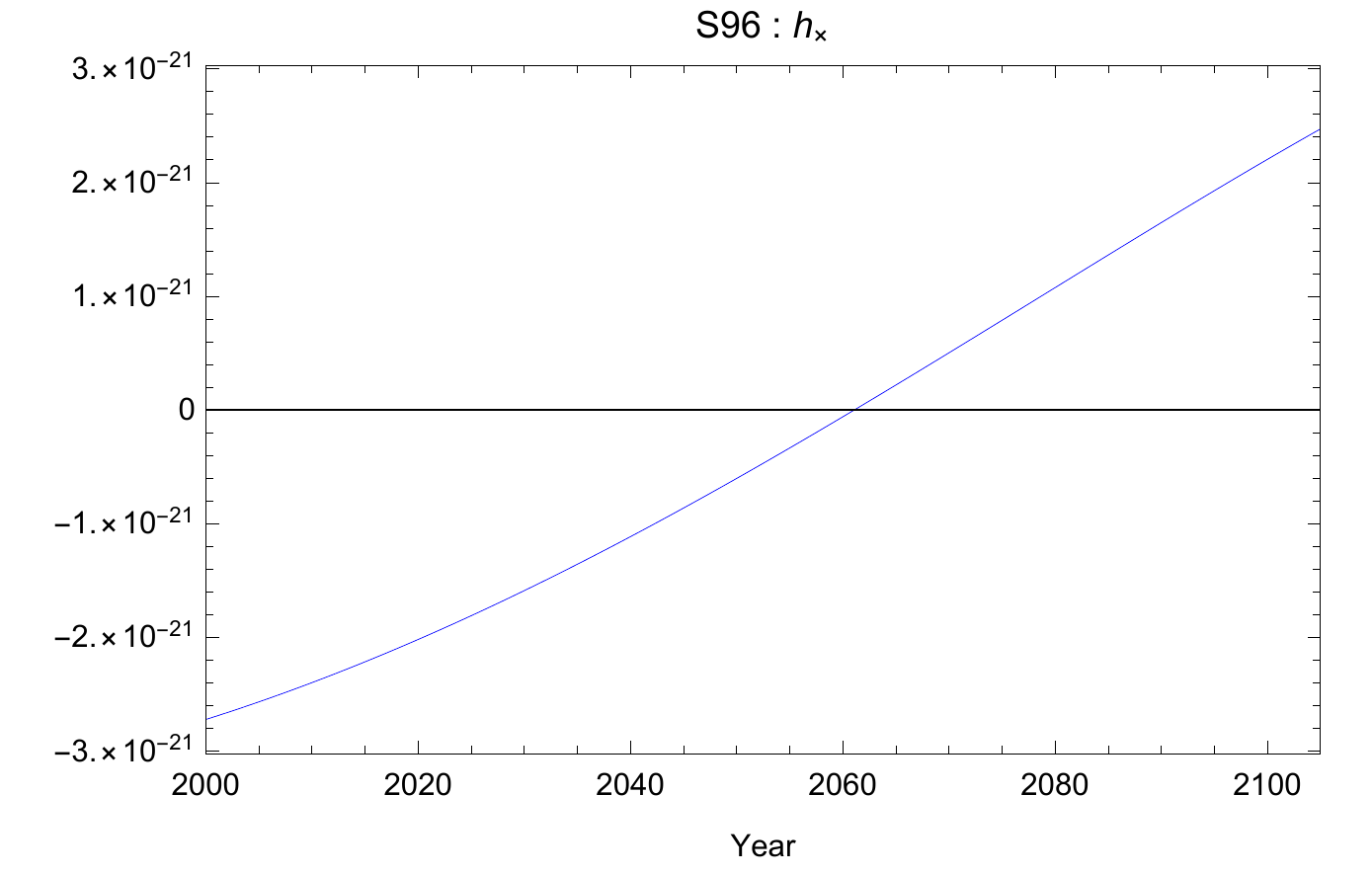}\\
\includegraphics[width=0.4\textwidth]{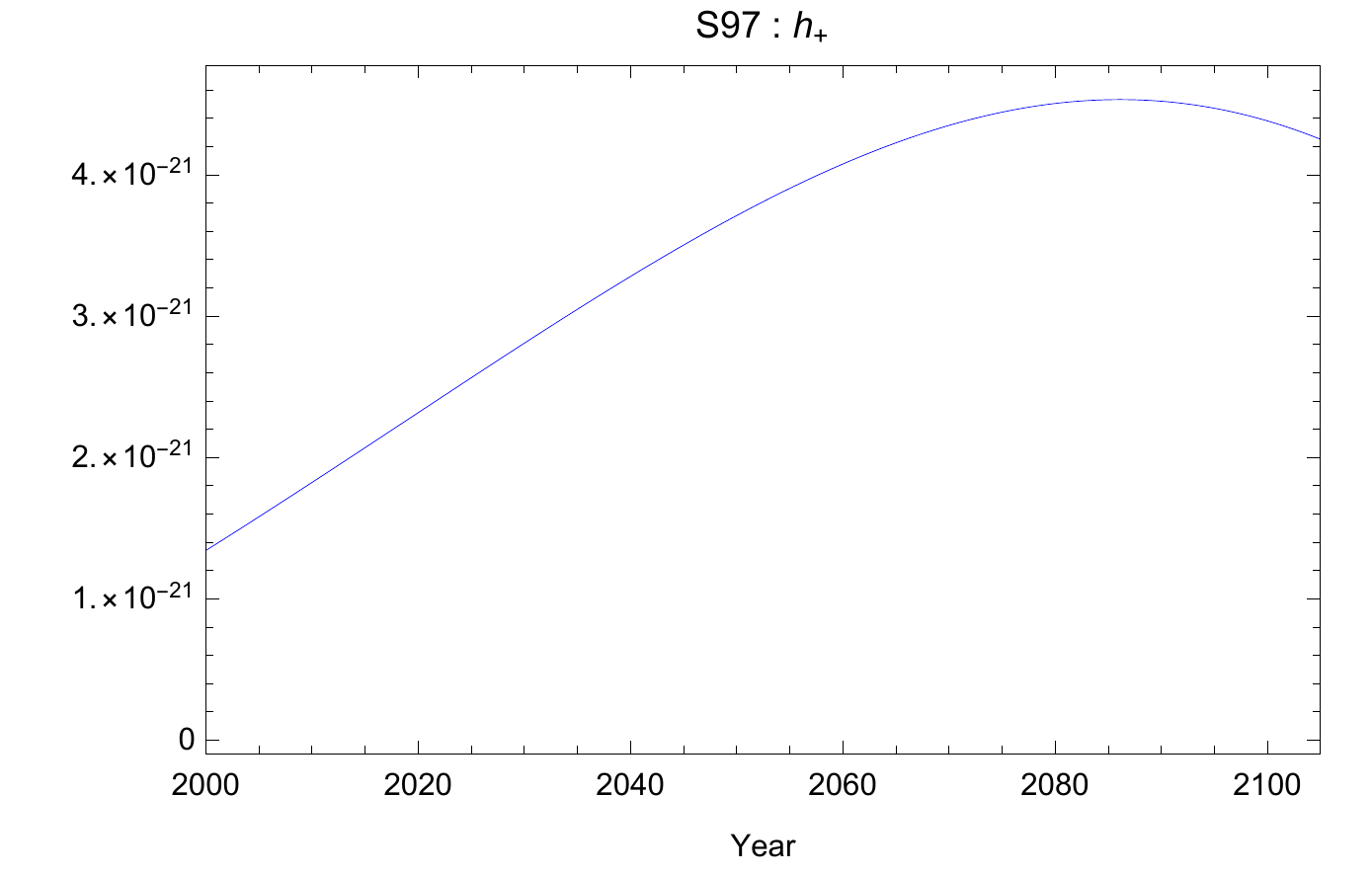}
\includegraphics[width=0.4\textwidth]{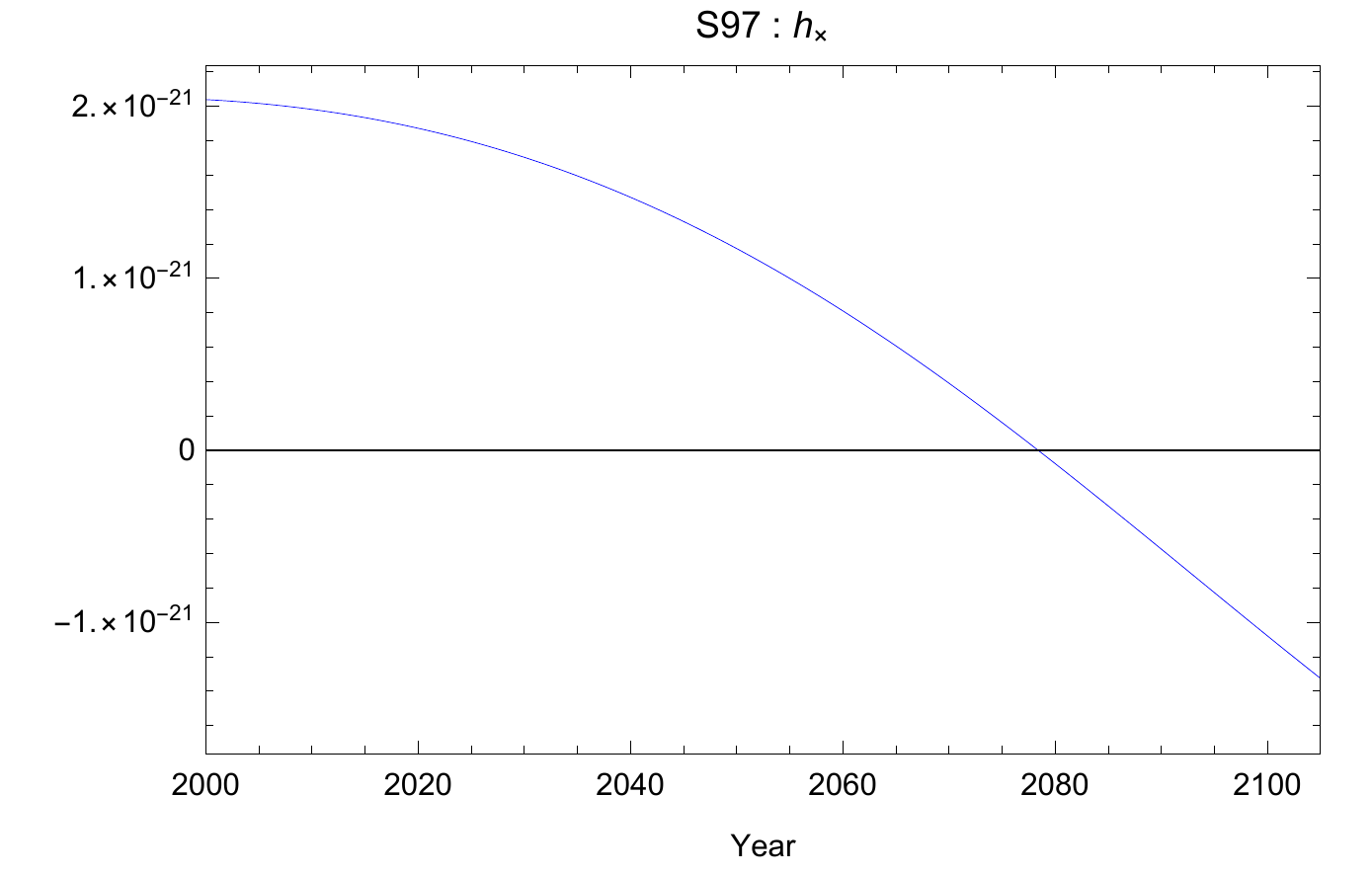}
\end{figure*} 

\begin{figure*}
\includegraphics[width=0.4\textwidth]{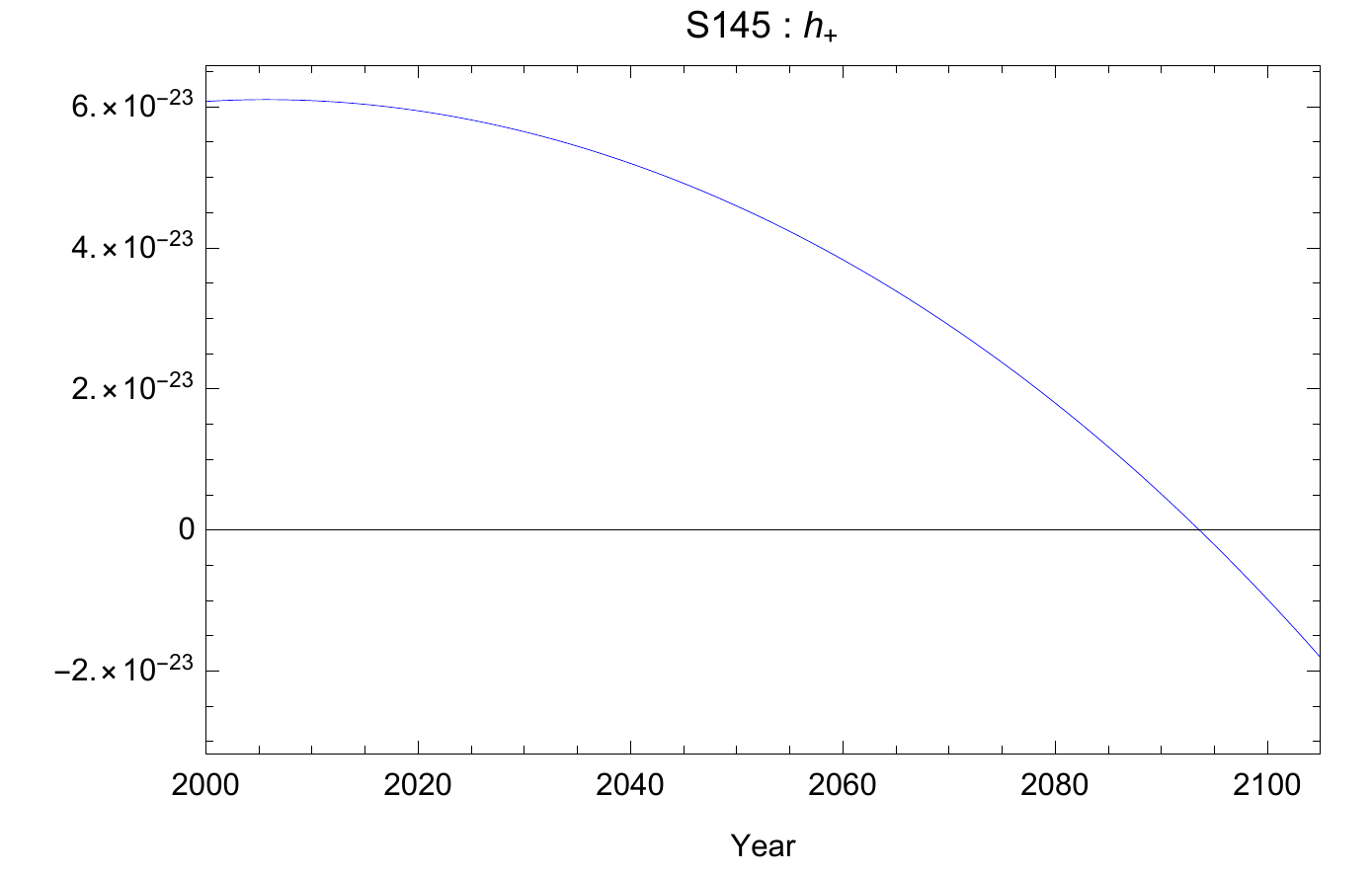}
\includegraphics[width=0.4\textwidth]{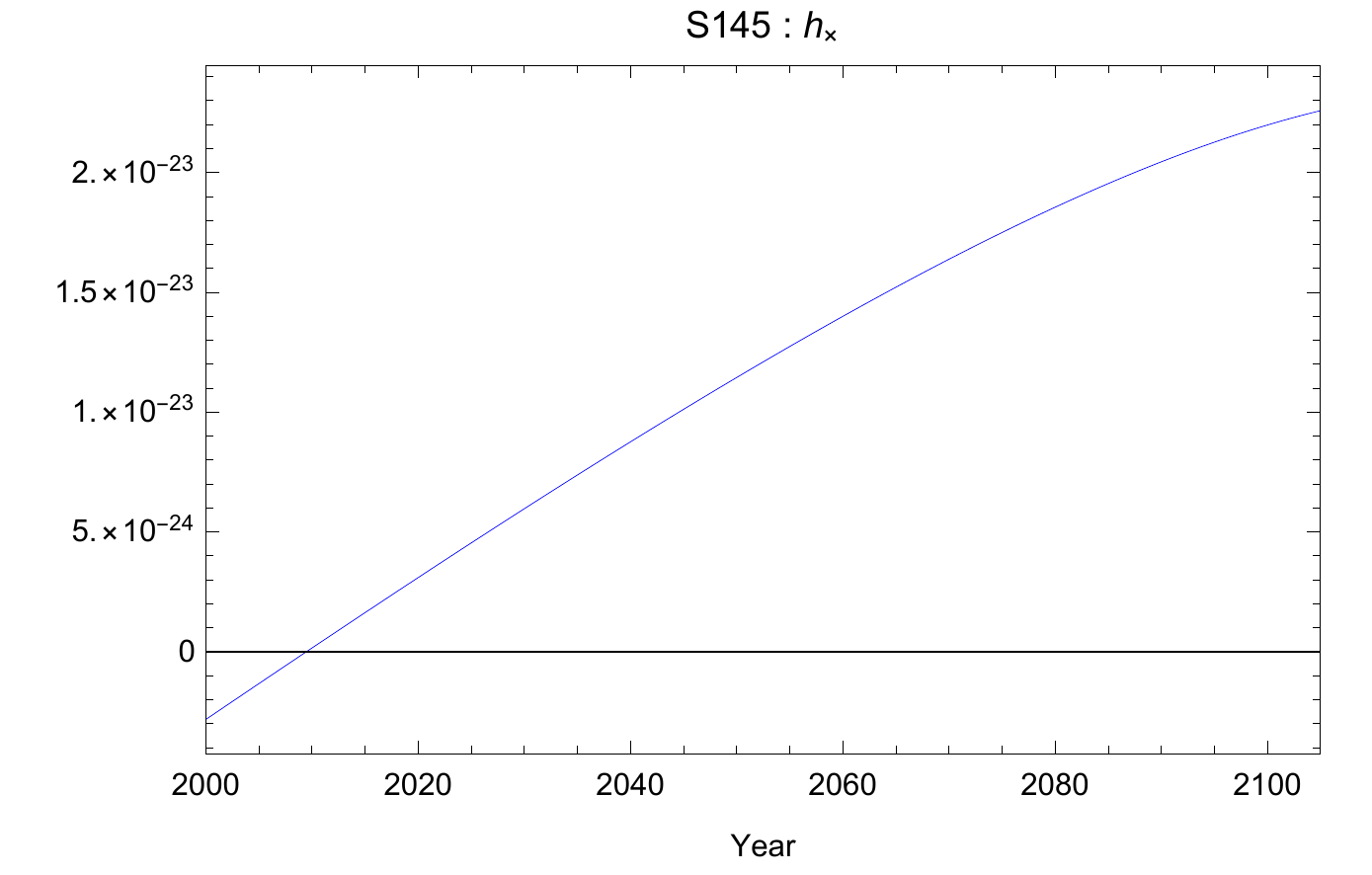}\\
\includegraphics[width=0.4\textwidth]{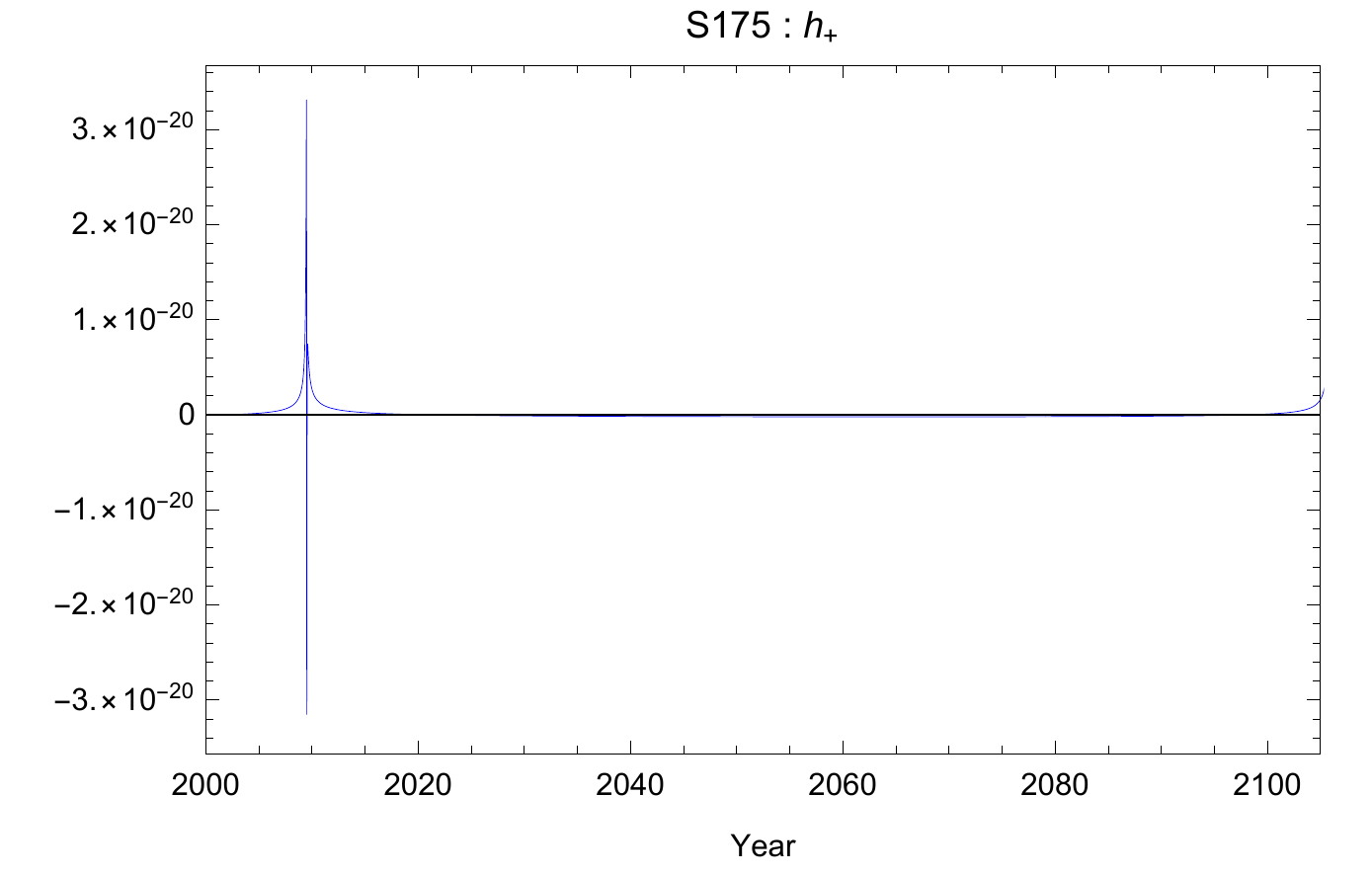}
\includegraphics[width=0.4\textwidth]{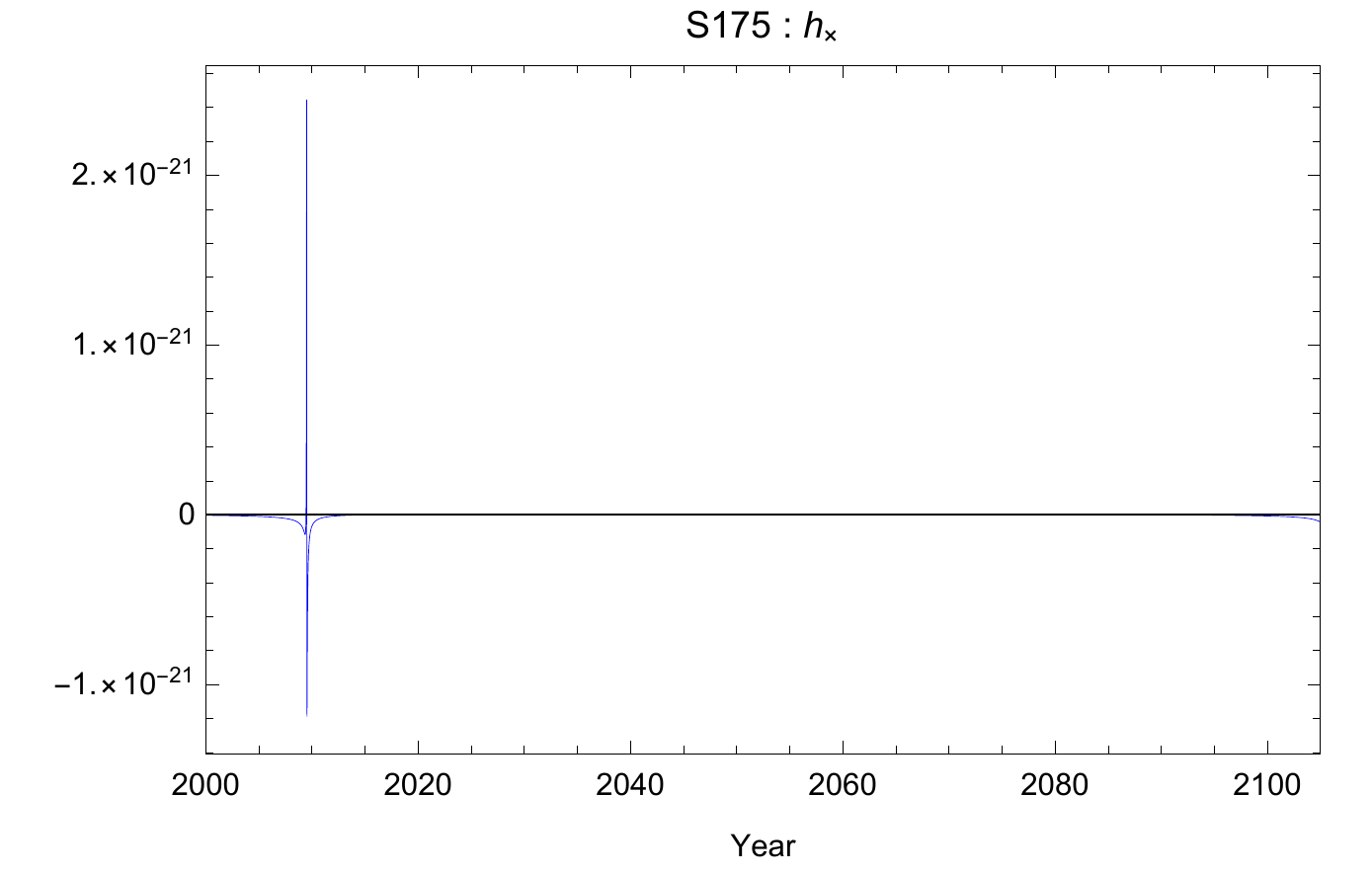}\\
\end{figure*} 

\bibliographystyle{utphys}
\bibliography{ref}

\end{document}